\documentclass[twocolumn,twocolappendix,trackchanges]{aastex631}

\newcommand{\pnt}[1]{{\scriptstyle#1}}
\usepackage{hyperref}
\usepackage{roboto}
\usepackage{threeparttable}
\usepackage{amsmath}
\usepackage{enumitem}
\usepackage{orcidlink}
\begin{document}

\title{Compact [O{\sc iii}] emission-line regions (``Green Seeds") in $\mathrm{H\alpha}$ emitters at Cosmic Noon from JWST Observations}

\author{Nuo Chen \orcidlink{0000-0002-0486-5242}}
\affiliation{Department of Astronomy, Graduate School of Science, The University of Tokyo, 7-3-1 Hongo, Bunkyo-ku,
Tokyo 113-0033, Japan; \url{nuo.chen@grad.nao.ac.jp}}
\affiliation{National Astronomical Observatory of Japan, 2-21-1 Osawa, Mitaka, Tokyo 181-8588, Japan} 

\author{Kentaro Motohara \orcidlink{0000-0002-0724-9146}}
\affiliation{Institute of Astronomy, Graduate School of Science, The University of Tokyo, 2-21-1 Osawa, Mitaka,
Tokyo 181-0015, Japan}
\affiliation{National Astronomical Observatory of Japan, 2-21-1 Osawa, Mitaka, Tokyo 181-8588, Japan}
\affiliation{Department of Astronomical Science, SOKENDAI, 2-21-1 Osawa, Mitaka, Tokyo 181-8588, Japan}

\author{Lee Spitler \orcidlink{0000-0001-5185-9876}}
\affiliation{Macquarie University Astrophysics and Space Technologies Research Centre,  Macquarie University, Sydney, NSW 2109, Australia}
\affiliation{Australian Astronomical Optics, Faculty of Science and Engineering, Macquarie University, Sydney, NSW 2113, Australia}

\author{Matthew A. Malkan \orcidlink{0000-0001-6919-1237}}
\affiliation{Department of Physics \& Astronomy, University of California, Los Angeles, 430 Portola Plaza, Los Angeles, CA 90095, USA}

\begin{abstract}
We present a rest-frame optical, spatially resolved analysis of more than 100 $\mathrm{H\alpha}$ emitters (HAEs) at $z\sim2.2$ in the ZFOURGE-CDFS field using NIRCam imaging from the JWST Advanced Deep Extragalactic Survey (JADES). The ultra-deep, high-resolution data gives us maps of the resolved emission line regions of HAEs with stellar mass ranging from $10^{8}\,M_{\odot}$ to $10^{10}\,M_{\odot}$. An [O{\sc iii}] emission-line map of each HAE is created from the flux excess in the F150W filter, leading to the discovery of a population of kiloparsec-scale compact emission line regions (``Green Seeds") with high equivalent widths ($\mathrm{EW}$). We obtain a sample of 128 Green Seeds from 68 HAEs with rest-frame $\mathrm{EW_{[O\pnt{III}]}}>200\mathrm{\AA}$. Moreover, 17 of them have extremely large $\mathrm{EW_{[O\pnt{III}]}}>1000\mathrm{\AA}$, suggesting the possible Lyman continuum (LyC) leakage from these emission line regions. Embedded within the host galaxy, many Green Seeds correspond to UV star-forming clumps and H{\sc ii} regions, indicating elevated starburst activity in them, with specific star formation rates (sSFR) several times higher than that of the host galaxy.
Based on theoretical frameworks, Green Seeds are expected to be formed through gravitational disk instability and/or galaxy mergers. Considering the stellar masses of Green Seeds, we speculate that high-mass Green Seeds may migrate toward the galactic center to build the central bulge, while low-mass Green Seeds are easily disrupted and short-lived. Besides, we propose that some Green Seeds could be the progenitors of globular clusters or ultracompact dwarf galaxies observed in the local universe.
\end{abstract}

\keywords{galaxies: structure – galaxies: high-redshift - galaxies: star-formation - galaxies: emission lines -  galaxies: evolution}
\let\clearpage\relax

\section{Introduction} 
\label{sec:intro}
The redshift $z \sim 2$, often referred to as ``Cosmic Noon", marks a critical epoch for galaxy evolution, during which the overall star formation activity was at its peak level and nearly 50\% of the cosmic stellar mass was formed \citep{Madau2014}. To further understand galaxy evolution, it is essential to characterize the spatial structures and morphologies of galaxies at Cosmic Noon, and determine how these properties evolve over cosmic time.

Before the launch of the James Webb Space Telescope (JWST), the Hubble Space Telescope (HST) was the most powerful high-resolution optical and near-infrared (NIR) imager available. Previous studies of galaxy spatial structures at high redshifts were primarily based on several well-known deep HST surveys \citep[e.g.,][]{Beckwith06, Grogin11, Brammer12}. One important finding was the prevalence of morphologically peculiar galaxies at $z > 2$ \citep[e.g.,][]{Conselice08,Conselice14}. However, due to the limited wavelength coverage and image depth of HST observations, we have not fully captured the rest-frame optical and NIR light of the galaxies at $z > 2$. The launch of JWST and the JWST Early Release Observations (EROs) open a new era in the study of galaxy spatial structure at high-redshift. For example, recent studies report a much higher fraction of elongated or disk-like morphologies in galaxies at $z > 2$ compared to previous HST observations \citep{Ferreira22, Ferreira23, Kartaltepe23, Jacobs23}. The improved sensitivity of JWST allows us to detect low surface brightness structures that were too faint to be captured by HST.

Due to their high star formation rates, a common feature of galaxies at $z \sim 2$ is strong optical emission lines, particularly $\mathrm{H\alpha}$ and [O{\sc iii}]$\lambda\lambda4959,5007$ \citep[e.g.,][]{Nakajima14, Shivaei15, Tang19, Terao2022}. These lines are powerful indicators of the physical and chemical conditions within galaxies, including their stellar populations, metallicities and ionization sources \citep[see][for a review]{Kewley19}. Over the past decade, spatially resolved analyses of emission lines have mainly been conducted through ground-based integral field unit (IFU) observation surveys \citep[e.g.,][]{ForsterSchreiber09, Wisnioski19}. Inevitably, due to limitations in their depth and resolution, these studies were restricted to galaxies with stellar mass larger than $10^{9.5}\ M_{\odot}$.

Recent observations have shown that extreme emission-line galaxies (EELGs; typically defined as $\mathrm{EW}>300\mathrm{\AA}$) exhibit clear flux excesses in broad-band photometry \citep[e.g.,][]{Malkan2017,  Forrest18,Onodera20,Terao2022,Chen24a}. The emission line fluxes extracted from these photometric flux excesses agree with those obtained from spectroscopy \citep[][hereafter C24]{Chen24a}. Notably, these EELGs tend to have lower stellar masses compared to other samples. Therefore, to achieve a comprehensive and unbiased analysis of spatially resolved emission lines at Cosmic Noon, it is essential to include these EELGs. Fortunately, with the arrival of high-resolution deep photometric data from JWST/NIRCam, it has become feasible to spatially resolve galaxy emission lines at even higher redshifts \citep[e.g.,][]{Arteaga23, Arteaga24,Shen24}. These studies successfully extracted emission line regions on a kiloparsec (kpc) scale with high EWs of several thousand angstroms, revealing intense starbursts. JWST NIRCam images also reveal that EELGs are likely undergoing strong interactions and major mergers \citep{Gupta23}. 

In this work, we present a comprehensive spatially-resolved analysis of a sample of 135 $\mathrm{H\alpha}$ emitters (HAEs) at $2.05 < z < 2.35$, using the JWST/NIRCam photometric data from the first and second data releases of the JWST Advanced Deep Extragalactic Survey \citep[JADES;][]{Eisenstein23,Rieke23,Eisenstein23b}. 
The parent HAEs are selected by the flux excesses in the deep broad-band $K_s$ filter from the FourStar galaxy evolution survey \citep[ZFOURGE;][]{Straatman16} which span a wide range of stellar mass from $10^8\,M_{\odot}$ to $10^{10}\,M_{\odot}$ \citep[][C24]{Terao2022}. From the F150W images, we identify a large number of resolved strong [O{\sc iii}] emission line regions (Green Seeds) within the galaxy structures and apply an algorithm to extract the properties of these regions.

The outline of this paper is as follows. In Section \hyperref[sec:samp]{2}, We introduce the HAEs from the ZFOURGE-CDFS field and their cross-matched sample in JADES field. Section \hyperref[sec:method]{3} describes the methodology used for extracting the resolved strong emission line regions and introduces the newly-discovered Green Seeds. Then, we present a comprehensive analysis of the properties of these Green Seeds in Section \hyperref[sec:result]{4}, compare them with other spatially-resolved structures in Section \hyperref[sec:dis1]{5}, and discuss  the potential triggers for these Green Seeds and their possible roles in galaxy evolution in Section \hyperref[sec:dis2]{6}. Finally we summarize our results and conclusions in Section \hyperref[sec:con]{7}. Throughout this paper, we adopt the AB magnitude system \citep{Oke83}, a $\mathrm{\Lambda CDM}$ cosmology with $H_0 = \mathrm{70\ km\ s^{-1}\ Mpc^{-1}}$, $\Omega_m = 0.3$, and $\Omega_{\Lambda} = 0.7$.

\section{Data and Sample Selection}
\label{sec:samp}
\subsection{The HAEs in ZFOURGE survey}
The full sample of HAEs used in this work is drawn from the C24 catalog, which comprises 1318 HAEs at $2.05 < z < 2.5$ from the ZFOURGE survey \citep{Straatman16}. These HAEs are included in three well-known extragalactic fields covered by ZFOURGE: the GOODS-South, COSMOS, and UDS fields, which overlap with the CANDELS survey \citep{Grogin11, Koekemoer11}. The survey areas of these fields are 128, 135, 189 $\mathrm{arcmin^2}$ respectively. 

In C24, each HAE was identified from its flux excess observed in the ZFOURGE-$K_s$ broad-band filter, relative to the stellar continuum estimated from the best-fit spectral energy distribution (SED) to the remaining filter photometry. The redshifts of HAEs are initially based on the photometric redshift ($z_{phot}$) in the ZFOURGE catalog, and then refined by incorporating spectroscopic redshifts ($z_{spec}$) from the MOSDEF survey \citep{Kriek15} and grism redshifts ($z_{grism}$) from the 3D-HST Emission-Line Catalogs \citep{Momcheva16}. The C24 catalog also includes important integrated galaxy properties, such as stellar mass ($M_*$), star formation rate (SFR), and dust attenuation ($A_V$). These HAEs cover the stellar mass from $10^8\,M_{\odot}$ to $10^{10}\,M_{\odot}$, and many of them are very low-mass, analogous to the Lyman-$\alpha$ emitters (LAEs) at similar redshift \citep{Nakajima14}, and ``Green Pea” galaxies in the local universe \citep{Cardamone09,Amorin10,Izotov11,Yang17, Li2018}.

\subsection{JADES imaging data}
\begin{figure}[t]
\includegraphics[width=1\linewidth]{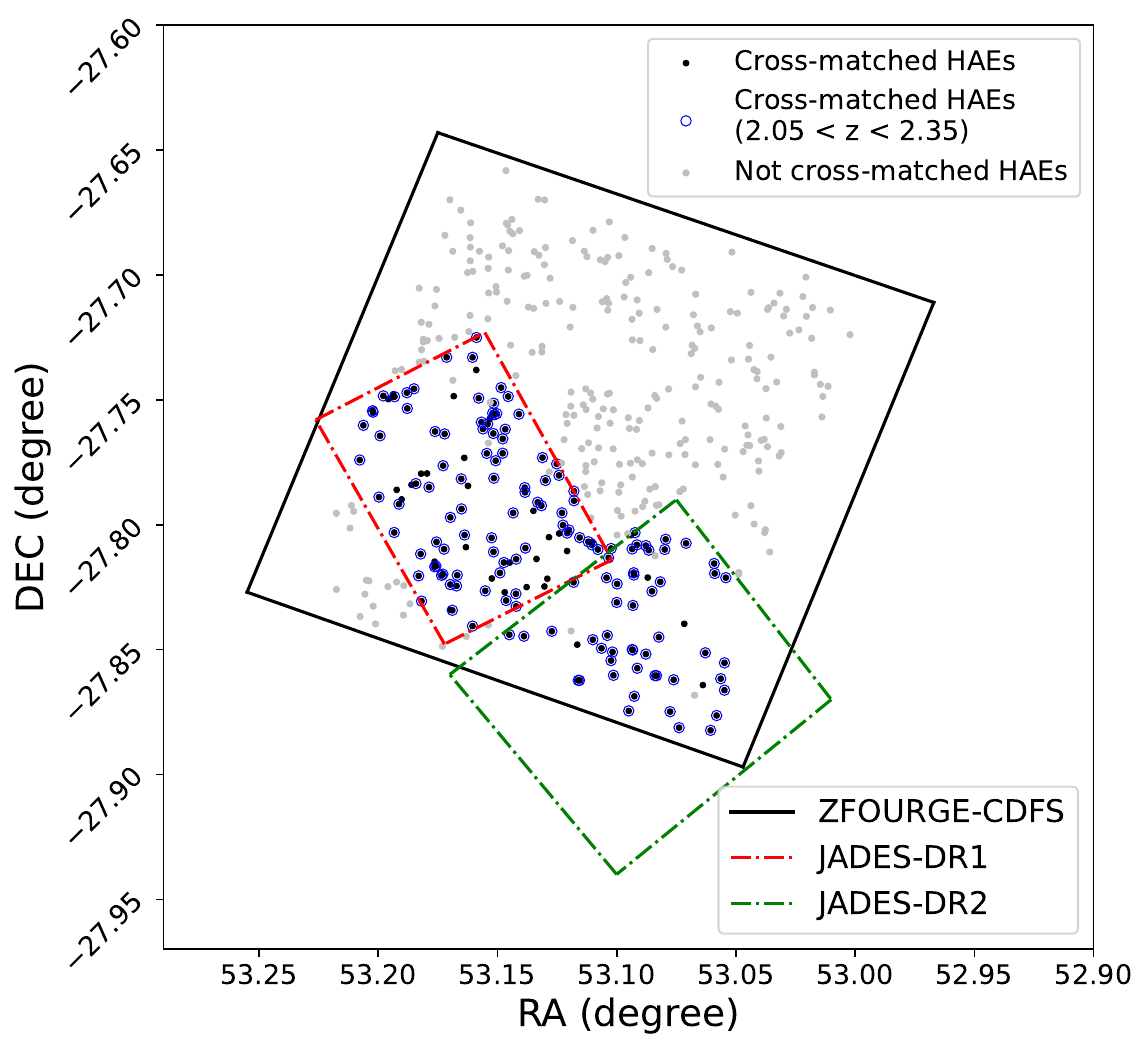}
    \label{fig:obs}
    \vspace{-0.3cm}
    \caption{Sky locations of the ZFOURGE survey and the JADES survey in the GOOD-South field. The ZFOURGE footprint covers an area of around $13'\times13'$, while JADES has a total coverage of 67.7 $\mathrm{arcmin^2}$. We make a cross-match between the C24 catalog, which contains HAEs at $2.05 < z < 2.5$, and the JADES catalog \citep{Rieke23}. The 170 cross-matched HAEs are initially marked as black dots, with the remaining HAEs shown as grey dots. Within the JADES footprint, seven HAEs are lost in detector gaps and artefacts around bright stars. Finally, the 135 HAEs are selected at $2.05 < z < 2.35$, where the [O{\sc iii}] emission line falls in the F150W filter, are highlighted with blue circles.}
\end{figure}

One of the aforementioned extragalactic fields: the GOODS-South field \citep{Giacconi02, Giavalisco04} was observed by the Cycle 1 JADES survey \citep{Rieke23, Eisenstein23b}, which includes infrared imaging with nine broad-bandwidth filters from JWST/NIRCam. With exposure times exceeding 10 hours for each filter, JADES provides the deepest ever near-infrared view of this field, reaching a $5\sigma$ depth of approximately $\mathrm{30\,AB}$ magnitude in each filter. JADES First and Second Data Release (DR1 \& DR2) provide mosaics images over a total area of 67.7 $\mathrm{arcmin^2}$, including the F090W, F115W, F150W, F200W, F277W, F335M, F356W, F410M, and F444W filters. The footprint of JADES and ZFOURGE-CDFS fields is presented in Figure \hyperref[fig:obs]{1}. Additionally, JADES released a 23-band space-based photometric catalog of 94,000 distinct objects, all PSF-matched to the F444W images.

In this work, we focus on the spatially-resolved structure of the HAE sample. Therefore, we utilize the F090W, F115W, F150W, F200W, F277W imaging, which provides better angular resolution. The empirical PSF FWHM of the F277W filter is $0.''092$.

\subsubsection{Astrometric correction}
The JADES images were astometrically corrected by registering bright images to the Guide Star Catalogue coordinates from Gaia DR2 \citep{Gaia18}, known as GSC 2.4. The astrometry of ZFOURGE GOODS-South images agrees almost perfectly with that of the CANDELS HST images \citep{Grogin11,Koekemoer11}, which were registered to the Guide Star Catalog II \citep[GSC 2.3;][]{Lasker08}. The different versions of astrometric reference stars lead to astrometric errors reported to be $0.''2-0.''7$, according to the Mikulski Archive for Space Telescopes (MAST).

To correct for the differences in astromertic alignment and to update the CANDELS and ZFOURGE catalogs to the newest astrometry, we cross-match the objects in the CANDELS catalog \citep{Guo13} and the JADES catalog with a maximum angular separation of $1''$. This results in 17,394 cross-matched objects. We measure a median deviation in right ascension ($\mathrm{RA_{JADES}-RA_{CANDELS}}$) of $\mathrm{\Delta\,RA_{med} = 0.''107}$; and a median deviation in declination ($\mathrm{DEC_{JADES}-DEC_{CANDELS}}$) of $\mathrm{\Delta\,DEC_{med}= -0.''244}$. A similar result is obtained from cross-matching the ZFOURGE catalog with the JADES catalog, with $\mathrm{Delta\,RA_{med} = 0.''122}$ and $\mathrm{\Delta\,DEC_{med}= -0.''247}$ from 12,375 matched objects.

Based on these results, we correct all the right ascensions by $0.''12$ (4 pixels in JADES image) and all declinations by $-0.''24$ (8 pixels in JADES image) for the original coordinates of the HAE sample in the GOODS-South field. 

\subsection{Sample selection}
\label{sec:sampselect}
\begin{figure*}[hbt!]
    \centering
    \includegraphics[width=0.29\textwidth]{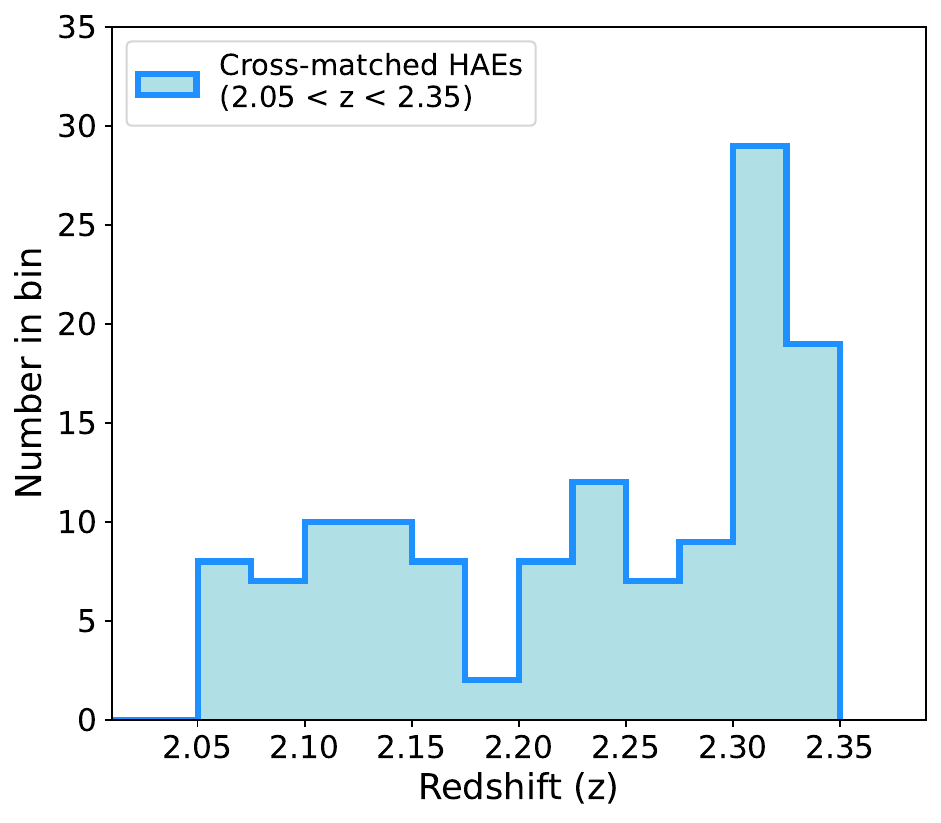}
    \includegraphics[width=0.4\textwidth]{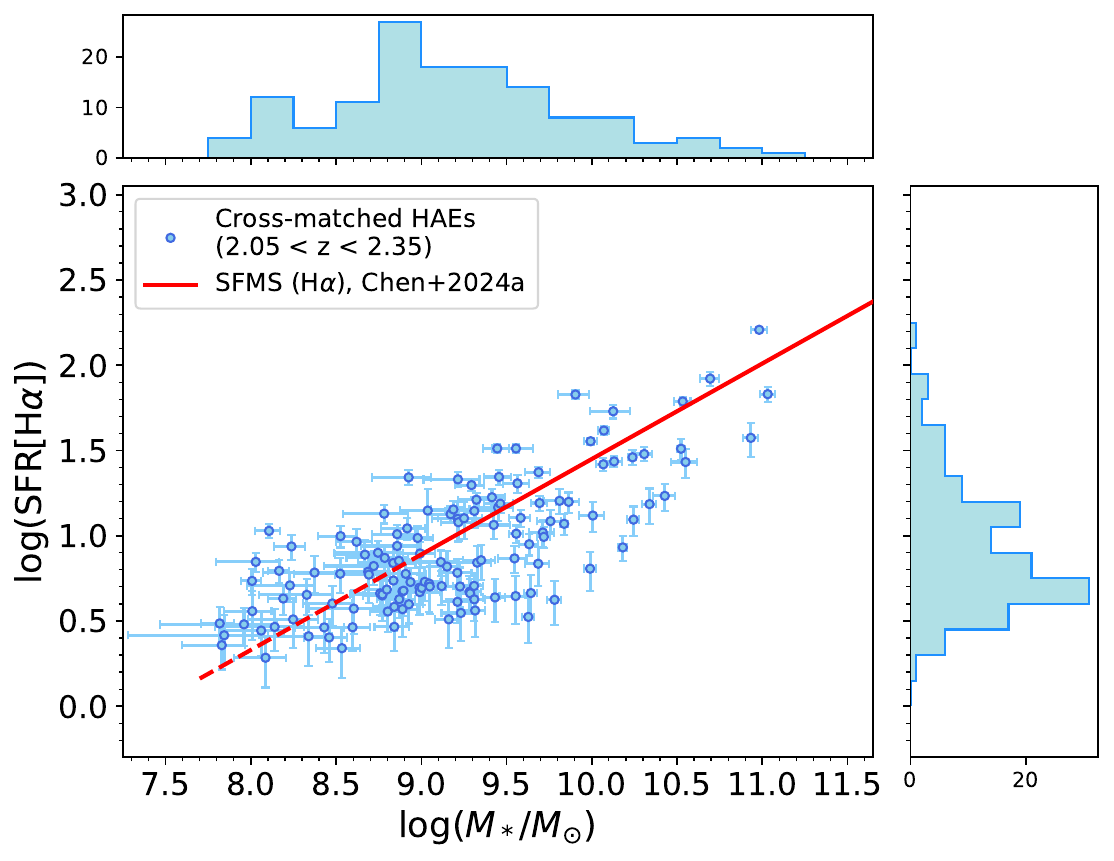}
    \includegraphics[width=0.29\textwidth]{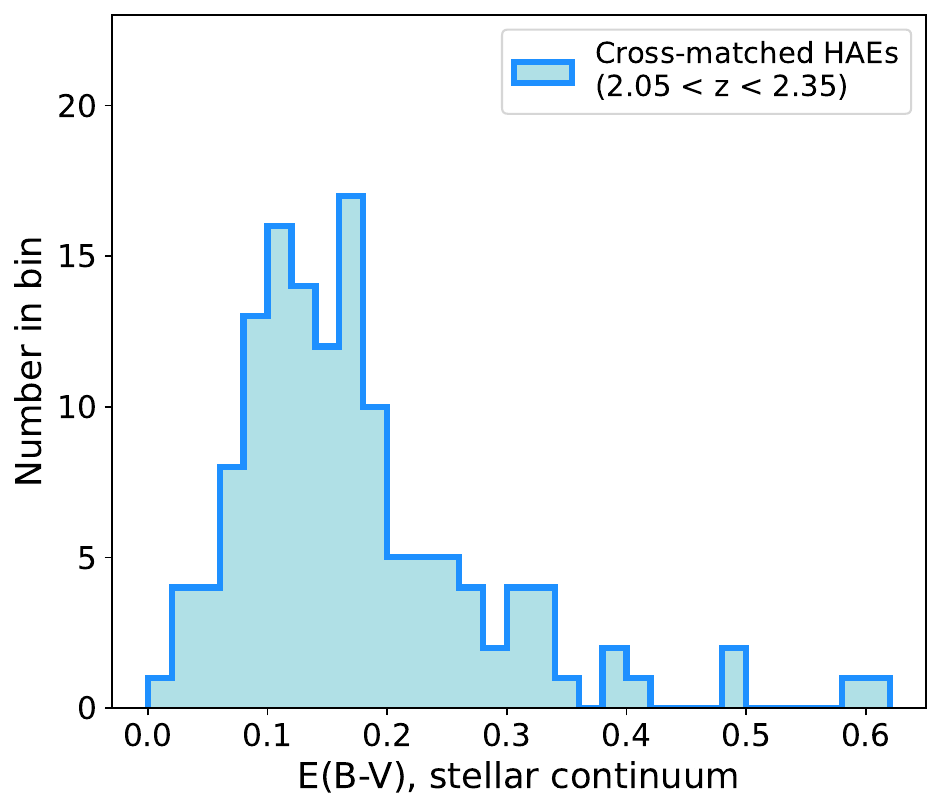}
    \label{fig:samp}
    \caption{\textbf{Left}: Redshift distribution of the 135 JADES cross-matched HAEs from C24. The median redshift of the HAEs is $z_{med} = 2.23$. \textbf{Middle}: The star formation main sequence (SFMS) of the HAEs in the ZFOURGE fields, based on the $\mathrm{H\alpha}$ emission line. The HAEs are represented by blue circles. The red solid line is the best linear fit of the $M_*$-SFR relation for the whole HAE sample from C24 with stellar mass larger than $10^9\,M_{\odot}$ (the mass completeness limit of the ZFOURGE survey). It is extrapolated to the lower mass regime with a red dashed line. \textbf{Right}: Distribution of stellar continuuum dust reddening of the HAEs. The layout is the same as in the left panel. The HAEs have a median gas reddening of $E(B-V)_{cont}=0.15$.}
    \vspace{0.3cm}
\end{figure*}

We cross-match between the C24 and JADES catalogs \citep{Rieke23}, requiring an angular separation of less than $0.''5$ and the availability of photometric data in all the F090W, F115W, F150W, F200W, and F277W filters. We find that 170 HAEs meet these criteria, with a median separation of $0.''06$. These HAEs are marked as black dots in Figure \hyperref[fig:obs]{1}. Given that the JADES images provide the deepest near-infrared view of this field, each HAE is expected to have a corresponding detection in the JADES catalog. However, within the JADES DR1 \& DR2 footprint (see Figure \hyperref[fig:obs]{1}), we find that three of these HAEs (ZF-14229, 19965, 23087) fall between the detector gaps of the F200W image, and four others (ZF-7627, 10297, 18528, 21984) overlap with the hexagon-shaped artefacts around bright stars in the JADES images, which were masked when processing photometry.

For our HAE sample, the [O{\sc iii}] emission line at $2.05 < z < 2.35$ is redshifted into the F150W filter, while the $\mathrm{H\alpha}$ emission line at $2.05 < z < 2.40$ is redshifted into the F200W filter. To spatially resolve these two emission lines simultaneously, we therefore require that HAEs fall within the redshift range of $2.05 < z < 2.35$. This results in a final sample of 135 HAEs, which are the primary targets of this work. Their sky distributions are added with blue circles around the corresponding black dots in Figure \hyperref[fig:obs]{1}.

In Figure \hyperref[fig:samp]{2}, we show a histogram of redshifts,
distribution of HAEs in star formation main sequence (SFMS: correlation betweeen SFR and stellar mass), and a histogram of best-fit stellar reddenings. Their median redshift is $z_{med} = 2.23$, along with a distribution peaks at $z\sim2.3$, which may be due to clustered regions in the field. These galaxies are scattered widely around the usual SFMS relation, showing only a slight elevation in SFR($\mathrm{H\alpha}$) from the SFMS, less than an average of 0.05 dex in the low-mass ($<10^9\,M_{\odot}$) regime. The median stellar continuuum dust reddening is $E(B-V)_{cont}=0.15$ mag, with only 15 of 135 galaxies exhibiting high dust attenuation of $E(B-V)_{cont}>0.3$ mag.

\begin{figure*}[hbt!]
    \centering
    \includegraphics[width=0.98\textwidth]{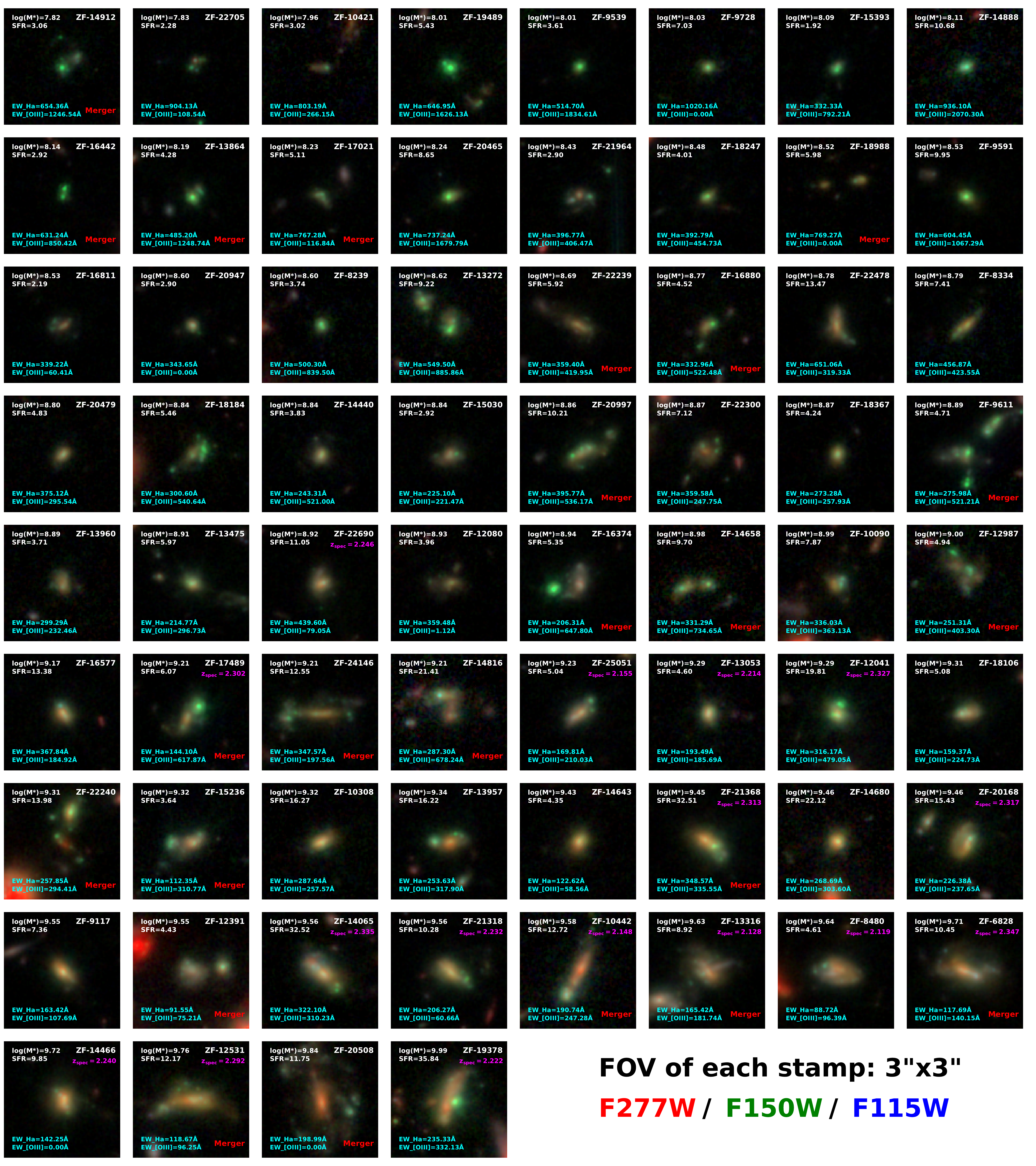}
    \label{fig:rgb_main}
    \caption{JWST/NIRCam cutout images of 68 selected HAEs, each with a size of 3” × 3”. The RGB images are built by combining the F115W (B), F150W (G), and F277W (R) PSF-matched images to the F277W filter. For each filter, cutouts with dimensions of 101 × 101 pixels (0.03” pixel scale) are made. The ZFOURGE ID \citep{Straatman16} of each cross-matched HAE is labeled at the upper-right corner of each cutout and the HAEs with spectroscopic redshift measurements from MOSDEF \citep{Kriek15} or 3D-HST \citep{Momcheva16} are listed below. The SED-derived integrated galaxy properties: total stellar mass, SFR, and the integrated EW from C24, are also labeled here. These cutout images are ordered by stellar mass, from the lowest to the highest. Notably, these HAEs exhibit several compact [O{\sc iii}]+$\mathrm{H\beta}$ regions (green substructures) embedded within the galaxies. In Section \hyperref[sec:dis2]{6.1}, we conduct a non-parametric morphological analysis of merger systems, with the identified mergers listed in the bottom-right corner.}
\end{figure*}

Figure \hyperref[fig:rgb_main]{3} displays cutout color images from a selection of the HAEs, created by combining the F115W, F150W, and F277W images. Point-spread functions (PSF) of the shorter wavelength images are matched to that of the F277W image with $0.''03$ per pixel (See Section \hyperref[sec:method]{3} for details). These three filters directly track the rest-frame blue, green, and far-red colors of the galaxies, roughly corresponding to $b$, $g$ and $z$ filters in the rest frame. Notably, the composite color images reveal a prominent number of resolved compact green components, representing strong [O{\sc iii}]+$\mathrm{H\beta}$ emission line regions with high $\mathrm{EW}$ more than several hundreds of angstroms \citep{Cardamone09}, and indicating extreme interstellar medium (ISM) properties. In the next section, we demonstrate our unique methodology for extracting and quantitatively analyzing these [O{\sc iii}]+$\mathrm{H\beta}$ regions. Note that the remaining HAEs that do not contain such a substructure in [O{\sc iii}]+$\mathrm{H\beta}$ regions are displayed in Appendix \hyperref[sec:noo3apx]{A}.

\section{Methodology}
\label{sec:method}
\begin{figure*}[hbt!]
    \centering
    \includegraphics[width=0.9\textwidth]{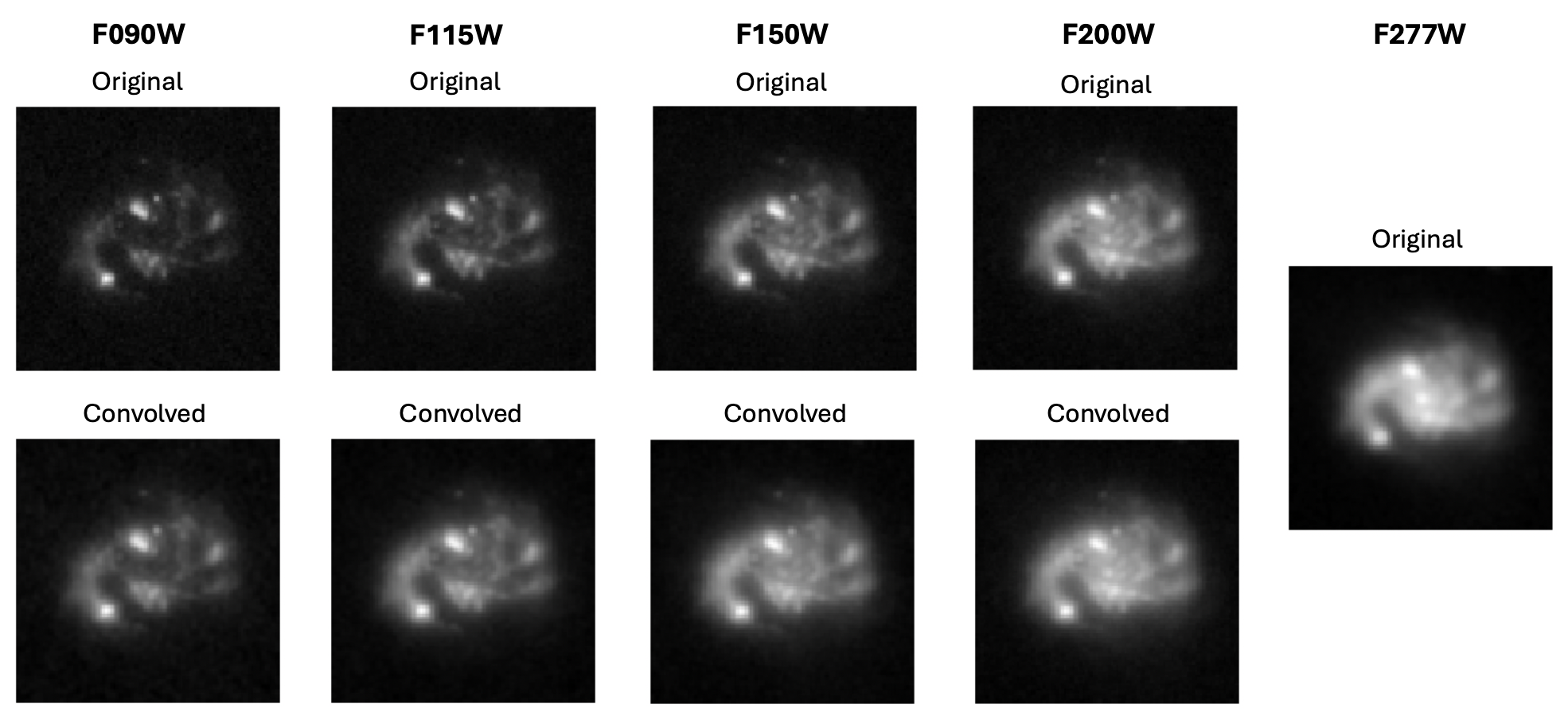}
    \label{fig:psf}
    \caption{Cutout images of a cross-matched HAE imaged in the F090W, F115W, F150W, F200W, and F277W filters. The top row displays the observed images at their original resolution, i.e., direct cutout stamps from the JADES images, while the bottom row shows the images convolved with a kernel that matches the PSF of the F277W filter.}
    \vspace{0.4cm}
\end{figure*}

We produce $3'' \times 3''$ ($101 \times 101$ NIRCAM pixel) cutout images of each cross-matched HAE from the F090W, F115W, F150W, F200W, and F277W images. The central pixel of each galaxy is determined as the source position of the JADES catalog from the \robotoThin{Sextractor} \citep{Bertin96,Barbary2016}. We match the PSFs of all images to that of the F277W image using the \robotoThin{WebbPSF} Python package \citep{Perrin12,Perrin14}. \robotoThin{WebbPSF} produces simulated PSFs for the selected JWST instrument and filter. 
We set the target model PSF as that of the F277W image, and then derive a convolution kernel for the other bands using the \robotoThin{scikit-image} Python package \citep{van2014scikit} through the Wiener filtering method. In Figure \hyperref[fig:psf]{4}, we present a sample galaxy (ZF-12763) from the cross-matched HAEs to visualize the PSF-matching process. The top row displays original images from NIRCam, and the bottom images convolved.

\subsection{Resolved continuum and emission line map}
Based on the best-fit SEDs of HAEs at $z\sim2$ from C24, we discover that strong optical emission lines, such as [O{\sc iii}] and $\mathrm{H\alpha}$, can cause noticeable flux excesses in broad-band photometric data, specifically in the F150W and F200W images in our sample. On the other hand, other broad-band photometric data can roughly trace the stellar continuum because there is no inclusion of strong emission lines. Therefore, we assume that the F115W and F277W photometric data represent the stellar continuum and that a linear relationship of the stellar continuum between them can be applied.

\begin{figure*}[hbt!]
    \centering
    \includegraphics[width=1\textwidth]{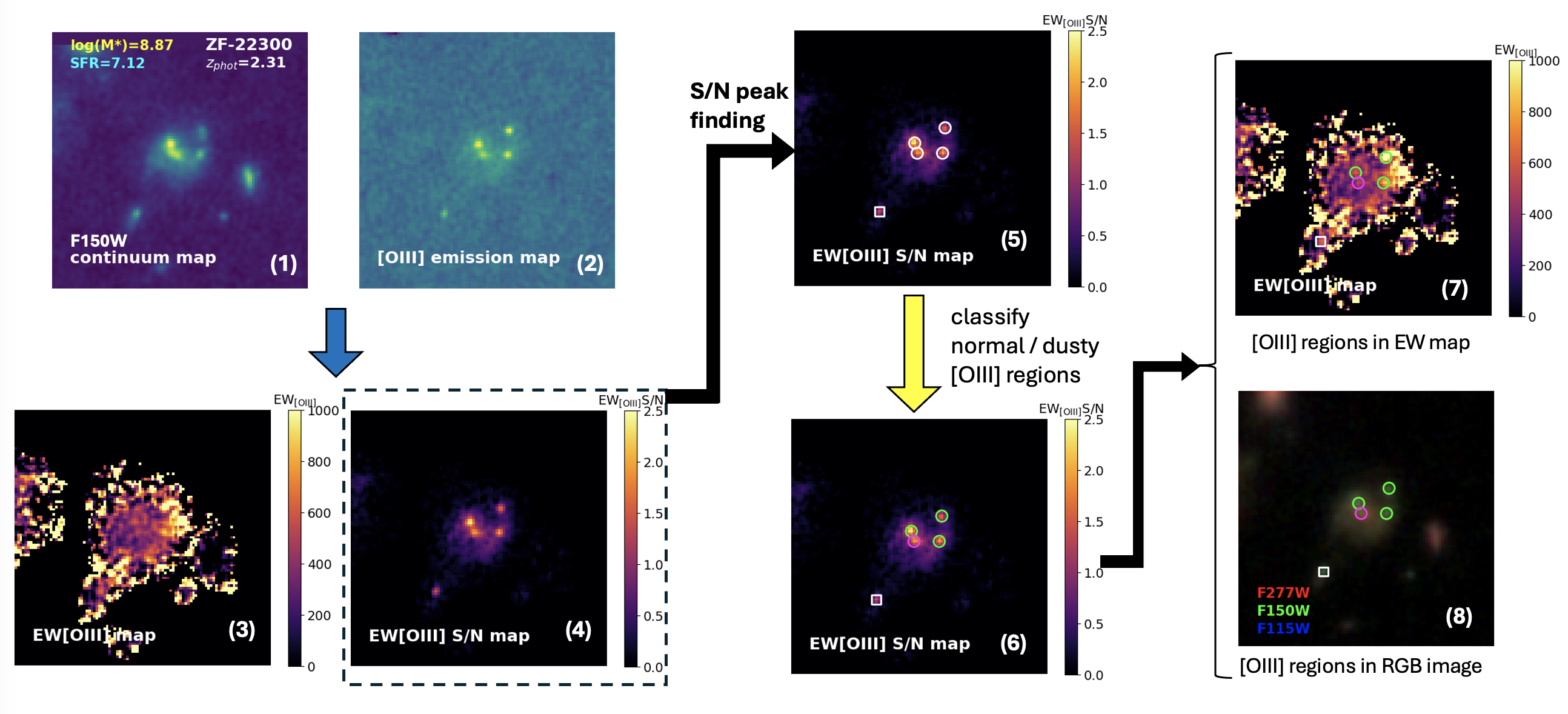}
    \label{fig:map}
    \vspace{-0.4cm}
    \caption{The flowchart of our methodology to identify ``Green Seeds" in a sample galaxy. The stellar continuum map at $1.5\,\mu m$ (1) is derived by assuming a linear relationship between the stellar continuum in the F115W and F277W filters. The residual image obtained by subtracting this continuum map from the F150W image is the [OIII] emission line map (2) in the unit of njy/pixel. From the continuum and emission line images, we generate the EW map (3) and the EW S/N map (4). The EW S/N map reveals several bubble-like structures with high $\mathrm{EW}$. We apply a peak finding algorithm to the EW S/N map to extract these resolved [O{\sc iii}]+$\mathrm{H\beta}$ emission line regions, requiring $\mathrm{S/N > 2.5}$,
    which are indicated by white circles in panel (5). White squares represent regions with $\mathrm{S/N < 2.5}$. We then classify the green (normal) and red (dusty and/or older) [O{\sc iii}]+$\mathrm{H\beta}$ regions through an SED-based color diagram, labeling the former ones as Green Seeds due to their appearance in the RGB images. Green Seeds are highlighted as green circles in panel (6), while Red Seeds are shown as magenta circles. Panel (7) \& (8) display these Green Seeds on the EW map and the RGB image, respectively.}
    \vspace{0.4cm}
\end{figure*}

At an arbitrary wavelength ($\lambda$) between F115W and F277W, the stellar continuum can be expressed as follows:
\begin{equation}
\label{equ:cont}
    \begin{aligned}
    f_{\lambda, cont}\, = & f_{F115W} + (f_{F277W} - f_{F115W}) \\ & \times \frac{\lambda\,(\mu m) - 1.15\,\mu m}{2.77\,\mu m  - 1.15\,\mu m},
    \end{aligned}
\end{equation}
where $f_{F115W}$ and $f_{F277W}$ are the photometric fluxes in the F115W and F277W filters, respectively. In most cases, the observed F150W and F200W photometric fluxes are higher than the stellar continuum at the corresponding wavelengths from Equation \hyperref[equ:cont]{1} due to strong optical emission lines like [O{\sc iii}] and $\mathrm{H\alpha}$.
This assumption is also applicable in the resolved view. Using the cutout images from F115W and F277W, it is possible to construct a pixel-by-pixel stellar continuum map for the F150W filter using Equation \hyperref[equ:cont]{1}, as shown in panel (1) of Figure \hyperref[fig:map]{5}. 

Since the strong optical emission lines can lead to significant flux excesses, it is reasonable to assume that:
\begin{equation}
\label{equ:emi}
    F_{\mathrm{EL}} \, S(\lambda_{\mathrm{EL}}) =  (f_{\lambda, \mathrm{obs}} - f_{\lambda, \mathrm{cont}}) \int_{\lambda_1}^{\lambda_2}  S(\lambda) \, d\lambda,
\end{equation}
where $F_{EL}$ represents the emission line flux fall in the filter, $f_{\lambda, obs}$ is the observed photometric data, and $f_{\lambda, cont}$ is the stellar continuum derived from Equation \hyperref[equ:cont]{1}. $S(\lambda)$ is the throughput function of the filter, which varies with wavelength, and $S(\lambda_{\mathrm{EL}})$ is the exact throughput where the emission line drop in.
In this work, we also apply Equation \hyperref[equ:emi]{2} to derive the resolved continuum-subtracted emission line map. For example, after constructing the stellar continuum map at $1.5\,\mu m$, we subtracted it from the observed F150W image. This resulting residual image represents the emission line map of [O{\sc iii}] (and $\mathrm{H\beta}$), which falls in the F150W filter at $z\sim2$. As shown in the panel (2) of Figure \hyperref[fig:map]{5}, we present the [O{\sc iii}]+$\mathrm{H\beta}$ emission line map of a sample galaxy. 
Note that, in Section \hyperref[sec:result]{4}, we apply statistical corrections to account for contamination from the weaker $\mathrm{H\beta}$ emission line in each emission line region.

\begin{figure*}[hbt!]
    \centering
    \includegraphics[width=0.49\textwidth]{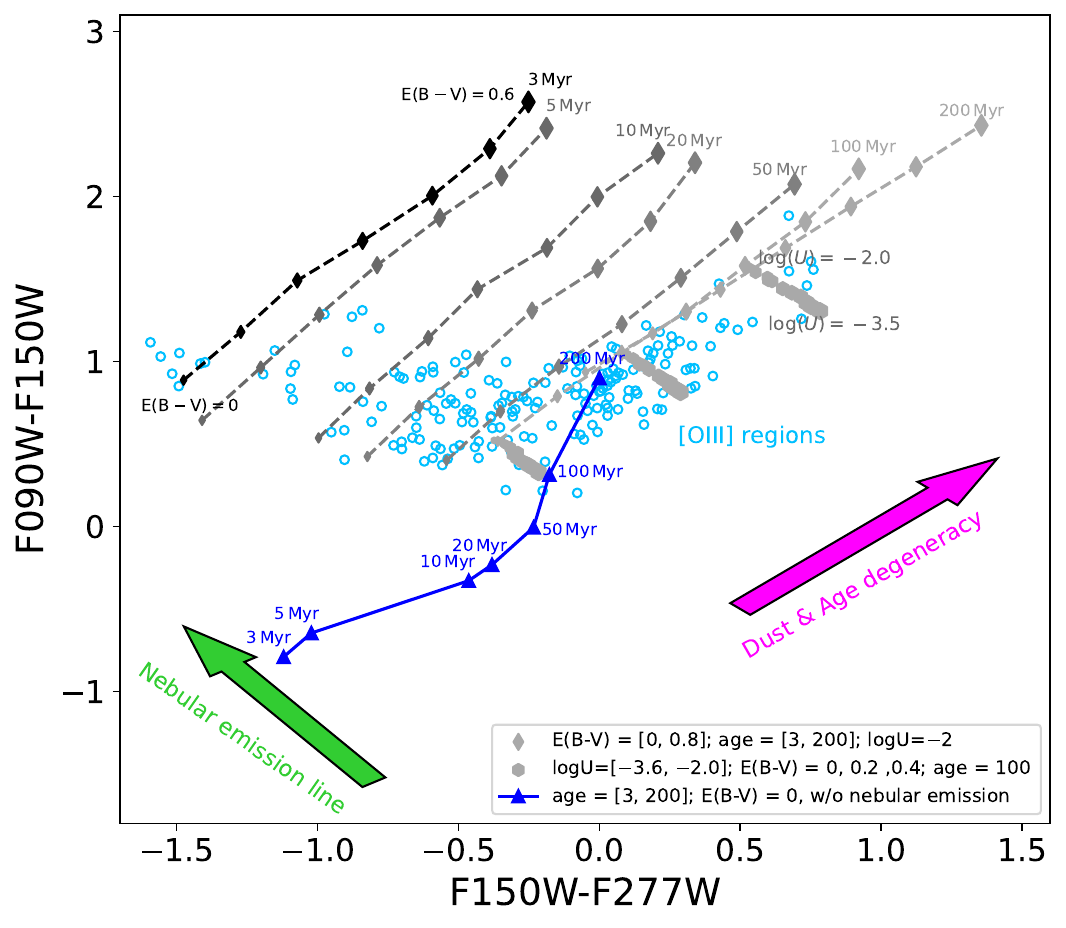}
    \includegraphics[width=0.48\textwidth]{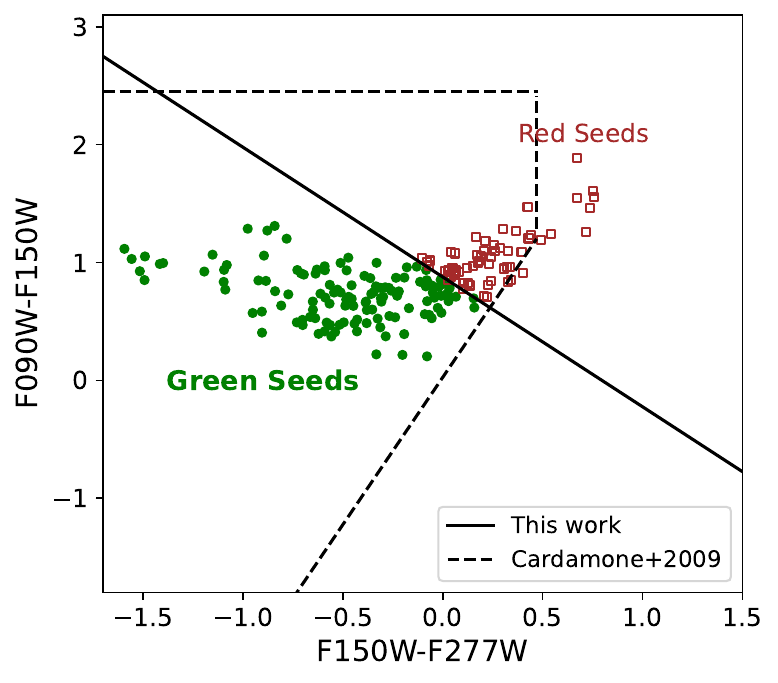}
    \label{fig:sedcolor}
    \vspace{-0.2cm}
    \caption{\textbf{Left}: The $\mathrm{F150W - F277W}$ versus $\mathrm{F090W - F150W}$ color-color diagram for the 187 [O{\sc iii}]+$\mathrm{H\beta}$ regions. A series of SED model grids are overlaid as varying symbols. The blue triangle represent the stellar continuum of a burst SFH from 3 to $\mathrm{200\,Myr}$ without emission lines and dust attenuation. Then, emission-line templates from CLOUDY \citep{Ferland98,Ferland13} and dusty attenuation are added to these pure SED models. The diamonds represent SED models with varying stellar ages and dust attenuation at a fixed ionization parameter of $\mathrm{log}\,U=-2$. The sizes of the diamond grids increase with dust attenuation from $E(B-V)=0$ to 0.8, as labeled. The grayscale of the diamond grids indicates the stellar ages of the SED models, with black representing $\mathrm{3\,Myr}$ and silver representing $\mathrm{200\,Myr}$. The hexagons show SED models with varying ionization parameters and dust attenuation at a fixed stellar age of $\mathrm{100\,Myr}$. From the bottom-left to upper-right, the grids have dust attenuation levels of $E(B-V)=0, 0.2, 0.4$. The sizes of the hexagon grids increase with decreasing ionization parameter from $\mathrm{log}\,U=-2$ to $-3.6$, evolving from upper-left to bottom-right, as labeled.
    \textbf{Right}: The color selection derived from the linear fitting of these SED model grids at a fixed $E(B-V)=0.15$ and $\mathrm{100\,Myr}$ with varying $\mathrm{log}\,U$ is shown as the black solid line.
    On the bottom-left side of this color selection, we classify 128 [O{\sc iii}]+$\mathrm{H\beta}$ regions with young age and little dust, and designate them as ``Green Seeds", referencing the local Green Pea galaxies \citep{Cardamone09}. These Green Seeds are marked as green filled circles. Red [O{\sc iii}]+$\mathrm{H\beta}$ regions are located on the other side, shown as brown open squares, named as ``Red Seeds". These regions are highly dust-attenuated and/or contain older stellar population. For comparison, the $r-z$ vs. $u-r$ color selection from \citep{Cardamone09} is also shown here, after being redshifted to z=2.2 and corrected from the SDSS magnitude to AB magnitude.}
    \vspace{0.3cm}
\end{figure*}

Since the prominent green color in the JWST RGB images results from high $\mathrm{EW_{[O\pnt{III}]+H\beta}}$, we apply an EW-limited selection to identify these regions; using the resolved stellar continuum map and emission line map, we can further extract the resolved equivalent width (EW) map for each HAE. The rest-frame EW map is constructed as follows:
\begin{equation}
\label{equ:ew}
    EW\, = \frac{F_{\mathrm{EL}}}{f_{\lambda, cont}}\times\frac{\Delta\lambda_{\mathrm{filter}}}{1+z},
\end{equation}
where $z$ is the redshift of the HAE, and $\Delta\lambda_{\mathrm{filter}}$ is the bandwidth of the corresponding filter, which are  $\Delta\lambda_{\mathrm{F150W}} \sim 3180\mathrm{\AA}$ and $\Delta\lambda_{\mathrm{F200W}} \sim 4630\mathrm{\AA}$.
Panel (3) of Figure \hyperref[fig:map]{5} shows the rest-frame EW map of a sample galaxy, masked by a segmentation map from the JADES data release \citep{Rieke23}. However, even after masking with the segmentation map, the EW map is still dominated by noise patterns, especially in the outer regions where the stellar continuum is quite low. 

To identify and select regions with sufficiently high underlying stellar continuum, we refer to the EW S/N map. In order to construct the EW S/N map, we first obtain the standard deviation of the background pixels. In this work, the pixel-by-pixel error map is referred from the ``ERR" uncertainty extensions of the JADES images \citep{Rieke23}.
Next, we calculate the uncertainties in EW on the pixel-by-pixel basis using the following formula:
\begin{equation}
\label{equ:ewerr}
    \begin{aligned}
    \sigma_{EW}\, = & \sqrt{\left(\frac{1}{f_{\lambda, cont}}\times\sigma_{\mathrm{pixel}}\right)^2 + \left(\frac{F_{\mathrm{EL}}}{f_{\lambda, cont}^2}\times\sigma_{\mathrm{pixel}}\right)^2} \\ & \times\frac{\Delta\lambda_{\mathrm{filter}}}{1+z},
    \end{aligned}
\end{equation}
where $\sigma_{\mathrm{pixel}}$ is noise of a pixel, originally obtained from the ERR uncertainty extensions of JADES F150W images.
Panel (4) of Figure \hyperref[fig:map]{5} displays the $\mathrm{EW_{[O\pnt{III}]+H\beta}}$ S/N map of a sample galaxy. This S/N map efficiently excludes the noise patterns observed in the $\mathrm{EW_{[O\pnt{III}]+H\beta}}$ map, typically shown as unconnected noise pixels. Note that the $\mathrm{H\alpha}$ emission line map, as well as the $EW_{\mathrm{H\alpha}}$ and S/N maps, are also derived using the combination of F115W, F200W, and F277W filters, where the $\mathrm{H\alpha}$ emission lines certainly fall in the F200W filter.

\subsection{Extracting [O{\sc iii}]+$\mathrm{H\beta}$ regions from S/N peak}
\label{sec:peakfinding}
When comparing panel (2) and (4) of Figure \hyperref[fig:map]{5}, we observe that the [O{\sc iii}]+$\mathrm{H\beta}$ regions are located at the same positions as those of the $\mathrm{EW_{[O\pnt{III}]+H\beta}}$ S/N peaks. To extract these emission line regions, we apply the \robotoThin{findpeaks} Python Package on the $\mathrm{EW_{[O\pnt{III}]+H\beta}}$ S/N map. 
After identifying and confirming the pixel coordinates of the $\mathrm{EW_{[O\pnt{III}]+H\beta}}$ S/N peaks, we place a circular aperture with $0.''15$ (5 pixels) diameter on each peak, corresponding to $\sim\mathrm{1.2\,kpc}$ at the median redshift $z_{med} = 2.23$, and to nearly twice the FWHM of the F277W image. We require an average $\mathrm{EW_{[O\pnt{III}]+H\beta}}$ S/N within the circular aperture to be greater than 2.5, as shown by the white circles in panel (5) of Figure \hyperref[fig:map]{5}. Those peaks identified with an average $\mathrm{S/N} < 2.5$ are marked with white squares.

This process is repeated iteratively for all 135 cross-matched HAEs to determine the $\mathrm{EW_{[O\pnt{III}]+H\beta}}$ S/N peaks with circular apertures. By applying the same circular aperture on the other scientific frames of the same galaxy, we successfully obtain the total luminosity, emission line flux, and $\mathrm{EW_{[O\pnt{III}]+H\beta}}$ for each emission line region. Among all the S/N peaks, 187 circular apertures have $\mathrm{S/N} > 2.5$ and $\mathrm{EW_{[O\pnt{III}]+H\beta}}>200\mathrm{\AA}$, which are considered as the [O{\sc iii}]+$\mathrm{H\beta}$ regions. In this work, we focus on these 187 [O{\sc iii}]+$\mathrm{H\beta}$ regions and analyze their properties quantitatively.


\subsection{Green Seeds and Red Seeds}
\label{sec:greenred}
\begin{figure*}[hbt!]
    \centering
    \includegraphics[width=1\textwidth]{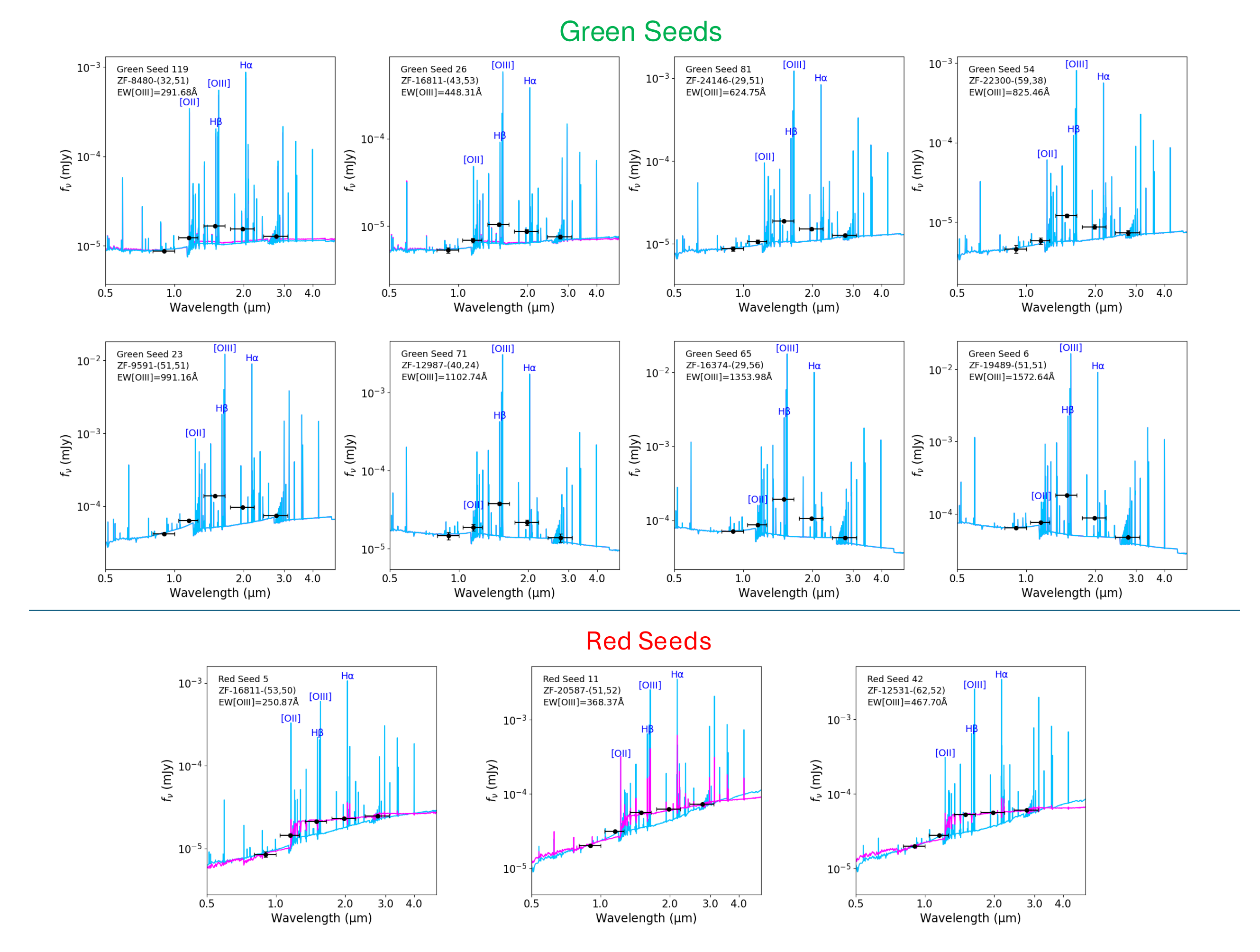}
    \label{fig:sedsamp}
    \vspace{-0.4cm}
    \caption{Eight example SEDs of Green Seeds and three example SEDs of Red Seeds are given in the small panels. The blue and magenta spectra represent the best-fit SEDs, with the blue spectra constrained by the largest stellar age of $\mathrm{100\,Myr}$. Green Seeds favor young stellar populations with low dust attenuation. In contrast, due to the limited photometric data, Red Seeds could either choose younger spectra with large dust content or older spectra with less dust, i.e., the age-dust degeneracy. These examples validate the feasibility of the color selection in Figure \hyperref[fig:sedcolor]{6} for identifying Green Seeds with young stellar populations and low dust attenuation. Also, the spectra demonstrate that assuming a linear relationship between the stellar continuum in the F115W and F277W filters does not introduce large uncertainties.}
    \vspace{0.4cm}
\end{figure*}

\begin{figure*}[hbt!]
    \centering
    \includegraphics[width=0.48\textwidth]{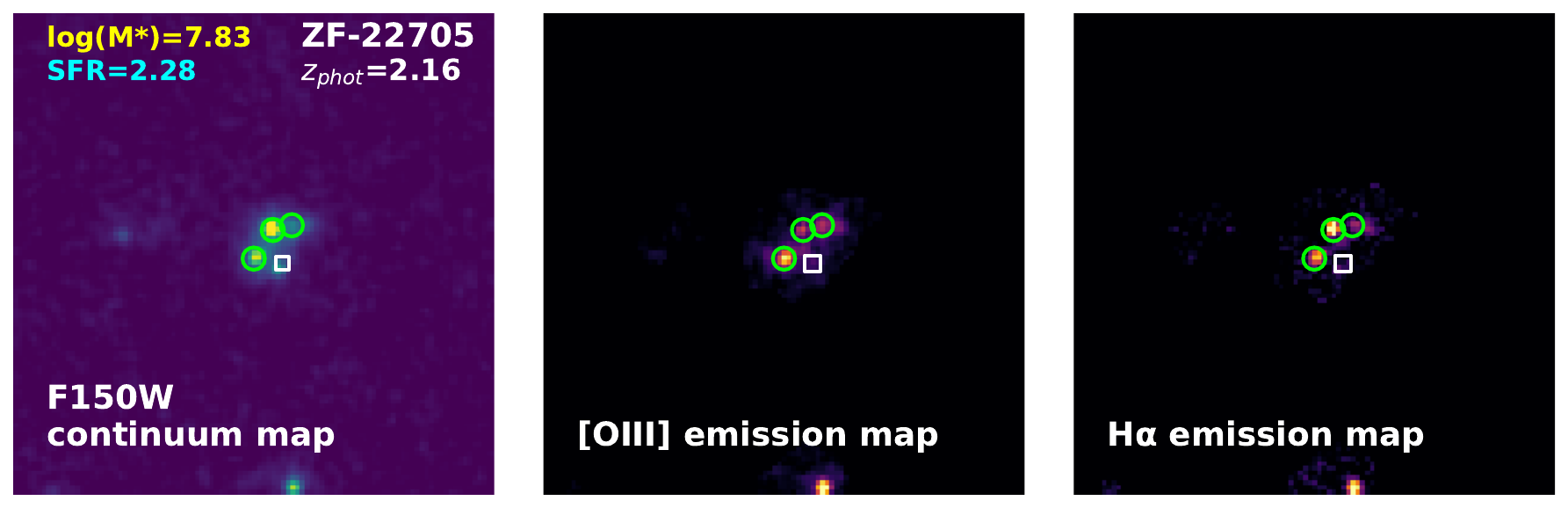}
    \hspace{0.2cm}
    \includegraphics[width=0.48\textwidth]{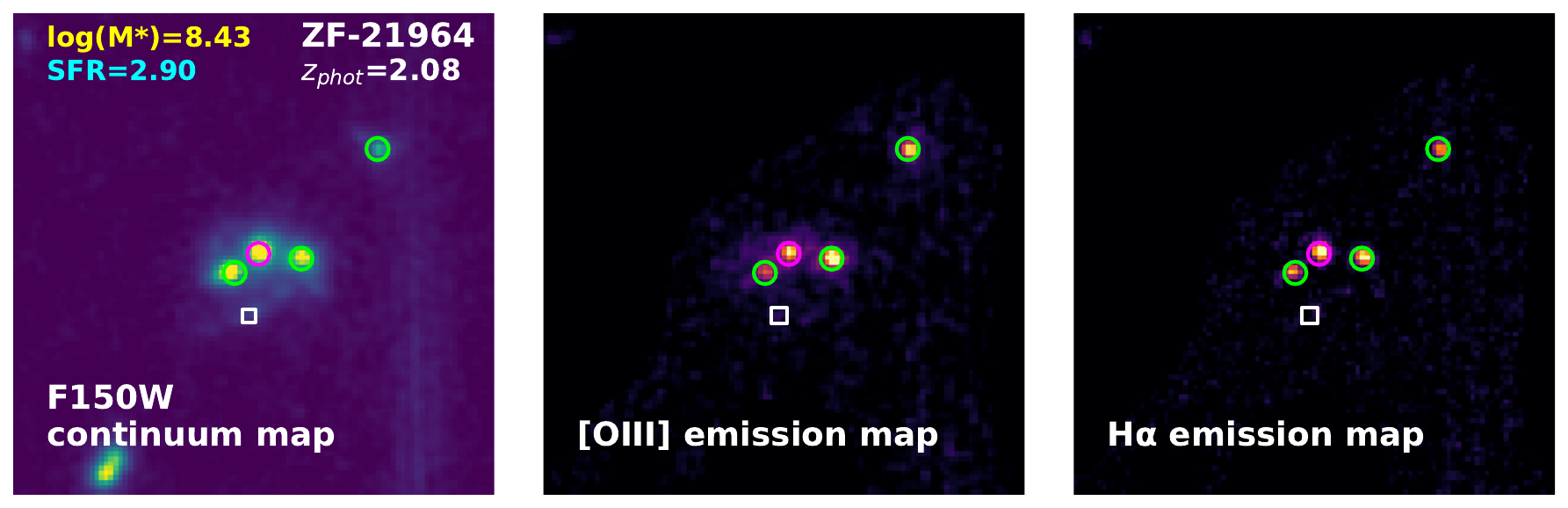}
    \hspace{0.2cm}
    \includegraphics[width=0.48\textwidth]{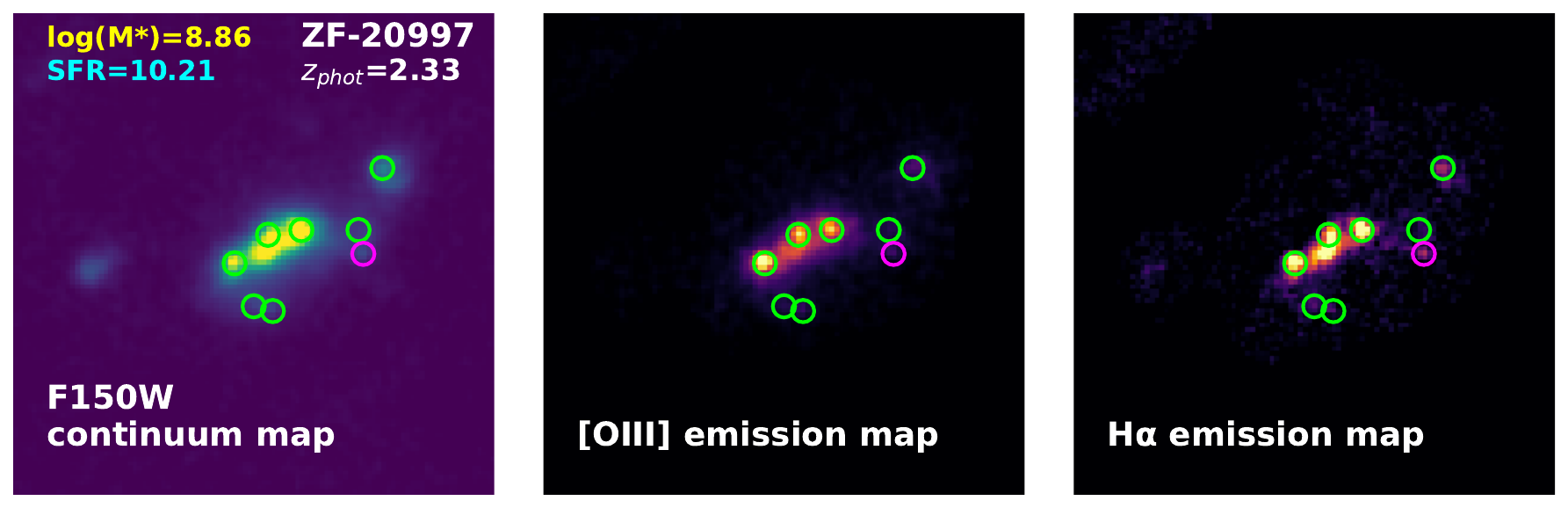}
    \hspace{0.2cm}
    \includegraphics[width=0.48\textwidth]{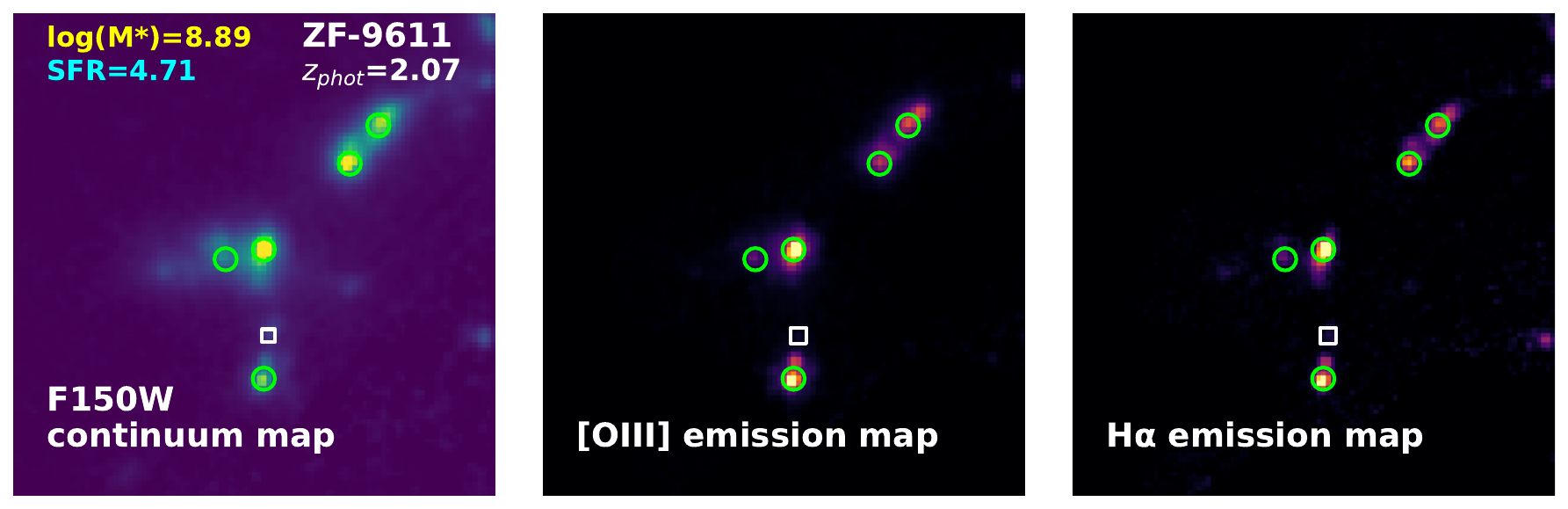}
    \hspace{0.2cm}
    \includegraphics[width=0.48\textwidth]{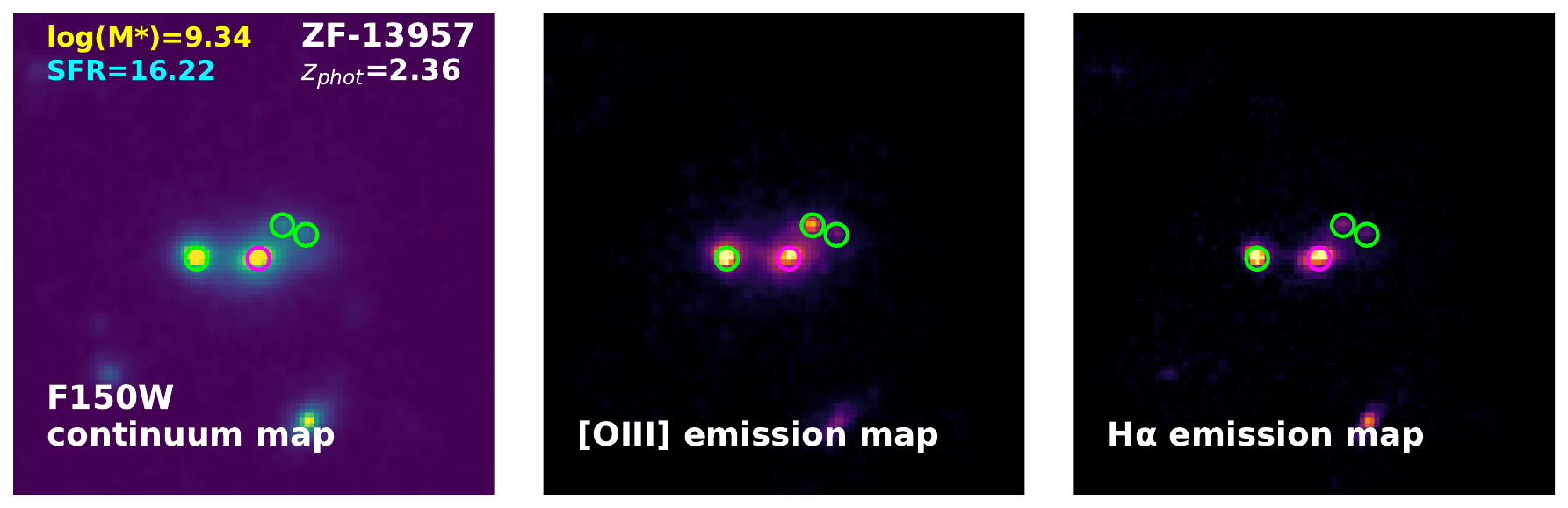}
    \hspace{0.2cm}
    \includegraphics[width=0.48\textwidth]{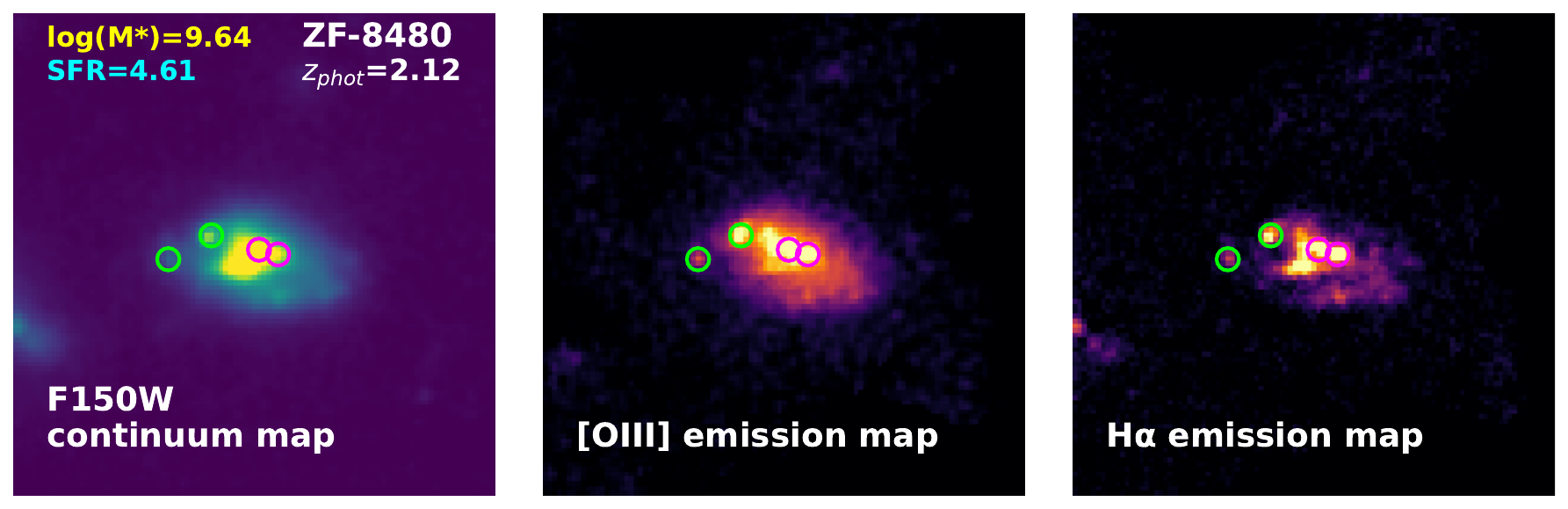}
    \label{fig:emimap}
        \caption{The stellar continuum map at $1.5\,\mu m$, the [O{\sc iii}]+$\mathrm{H\beta}$ emission line map, and the $\mathrm{H\alpha}$ emission line map for six HAEs that contain Green Seeds. These galaxies are ordered by their stellar mass. Green Seeds are marked as green circles, while Red Seeds are shown as magenta circles. Also, white squares represent regions with $\mathrm{S/N < 2.5}$, similar to Figure \hyperref[fig:map]{5}.}
    \vspace{0.3cm}
\end{figure*}

In the next step, we visually inspect the [O{\sc iii}]+$\mathrm{H\beta}$ regions on their highly sensitive NIRCam RGB (F115W+F150W+F277W) images. Upon inspection, we figure a continuous range of the properties of these regions. On one end is a population showing greenish color in the RGB images, which
are thought to be the most extreme starbursts having the largest $\mathrm{EW_{[O\pnt{III}]+H\beta}}$ values (greater than $500\,\mathrm{\AA}$) and the bluest underlying stellar continuum from the largest proportion of very young stars.
On the other end is a population of redder color,  which are expected to have dusty spectra with prominent $\mathrm{EW_{[O\pnt{III}]+H\beta}}$ or older populations having Balmer break and no emission line.

We quantify this sequence in a color-color diagram (F090W$-$F150W vs. F150W$-$F277W) and compare these observed colors with simple models of stars and gas in Figure \hyperref[fig:sedcolor]{6}. First, we generate a series of SED models using BC03 \citep{Bruzual03} with a \citet{Chabrier03} IMF. These models assume a instantaneous burst star formation history (SFH), which happened at 3, 5, 10, 20, 50, 100, 200 Myrs before.
The sequences of pure stellar light are shown in solid blue line. Emission-line templates generated by CLOUDY \citep{Ferland98,Ferland13} are then added to these SED models to show  their effects on broad-band photometry. The stellar metallicity and gas-phase metallicity are fixed at $Z=0.008$ (0.6 of Solar metallicity), typical for galaxies at Cosmic Noon \citep[e.g.,][]{Erb06, Steidel14, Sanders15, Sanders21}. The free parameters in these SED models are stellar age, dust attenuation and ionization parameters. Older stellar ages and higher dust attenuation lead to redder spectra, while higher $\mathrm{log}\,U$ increases emission line fluxes in the models, resulting in more pronounced flux excesses in certain filters (e.g., F150W and F200W in this work). We then convolve the JWST filter transmission curves to these SED models to obtain the corresponding color indices.
 
We overlay the color indices from the SED models as grids on the left panel of Figure \hyperref[fig:sedcolor]{6}. The diamond grids with dashed lines exhibit how the color indices change with dust attenuation and stellar age at a fixed $\mathrm{log}\,U=-2$. The sizes of the diamond grids increase with dust attenuation level from $E(B-V)=0$ to 0.8. The grayscale of the diamonds grids represents the stellar ages, with the darkest shading indicating the youngest age of $\mathrm{3\,Myr}$ and the lightest the oldest age of $\mathrm{500\,Myr}$. As suggested by blue solid line and black dashed line, both increasing the age of a stellar population and increasing the amount of dust can redden the spectra in $\mathrm{F090W - F150W}$ and $\mathrm{F150W - F277W}$, known as the age-dust degeneracy occurring at the rest-frame optical light. Also, spectra with younger stellar populations exhibit more obvious flux excesses in F150W from [O{\sc iii}]+$\mathrm{H\beta}$ emission line at fixed $\mathrm{log}\,U$ because of their lower stellar continuum. 

The age-dust degeneracy raise an issue on whether the identified redder [O{\sc iii}]+$\mathrm{H\beta}$ regions are dusty young starburst or containing older stellar population. Actually, the stellar continuum could become discontinuous by Balmer break, and the amount of the break depends on the age of the galaxy. This may raise uncertainties on the estimation of [O{\sc iii}]+$\mathrm{H\beta}$ emission line fluxes. Thus, we develop a new color selection criterion by fitting the SED model grids. For young galaxies with little dust content, $E(B-V)$ values are typically small, ranging from 0 to 0.15 magnitudes \citep[e.g.,][]{Nakajima16,Tang19}. Thus, we set $E(B-V)=0.15$ and stellar age equals to $\mathrm{100\,Myr}$ as the threshold to classify two populations. We apply the linear fits to the SED model grids based on ionization parameters from $\mathrm{log}\,U=-2$ to $-3.6$. The resulting color limit is:
\begin{equation}
\label{equ:seddust}
    \mathrm{F090W - F150W} =  -1.1\times(\mathrm{F150W - F277W}) + 0.9
\end{equation}
We display this color selection in the right panel of Figure \hyperref[fig:sedcolor]{6}, separating all the [O{\sc iii}]+$\mathrm{H\beta}$ regions into 128 normal ones and 59 older and/or dusty ones, shown as green filled circles and brown open squares, respectively. Those 128 [O{\sc iii}]+$\mathrm{H\beta}$ regions are named as ``Green Seeds", in reference to the famous Green Pea galaxies at $z\sim0.2$ \citep{Cardamone09}. The Green Pea galaxies were discovered in the citizen science project Galaxy Zoo, exhibiting peculiar bright green colors in Sloan Digital Sky Survey \citep[SDSS;][]{York00,Kauffmann03} imaging due to very strong [O{\sc iii}]+$\mathrm{H\beta}$ emission lines with EWs up to $\sim1000\,\mathrm{\AA}$. Green Seeds in this work have a similar appearance on the JWST RGB images. On the other hand, other emission line regions are named as ``Red Seeds".

\begin{table}[t]
    \small
    \centering
    \label{tab:sample}
    \caption{The selection process from the parent HAE sample to the identification of substructures within galaxies, along with the corresponding number of selected objects at each stage}
    \begin{tabular}{cc}
        \hline\hline
        $\ $ Selection Step $\ $ & $\ $ \# of $\mathrm{H\alpha}$ emitters $\ $ \\
         & (applicable) \\
        \hline
         Parent sample &  422 \\
        in ZF-CDFS & \\
        \hline
        Cross-match w/ &  170 \\
        JADES catalogs &   \\     
        \hline
        Redshift constraint$\mathrm{\,^{a}}$ & 135  \\
        &  \\
        \hline
         HAEs contain resolved & 93 \\
         $\mathrm{[O\pnt{III}]}$  regions$\mathrm{\,^{b}}$ & \\
         \hline 
        HAEs contain Green Seeds & 68  \\
        HAEs contain Red Seeds$\mathrm{\,^{b}}$ & 44  \\
        \hline\hline
    \end{tabular}
    
    \vspace{0.5cm}
    
    \begin{tabular}{c|c}
    \hline\hline
       [O{\sc iii}] regions with  & \# of resolved regions \\
        $\mathrm{EW_{[O\pnt{III}]+H\beta}}>200\mathrm{\AA}$  & \\
       \hline
       Green Seeds & 128 \\
       \hline
       Red Seeds & 59 \\
       \hline
       Total & 187\\
       \hline\hline
    \end{tabular}
    
    \vspace{0.2cm}
    
    \begin{tablenotes}
    \item \textbf{Notes.} ${ }^{\text {a }}$ The [O{\sc iii}] (H$\alpha$) emission line at $2.05 < z < 2.35$ redshifts into the JWST/NIRCam F150W (F200W) filter. \\
    ${ }^{\text {b }}$ In Section \hyperref[sec:peakfinding]{3.2}, a total of 93 HAEs are found to exhibit signicant $\mathrm{EW_{[O\pnt{III}]+H\beta}}$ S/N peaks. Then, in Section \hyperref[sec:greenred]{3.3}, 68 HAEs are identified as having Green Seeds, and 44 having Red Seeds, respectively.\\
    \end{tablenotes}
    \vspace{0.3cm}
\end{table}

\citet{Cardamone09} defined an $r-z$ vs. $u-r$ color selection criteria by comparing Green Pea galaxies to a large sample of local star-forming galaxies (SFGs) and quasars. Since the three SDSS filters ($u$, $r$, $z$) have similar rest-frame wavelengths to the JWST filters (F090W, F150W, F277W), we also incorporate their color selection criteria in the right panel of Figure \hyperref[fig:sedcolor]{6}. The color selection from \citet{Cardamone09} does not clearly distinguish Green Seeds from Red Seeds, as identified by our color selection (Equation \hyperref[equ:seddust]{5}). In Figure \hyperref[fig:sedsamp]{7}, we present a series of SED samples for both Green Seeds and Red Seeds using the SED fitting code \robotoThin{CIGALE} \citep{Noll09,Boquien19}. 
The blue and magenta spectra in each panel represent both the best-fit SED, but for the blue ones,  maximum stellar age of the SED fitting is limited to $\mathrm{100\,Myr}$. It is clearly illustrated that Green Seeds tend to favor younger spectra with less dust content in the both fitting, and their stellar continuum can be well estimated by a linear interpolation between the F115W and F277W filters. On the other hand, Red Seeds suffer from the age-dust degeneracy, where it is unclear whether younger stellar population and strong emission lines with high dust attenuation or older population with low dust attenuation represent the true SED, given only five JWST photometric data points.

In Figure \hyperref[fig:emimap]{8}, we display the stellar continuum map at $1.5\,\mu m$, the [O{\sc iii}]+$\mathrm{H\beta}$ emission line map, and the $\mathrm{H\alpha}$ emission line map for six samples among the total 68 HAEs that contain Green Seeds. They are ordered by stellar mass, from lower to higher. The 68 HAEs have stellar mass from $10^{7.8}\,M_{\odot}$ to $10^{10}\,M_{\odot}$. The continuum and emission line maps of the remaining HAEs with Green Seeds are displayed in Appendix \hyperref[sec:appendixsample]{B}. In this paper, we mainly focus on the properties, origins, and evolution of Green Seeds, in conjunction with other populations, including Red Seeds and H{\sc ii} regions.
In Table \hyperref[tab:sample]{1}, we summarize the number of HAEs, Green Seeds, and Red Seeds identified at each selection step from the parent sample, as discussed in Section \hyperref[sec:samp]{2} and Section \hyperref[sec:method]{3}.

\section{The properties of Green Seeds}
\label{sec:result}
\subsection{Stellar mass}
To derive the stellar mass of the 128 individual Green Seeds, we use the SED fitting code \robotoThin{CIGALE} with emission-line templates. The parameter settings are as follows: stellar population models from BC03 \citep{Bruzual03} with a \citet{Chabrier03} IMF; a latest starburst to represent SFH with the age of 3, 5, 10, 20, 50, 100, 200, 500 Myrs; metallicity $Z$ of 0.008; and ionization parameters of $\mathrm{\log\,U=-3.5 ,-3, -2.5, -2, -1.5}$. 
Dust attenuation is modeled using the \citet{Calzetti00} curve for stellar continuum, $E(B-V)_{cont}$ and the Milky Way curve \citep{Cardelli89} for emission lines, $E(B-V)_{neb}$. These parameter settings are also used in Figure \hyperref[fig:sedsamp]{7} to derive the best-fit SED in magenta color. Moreover, we adopt a Bayesian-like fitting approach, assigning weights to all models based on their $\chi^2$ values, to obtain more accurate estimates of physical properties, such as stellar mass, with reduced uncertainties.

\begin{figure}[t]
    \includegraphics[width=1\linewidth]{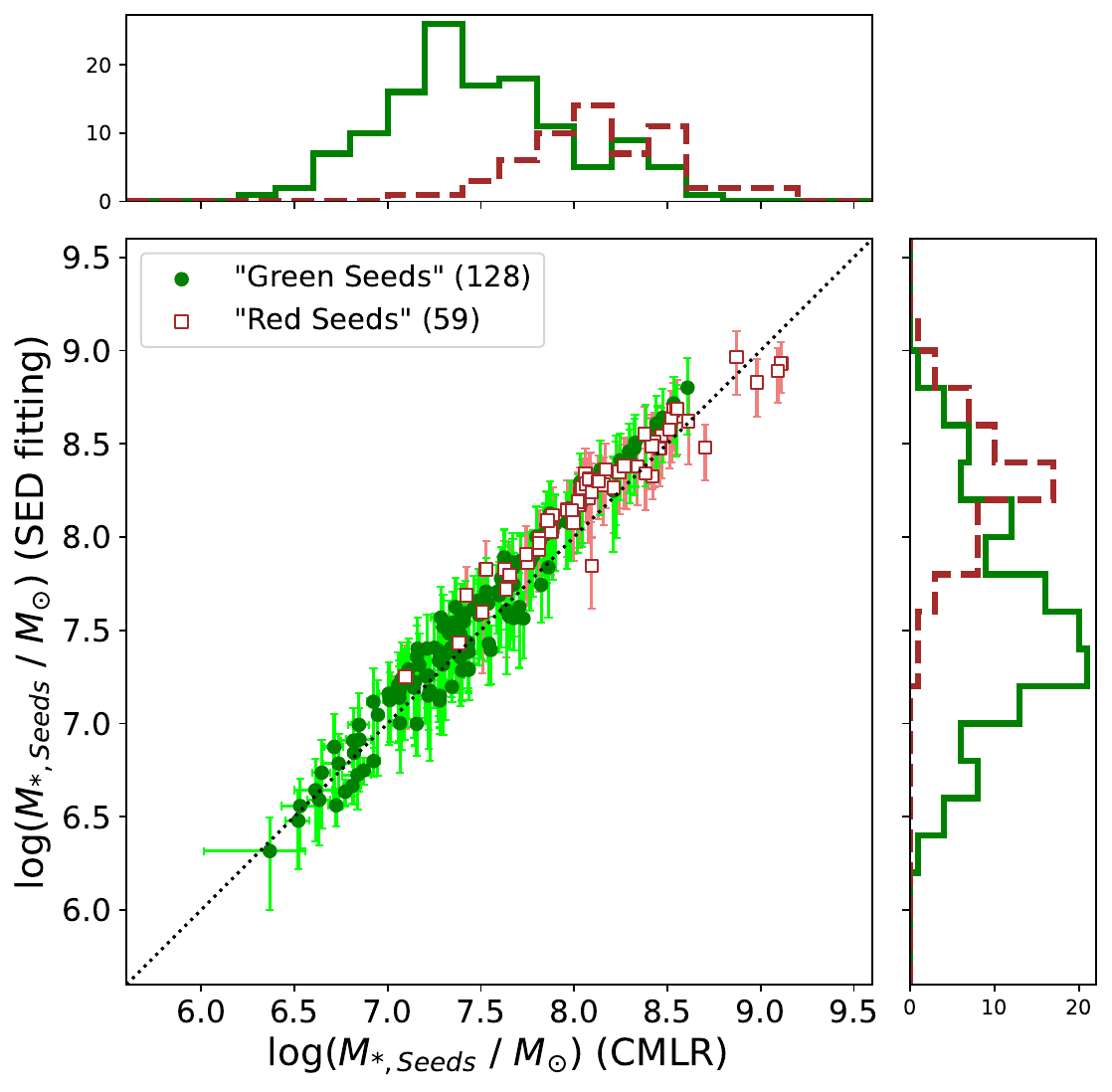}
    \label{fig:mass_histo}
    \vspace{-0.4cm}
    \caption{Comparison between the stellar masses of the resolved emission line regions estimated from SED fitting and the color-mass-light ratio (CMLR). The 128 Green Seeds are marked as green filled circles, while the 59 Red Seeds are shown as brown open squares. The overall estimation of stellar mass from SED fitting agrees well with the CMLR measurements. The stellar masses of Green Seeds and Red Seeds are also distributed as histograms on both axes in green and brown, respectively.}
\end{figure}

Given that we only utilize five JWST filters from JADES to achieve better angular resolution in this work, the limited number of photometric fluxes may lead to large uncertainties in stellar mass estimation. Thus, in addition to applying the standard SED fitting techniques, we also estimate the stellar masses of Green Seeds from the mass-to-light ratio method. \citet{McGaugh14} introduced an approach to estimate stellar mass using a color-mass-to-light ratio (CMLR), constructed by population synthesis models with the following relation:
\begin{equation}
\label{equ:cmlr}
    \log \Upsilon_*^k\,=\,a_k+b_k\left(B-V\right),
\end{equation}
where $k$ refers to the selected filters and the footnote $*$ equals to the color of $B-V$. In this work, we prefer using the rest-frame $I$ band (F277W), denoted as $\Upsilon_*^I$, to minimize the effects of underlying emission lines. The $B-V$ color can be obtained from the stellar continuum in F115W and F150W. The stellar mass of each Green Seeds is then derived from the rest-frame $I$-band luminosity by $M_*=\Upsilon_*^I\,L_I$. In Figure \hyperref[fig:mass_histo]{9}, we compare the stellar mass distributions of Green Seeds inferred from SED and CMLR, displaying the results as histograms. We demonstrate that these two methods provide closely aligned estimations of stellar masses. 

Most Green Seeds have stellar masses in the range of $10^{6.5}\,M_{\odot}$ to $10^{8.5}\,M_{\odot}$, with a median stellar mass of $10^{7.4}\,M_{\odot}$. When compared to the total stellar mass of their host galaxies, the stellar mass of Green Seeds contributes to $\sim3\%$ of the total mass (with a median of $\mathrm{1.5\,dex}$ lower). 
We also show the stellar mass distribution of Red Seeds in the same panel, which exhibits a $\sim\mathrm{1\,dex}$ higher median stellar mass than that of Green Seeds.

\subsection{Star formation rate}
\label{sec:sfr}

\begin{figure*}[hbt!]
    \centering
    \includegraphics[width=0.48\textwidth]{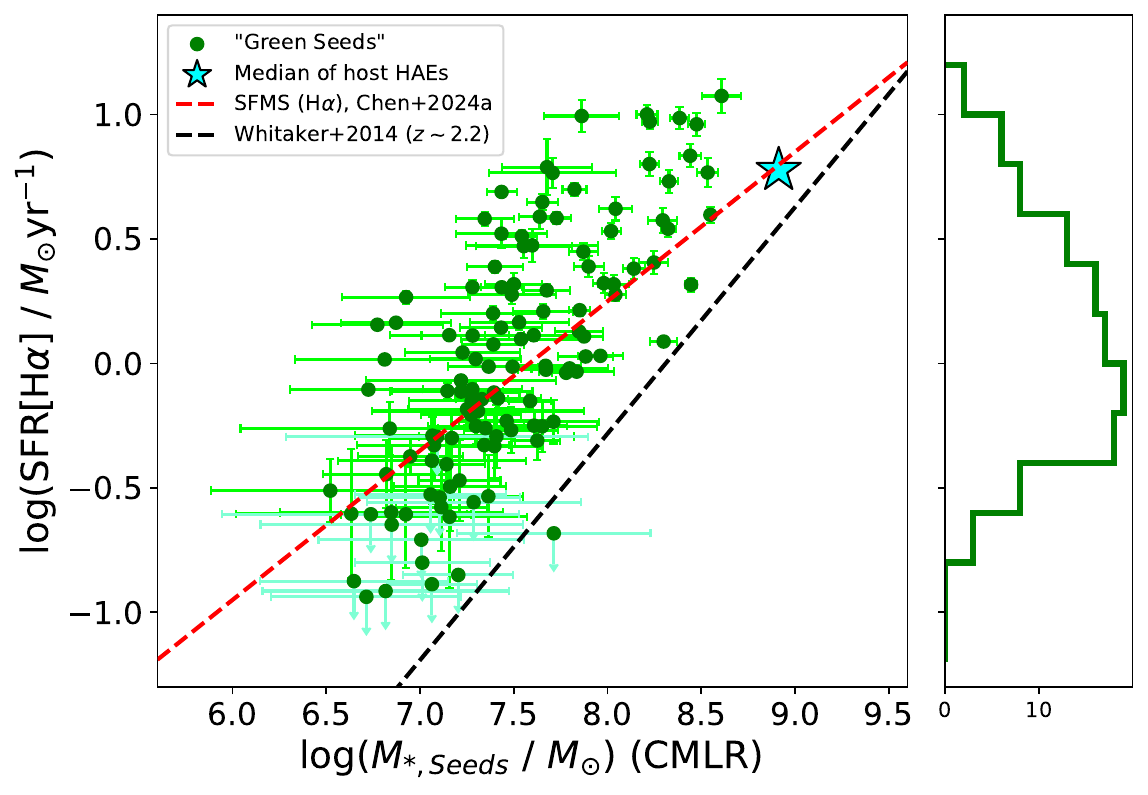}
    \includegraphics[width=0.48\textwidth]{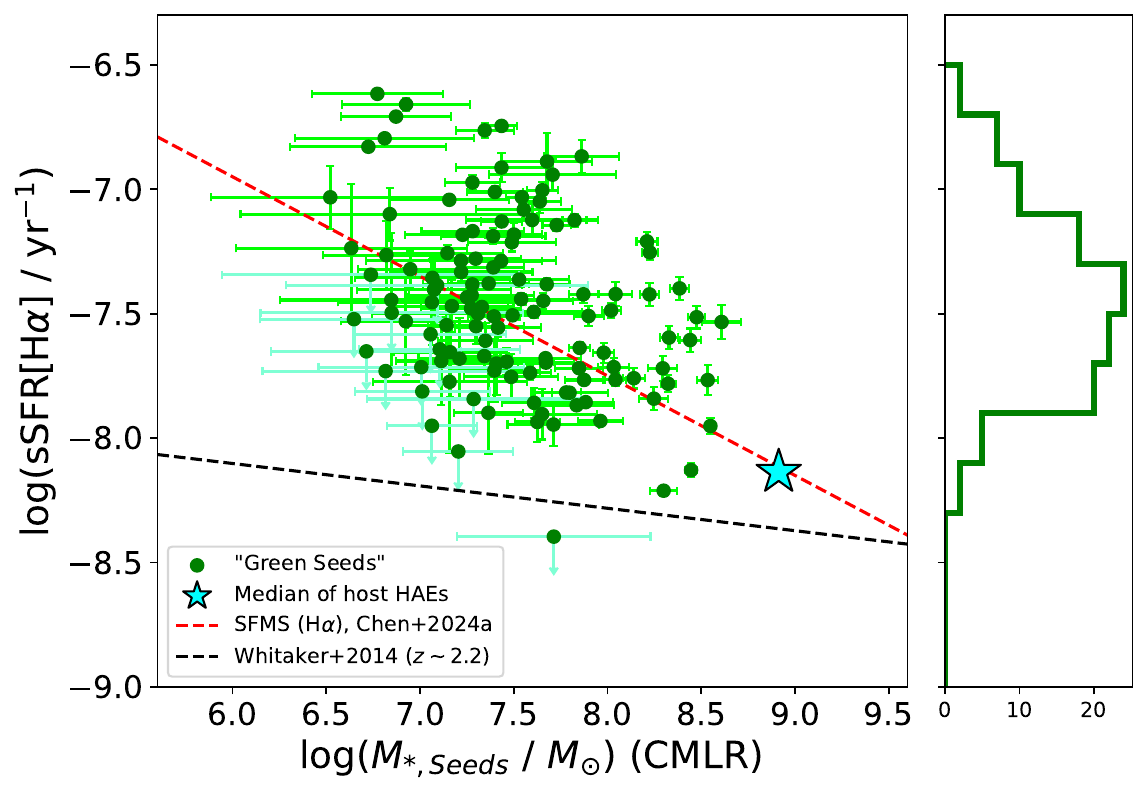}
    \label{fig:sfrbean}
    \caption{Based on the F200W photometry, we require $\mathrm{H\alpha}$ emission lines show flux excesses with $\mathrm{S/N>2.5}$ in every Green Seeds, and nearly 90\% of them meet this requirement. \textbf{Left}: The SFR($\mathrm{H\alpha}$) as a function of stellar mass, i.e., the ``SFMS" in Green Seeds distributed as green filled circles. Downward arrows indicate the $2.5\sigma$ upper limits for the $\mathrm{H\alpha}$-undetected regions. The median $M_*$ and integrated SFR of the host galaxies that contain Green Seeds is indicated by the cyan star. The red dashed line is the extrapolated best-fit SFMS from C24, while the black dashed one is from \citet{Whitaker2014} for $z\sim2.2$ galaxies. The mass completeness of both literatures are $\sim10^9\,M_{\odot}$. \textbf{Right}: The sSFR($\mathrm{H\alpha}$) as a function of stellar mass in Green Seeds, with the same outlines as in the left panel. The cyan star shows the median value of the integrated sSFR of the host galaxies. Green Seeds have a much higher sSFR by $\sim0.6\,\mathrm{dex}$ compared to their host galaxies. Also, they are generally above the extrapolated SFMS($\mathrm{H\alpha}$) from \citet{Whitaker2014}.}
    \vspace{0.2cm}
\end{figure*}

To determine the SFRs in Green Seeds, we measure the $\mathrm{H\alpha}$ luminosity within the same aperture by using the $\mathrm{H\alpha}$ emission line map, as shown in Figure \hyperref[fig:emimap]{8}. By analyzing the flux excess in the F200W filter, we calculate the total $\mathrm{H\alpha}$ luminosity and convert it to SFR by the calibration from \citet{Kennicutt12}, with a correction applied for the \citet{Chabrier03} IMF: 
\begin{equation}
    \mathrm{log\,SFR}(\mathrm{H} \alpha) = \mathrm{log}\,L_\mathrm{H\alpha} - 41.34.
\end{equation}
The correction for dust attenuation is based on the median $E(B-V)_{neb}=\mathrm{0.1}$ obtained from the SED fitting results of Green Seeds. In the F200W filter, the main contaminants include [N{\sc ii}] and [S{\sc ii}] emission lines. We have applied a correction of contamination, assuming the contamination ratio of 0.1 which are median values of the best-fit SED templates.

Following the methodology described in Section \hyperref[sec:method]{3}, we measure the $\mathrm{H\alpha}$ luminosity in the same aperture as used for Green Seeds, derived from the $\mathrm{H\alpha}$ emission line map in the F200W filter. A majority (112 out of 128) of Green Seeds show $\mathrm{H\alpha}$ emission line detection, contributing to clear flux excesses of $\mathrm{S/N > 2.5}$ in F200W. For those Green Seeds without strong flux excesses, we take the $2.5\sigma$ upper limit fluxes for the $\mathrm{H\alpha}$ measurements.

In the left panel of Figure \hyperref[fig:sfrbean]{10}, we show the SFR($\mathrm{H\alpha}$) as a function of the stellar mass of Green Seeds. Generally, SFGs exhibit a correlation between stellar mass ($M_*$) and SFR, known as the SFMS, which holds true up to at least $z \sim 3$ \citep[e.g.,][]{Whitaker2014,Speagle2014}. Green Seeds also appear to follow the SFMS with similar slopes to that from \citet{Whitaker2014}. However, the normalization of this relation is much higher for Green Seeds compared to their host galaxies, indicating a higher specific SFR (sSFR) for Green Seeds. We further illustrate this by displaying the correlation between sSFR and the stellar mass in the right panel of Figure \hyperref[fig:sfrbean]{10}. The inverse of sSFR (1/sSFR) directly indicates the time taken to double the stellar mass in the system. The median sSFR for Green Seeds reaches $-7.6\,\mathrm{yr^{-1}}$, which is $\sim0.6\,\mathrm{dex}$ higher than that of their host galaxies, suggesting $\sim4\times$ more rapid stellar mass assembly in Green Seeds. This supports the idea that starburst-like activities are occurring in Green Seeds.

\subsection{Equivalent width}
Based on multiple emission line analysis of a large number of galaxies at Cosmic Noon, previous studies suggested that the [O{\sc iii}] and $\mathrm{H\alpha}$ EW exhibit significant correlations with various galaxy properties \citep[e.g.,][]{Reddy18}; galaxies with larger [O{\sc iii}] and $\mathrm{H\alpha}$ EWs tend to have lower stellar mass, lower metallicity, higher ionization parameters, and higher ionizing photon production efficiencies \citep[$\xi_{ion}$;][]{Chen24b}. Also, the [O{\sc iii}] EW correlates most strongly with these galaxy properties compared to other emission lines like [O{\sc ii}].

Following Equation \hyperref[equ:ew]{3}, we construct the rest-frame $\mathrm{EW_{[O\pnt{III}]+H\beta}}$ map of each HAE. By applying the circular apertures derived from peak finding, we obtain the average $\mathrm{EW_{[O\pnt{III}]+H\beta}}$ of each [O{\sc iii}]+$\mathrm{H\beta}$ emission line region. To ensure accurate measurements of [O{\sc iii}] line fluxes, we assume that the total flux excesses in the F150W filter are contaminated by $\mathrm{H\beta}$, following the approach in C24. Assuming Case-B recombination with $T_e=10,000K$ and $n_e = 100\, \mathrm{cm}^{-3}$,  we derive the intrinsic $\mathrm{H\beta}$ fluxes from the intrinsic $\mathrm{H\alpha}$ fluxes using $F_{\mathrm{H\beta},int} = F_{\mathrm{H\alpha},int}/2.86$, where the $\mathrm{H\alpha}$ flux is obtained as described earlier.

\begin{figure}[t]
    \includegraphics[width=1\linewidth]{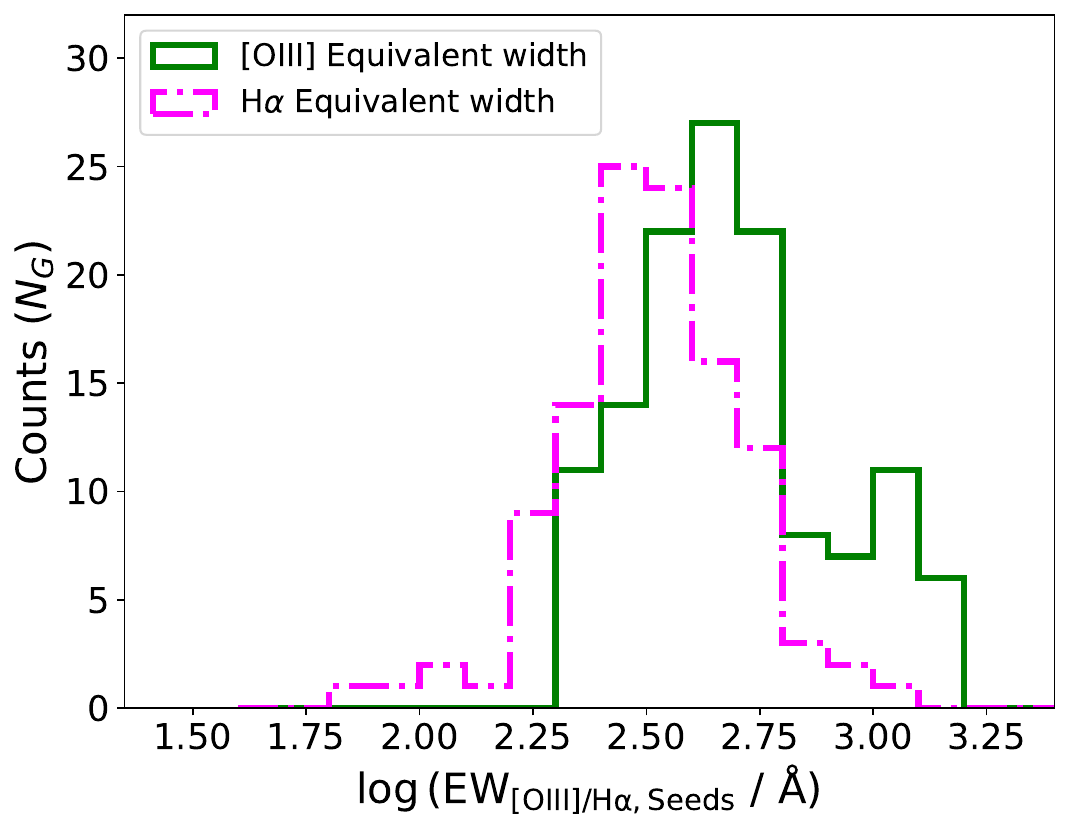}
    \label{fig:ew_histo}
    \vspace{-0.3cm}
    \caption{The distribution of [O{\sc iii}] and $\mathrm{H\alpha}$ EWs for Green Seeds, shown as histograms in green and magenta, respectively. This analysis of $\mathrm{H\alpha}$ EWs only includes those Green Seeds with $\mathrm{H\alpha}$ flux excesses of $\mathrm{S/N > 2.5}$, which contains a sample of 111 regions. The median equivalent width of $\mathrm{EW_{[O\pnt{III}]},med} = 452\mathrm{\AA}$ and $\mathrm{EW_{H\alpha},med} = 320\mathrm{\AA}$, respectively.}
\end{figure}

Figure \hyperref[fig:ew_histo]{11} displays the rest-frame $\mathrm{EW_{[O\pnt{III}]}}$ distribution for Green Seeds. Green Seeds have quite high equivalent width reaching to a median value of $\mathrm{EW_{[O\pnt{III}]},med} = 452\mathrm{\AA}$. Since $\mathrm{EW_{[O\pnt{III}]}}$ serves as an indicator of the ionization state of the ISM \citep[e.g.,][]{Reddy18}, this implies a very intense ionizing radiation field in Green Seeds. Similar results are also found for $\mathrm{EW_{H\alpha}}$, as shown as magenta dot-dashed line in figure \hyperref[fig:ew_histo]{11}. Green Seeds have a median equivalent width of $\mathrm{EW_{H\alpha},med} = 320\mathrm{\AA}$. More directly, the intensity of $\mathrm{H\alpha}$ emission line traces the recent star formation activities within the past $\mathrm{10\,Myr}$. The higher $\mathrm{H\alpha}$ EWs reflect the younger stellar populations in Green Seeds and explain the higher sSFRs observed in the right panel of Figure \hyperref[fig:sfrbean]{10}.

We find that 17 Green Seeds exhibit extremely large $\mathrm{EW_{[O\pnt{III}]}>1000\AA}$, a phenomenon previously reported only in a resolved case study at $z\sim2$ \citep{Zanella15}. On the other hand, such large [O{\sc iii}] EWs are more commonly observed in integrated studies of EELGs at similar redshift \citep[e.g.,][]{Nakajima14,Tang19}. These studies also suggested a close connection between extreme [O{\sc iii}] emission and Lyman continuum (LyC) leakage. Additionally, \citet{Chen24b} demonstrated that low-mass HAEs, which are widely included in this work, are potential lower-redshift analogs of the galaxies that reionized the universe during cosmic reionization at $z>6$. Our high-resolution observations from JWST may be capturing possible resolved LyC leakage regions in galaxies at $z\sim2$, which has so far only been reported in the local universe \citep{Izotov21}. These Green Seeds with $\mathrm{EW_{[O\pnt{III}]}>1000\AA}$ are likely dominated by extremely hot and massive stars, leading to a more intense radiation field, and creating so-called "density-bounded" H{\sc ii} regions \citep{Nakajima14} from which ionizing photon have a higher probability of leaking into the IGM.

\section{Implication of Green Seeds}
\label{sec:dis1}
\subsection{Relevance between Green Seeds and H{\sc ii} regions}
In the previous section, we estimated the SFRs in Green Seeds through $\mathrm{H\alpha}$ luminosity. We figure out that nearly 90\% of Green Seeds exhibit strong flux excesses with $\mathrm{S/N>2.5}$ in the F200W filter, primarily driven by $\mathrm{H\alpha}$ emission lines. Interestingly, 16 Green Seeds do not show strong flux excesses in F200W, but still have the excesses with at least $\mathrm{S/N>1}$ for the $\mathrm{H\alpha}$ emission lines. The intrinsically weaker $\mathrm{H\alpha}$ emission lines (compared to [O{\sc iii}]) at $z\sim2$ \citep{Sanders18,Reddy18} could be the main reason for the lower $\mathrm{S/N}$. Nevertheless, it is reasonable to assert the prevalence of both [O{\sc iii}] and $\mathrm{H\alpha}$ emission lines in the detected emission line regions.

Although not shown in Figure \hyperref[fig:map]{5}, we also apply the same peak finding algorithm to the $EW_{\mathrm{H\alpha}}$ S/N maps to extract the $\mathrm{H\alpha}$ emission line regions (H{\sc ii} regions) at $z\sim2.2$. We identify 164 regions with $\mathrm{S/N} > 2.5$ and $EW_{\mathrm{H\alpha}}>100\mathrm{\AA}$. Among these H{\sc ii} regions, we find that 6 regions do not have flux excesses $\mathrm{S/N>2.5}$ in the F150W filter, reflecting weak [O{\sc iii}] emission lines in them. Note that five of these 6 regions are located in massive galaxies that do not contain Green Seeds. This finding further suggests the ubiquity of both strong [O{\sc iii}] and $\mathrm{H\alpha}$ emission lines in the emission line regions at Cosmic Noon.

\subsection{Comparison with UV star-forming clumps}
\label{sec:compareuv}
The properties of these kpc-scale Green Seeds suggest that they are star-burst regions with hard ionizing radiation fields, recalling the star-forming clumps. Previous observations with the Hubble Space Telescope (HST) have revealed that many galaxies at $z\simeq1-3$ host discrete rest-frame ultraviolet (UV) star-forming clumps of similar scale \citep[e.g.,][]{Conselice04,Elmegreen05,Elmegreen07,ForsterSchreiber11b,Wuyts12,Guo12,Guo15,Guo18,Tadaki14,Livermore15,Garland15,Shibuya16,Calabro19,Claeyssens23}. These UV-bright clumps have been shown to have enhanced sSFRs, higher than those of their surrounding areas by a factor of several. However, these previous studies mostly focused on clumps in massive galaxies with stellar mass larger than $10^{10}\,M_{\odot}$ and within larger apertures than used in this work, such as $0.''30$ in \citet{Guo12}, due to the limitations in resolution and depth of HST observations. UV star-forming clumps in low-mass hosting galaxies have rarely been studied, with most investigations relying on the gravitational lensing effect \citep[e.g.,][]{Vanzella22}. With the unprecedented resolution and depth provided by JWST/NIRCam, it is worth exploring UV star-forming clumps in HAEs, especially those in low-mass galaxies that have only been resolved with the aid of gravitational lensing \citep{Vanzella23}.

\begin{figure*}[hbt!]
    \centering
    \includegraphics[width=0.8\textwidth]{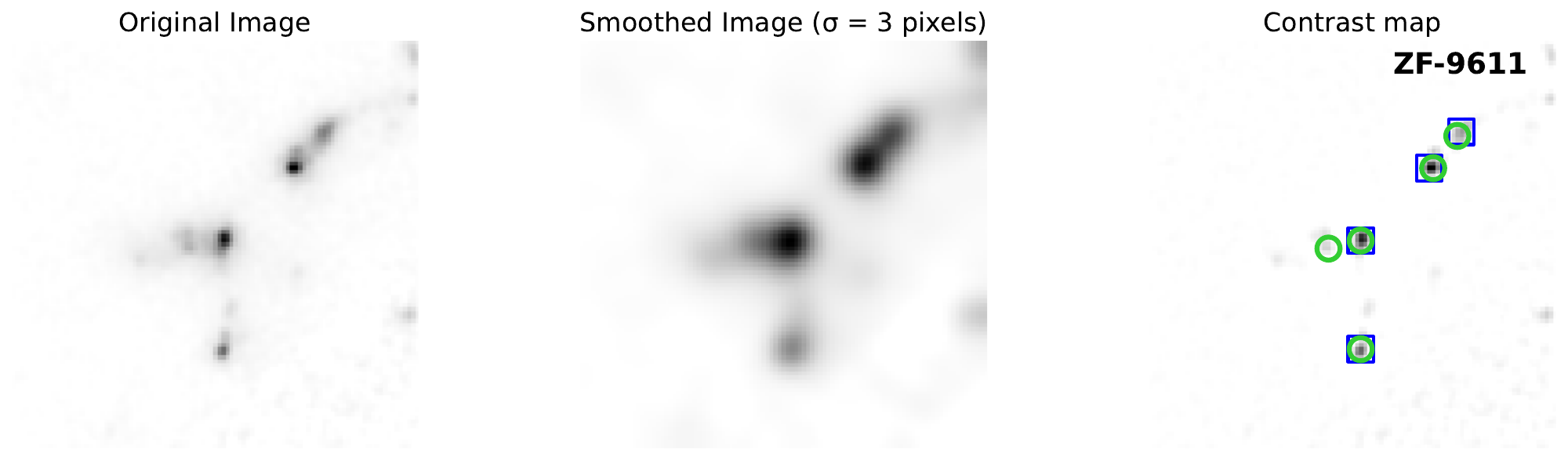}
    \hspace{2cm}
    \includegraphics[width=0.19\textwidth]{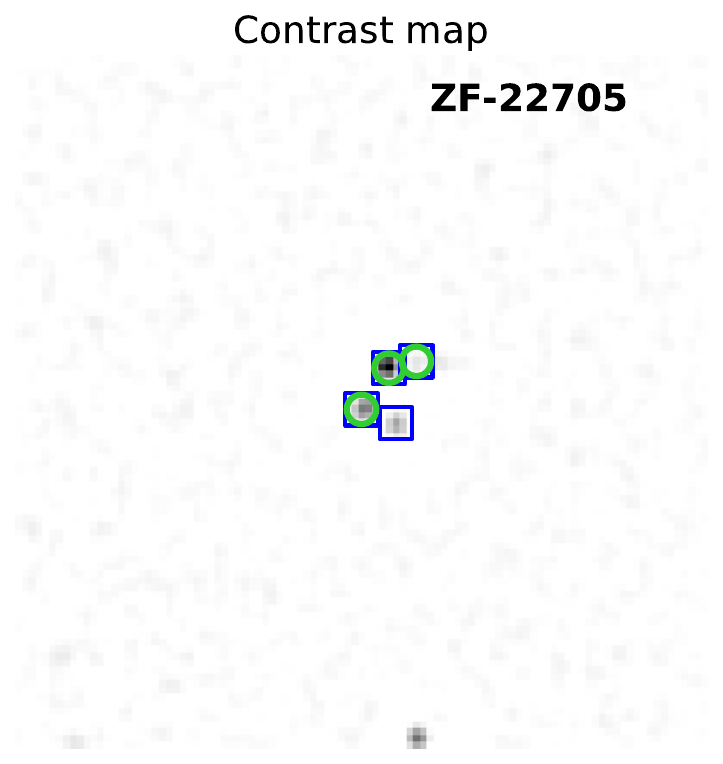}
    \includegraphics[width=0.19\textwidth]{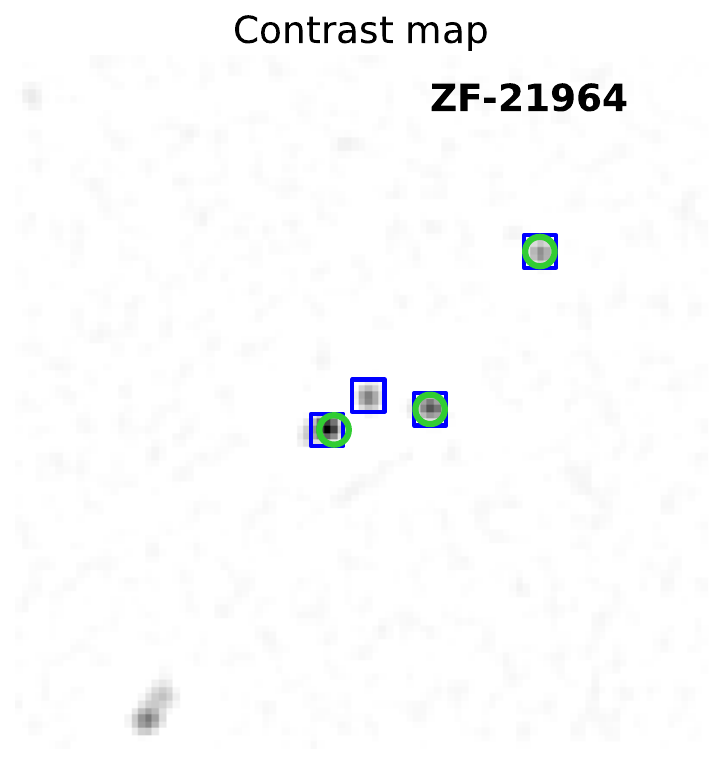}
    \includegraphics[width=0.19\textwidth]{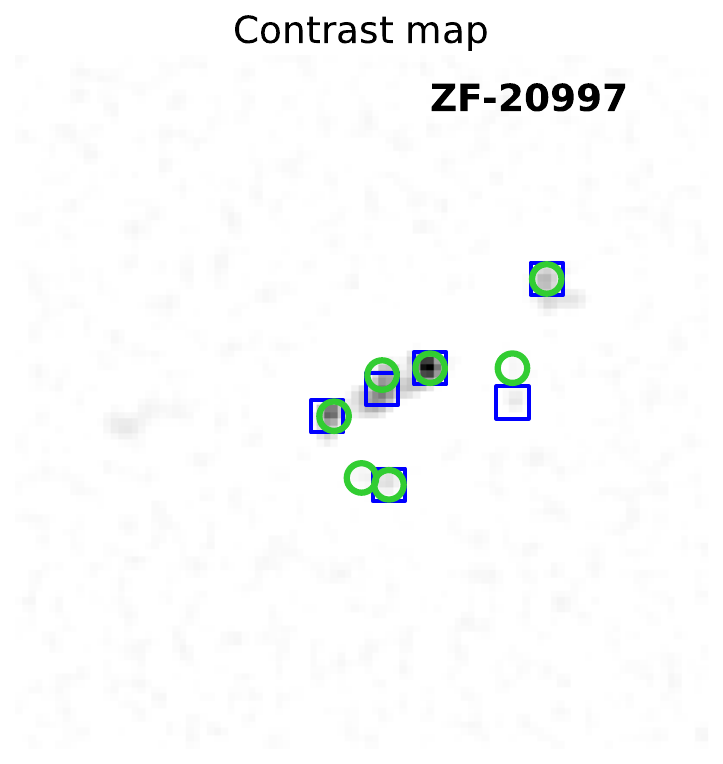}
    \includegraphics[width=0.19\textwidth]{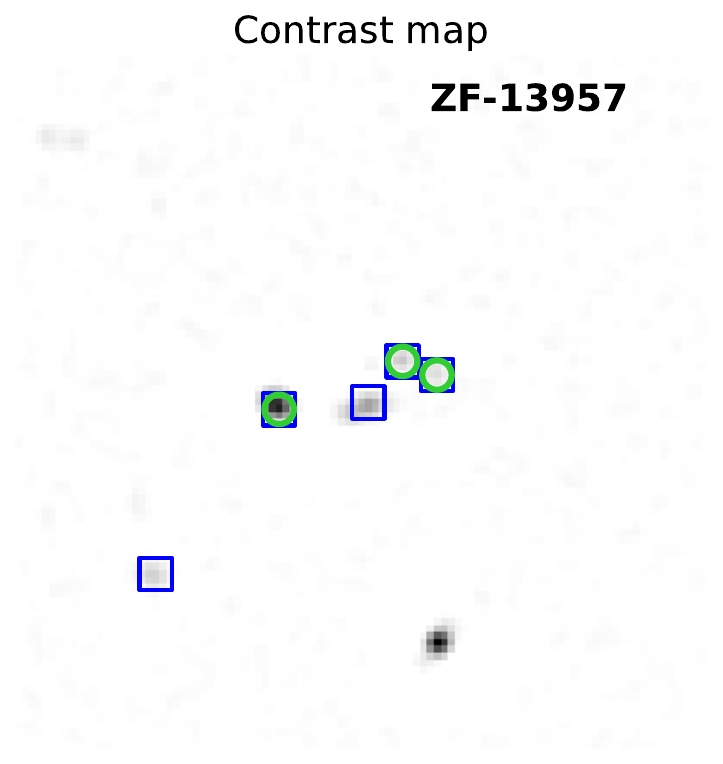}
    \includegraphics[width=0.19\textwidth]{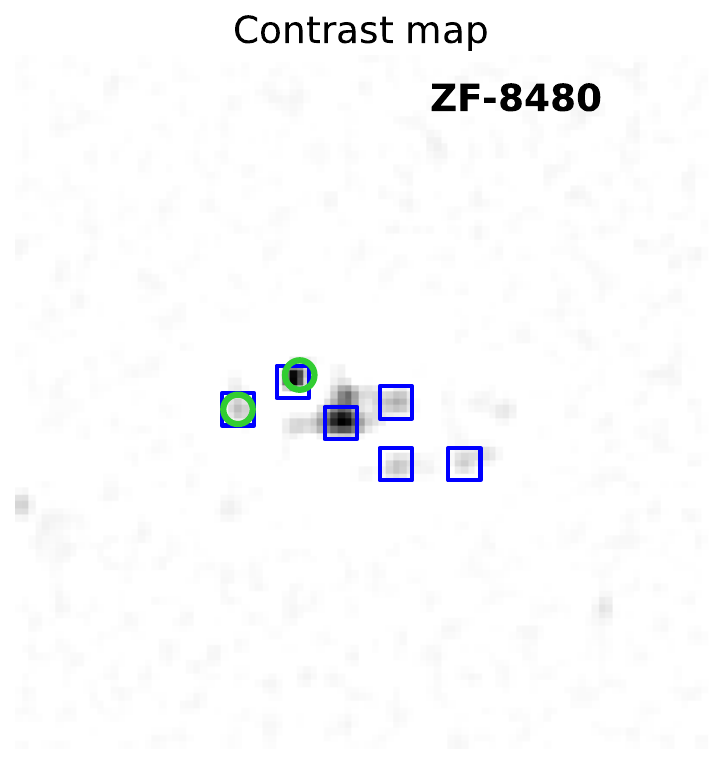}
    \label{fig:clumpmap}
        \caption{Illustration of the clump-detection process for HAEs. First, the original F090W image is smoothed using a Gaussian filter. The smoothed image is then subtracted from the original image to create a contrast image. The UV stellar continuum clumps are identified from the contrast image, which are marked as blue squares. For reference, Green Seeds are shown as green circles. The results shown here correspond to the six HAEs presented in Figure \hyperref[fig:emimap]{8}, with additional samples listed in Appendix \hyperref[sec:appendixsample]{C}.}
    \vspace{0.3cm}
\end{figure*}

We utilize the convolved F090W images and apply an automated clump-detection algorithm, following methods widely used for detecting clumps \citep[e.g.,][]{Conselice03,Guo15,Kalita24}. We first smooth the F090W images by a Gaussian filter with size of $\sigma=3$ pixels. Then, we subtract the smoothed image from the original image to create a contrast map. Next, we again run the \robotoThin{findpeaks} Python Package on the contrast map, similar to how we extracted the emission line regions in Section \hyperref[sec:method]{3}. This method successfully selects regions with peaked stellar continuum. In Figure \hyperref[fig:clumpmap]{12}, we illustrate the clump-detection method using the same six HAEs as in Figure \hyperref[fig:emimap]{8}. Previous works used Gaussian filters ranging from $3-6$ pixels \citep[e.g.,][]{Kalita24}. In our case, we find that a 3-pixel Gaussian filter works best for detecting clumps through visual inspection, and most UV star-forming clumps with typical sizes of $\sim\mathrm{1\,kpc}$ can be identified in the contrast image. The remaining contrast maps with UV star-forming clumps are displayed in Appendix \hyperref[sec:appendixsample]{C}.

Based on these contrast maps, we compare Green Seeds with the UV star-forming clumps selected from JWST observations. We find that many Green Seeds are also identified as UV stellar continuum clumps. This result is not surprising because UV clumps observed with HST also differ from their surrounding areas in terms of younger stellar age, higher sSFR and lower dusty attenuation \citep{Guo12}, similar to the appearance of Green Seeds. Moreover, compared to the UV clumps observed with HST at $z\sim2$, which are mostly more massive than $10^8\,M_{\odot}$, the clumps observed with JWST have a wider mass range, covering a large number of low stellar mass components down to less than $10^7\,M_{\odot}$.

From the contrast map in Figure \hyperref[fig:clumpmap]{12} and Appendix \hyperref[sec:appendixsample]{C}, we find more than half (70 out of 128) of the Green Seeds are co-located with the UV star-forming clumps. Also, \citet{Faisst24} combined JWST and ALMA data, finding that galaxies hosting UV star-forming clumps are likely to have larger gas fraction and higher star formation efficiency, suggesting a connection between UV star-forming clumps and gas inflows replenishment.

On the other hand, more than one-third of the Green Seeds are ``off-peak" to the UV star-forming clumps, with separations ranging from one pixel ($\sim\mathrm{200\,pc}$) to several pixels in the cutout stamps. Physically, the ``off-peak" appearance of emission line regions and UV stellar continuum could partly be explained by the different timescales of SFR indicators: UV light traces star formation over the past $\sim$100 Myr, while emission lines such as $\mathrm{H\alpha}$ trace a shorter timescale of $\mathrm{\sim10\,Myr}$ \citep[e.g.,][]{Kennicutt12,Theios19}. In our case, compared to UV clumps, a higher fraction ($\sim75\%$) of Green Seeds are co-located with $\mathrm{H\alpha}$ emission line peaks. This finding suggests that Green Seeds are more likely to track the star formation on shorter timescales. 
Another possible explanation could involve differences in dust attenuation. UV clumps typically experience stronger dust extinction compared to emission line regions in the galactic disk. This variation in dust extinction across the galactic disk may also contribute to the “off-peak” alignment between UV clumps and emission line regions.
The third possible explanation for the ``off-peak" appearance could be outflows driven by star formation feedback \citep[e.g.,][]{Rich10,Heckman17,Cresci17,ForsterSchreiber19}. Large amounts of ionized gas in galaxies could be pushed outward by the energy imparted from star formation activity. Observations of local SF-driven winds imply more efficient outflows in lower-mass galaxies \citep[e.g.,][]{Heckman15}. While the complex nature of these winds make them harder to be identified at high redshift. The outflow velocity maps from future high resolution IFU observations will be necessary to further constrain this viewpoint quantitatively. Note that there is no indication of AGN activity in the HAEs that contain Green Seeds (see Appendix \hyperref[sec:noo3apx]{A}).

\subsection{Comparison with Local H{\sc ii} regions}
Local H{\sc ii} regions, to some extent, share some similarities with Green Seeds, such as higher sSFR and lower chemical abundance compared to their host galaxies. Howevere, given the significant differences on the ISM properties between the local universe and Cosmic Noon \citep[e.g.,][]{Erb06,Liu08,Shapley15,Sanders18,Reddy18}, it is intriguing to compare local H{\sc ii} regions with Green Seeds at higher redshift. This comparison may reveal how galaxy and gas properties have evolved from early epochs to the present day. For this comparison, we refer to the recent PHANGS-MUSE survey \citep{Emsellem22}, which conducted IFU observations on 19 star-forming disc galaxies in the local universe, studying the physical properties of H{\sc ii} regions inside these galaxies with a physical resolution down to $\mathrm{100\,pc}$. These local H{\sc ii} regions are constructed from the $\mathrm{H\alpha}$ emission line map, with a detection threshold to $3\sigma$ above the background. The PHANGS-MUSE H{\sc ii} region catalogs \citep{Santoro22,Groves23} provide flux measurements and kinematic information for multiple optical emission lines in each H{\sc ii} region, along with the region area and physical properties.

\begin{figure*}[hbt!]
    \centering
    \includegraphics[width=0.48\textwidth]{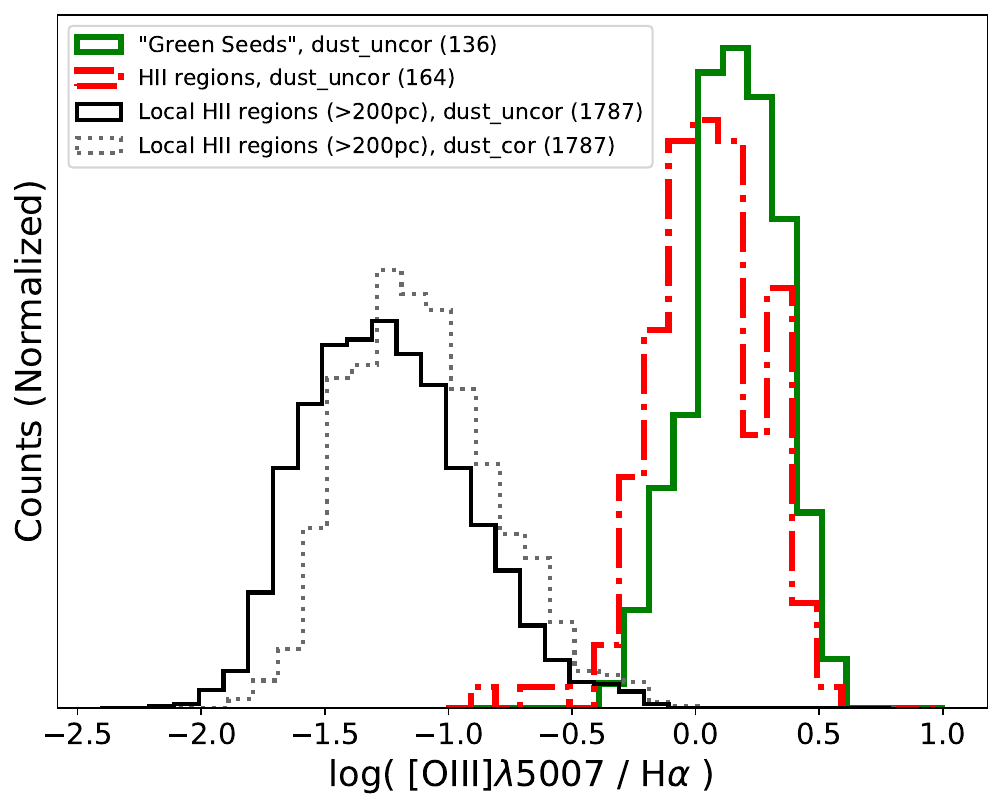}
    \includegraphics[width=0.48\textwidth]{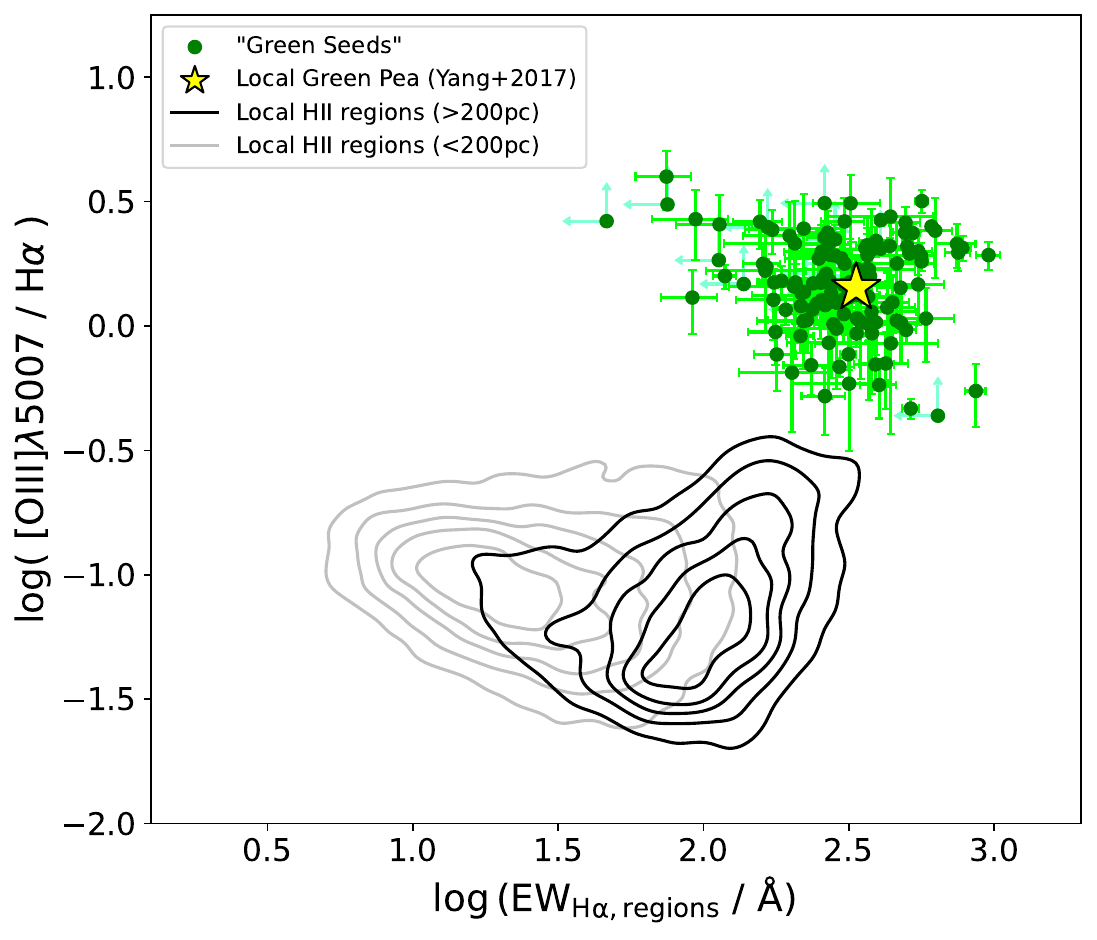}
    \label{fig:o3ha_histo}
    \caption{\textbf{Left}: The $\mathrm{[O\pnt{III}]\lambda5007}/\mathrm{H\alpha}$ ratios for Green Seeds, H{\sc ii} regions at $z\sim2$ from Section \hyperref[sec:dis1]{5.1}, and local H{\sc ii} regions ($>\mathrm{200\,pc}$) from the PHANGS-MUSE survey \citep{Groves23} are normalized distributed in a histogram. Green Seeds are represented by the green solid line, while H{\sc ii} regions at $z\sim2$ are shown as a red dotted-dashed line. For Green Seeds with weak $\mathrm{H\alpha}$ emission, we use the $2.5\sigma$ upper limit fluxes (see Section \hyperref[sec:sfr]{4.2}). We do not correct for dust attenuation in Green Seeds or H{\sc ii} regions at $z\sim2$, presenting the observed $\mathrm{[O\pnt{III}]\lambda5007}/\mathrm{H\alpha}$ ratios. For comparison, both dust-corrected and uncorrected line ratios of the local H{\sc ii} regions are exhibited as grey and black lines, respectively. While Green Seeds and H{\sc ii} regions at $z\sim2$ exhibit similar $\mathrm{[O\pnt{III}]\lambda5007}/\mathrm{H\alpha}$ ratio, there is a clear discrepancy when compared to local H{\sc ii} regions. \textbf{Right}: Relationship between $\mathrm{EW_{H\alpha}}$ and $\mathrm{[O\pnt{III}]\lambda5007}/\mathrm{H\alpha}$ ratios for Green Seeds distributed as green circles. Local H{\sc ii} regions with sizes $>\mathrm{200\,pc}$ ($<\mathrm{200\,pc}$) are represented by black (grey) contours. Despite having comparable $\mathrm{EW_{H\alpha}}$, Green Seeds display the highest $\mathrm{[O\pnt{III}]\lambda5007}/\mathrm{H\alpha}$ ratios. Additionally, the $\mathrm{EW_{H\alpha}}$ and $\mathrm{[O\pnt{III}]\lambda5007}/\mathrm{H\alpha}$ ratios in Green Seeds closely correspond to those of local Green Pea galaxies from \citet{Yang17}, with the median value of their 43 samples indicated by a yellow star.}
    \vspace{0.2cm}
\end{figure*}

In the PHANGS-MUSE survey, the median physical resolution is $7-20$ times higher than that of JWST observations at $z\sim2$ using the F277W filter ($\mathrm{0.7\,kpc}$) in this work. As a result, the catalog contains local H{\sc ii} region with areas ranging from $\mathrm{40\,pc}$ to $\mathrm{800\,pc}$, with an average size of $\mathrm{\sim110\,pc}$. This is smaller than the $\mathrm{1.2\,kpc}$ circular aperture ($0.''15$) used to identify Green Seeds in this work. To obtain a sample comparable to Green Seeds, we select local H{\sc ii} regions with sizes larger than $\mathrm{200\,pc}$ (assuming a circular shape inferred from the \robotoThin{region\_area} in the catalog), resulting in a sample of 1,787 local H{\sc ii} regions. 

Based on the estimation of [O{\sc iii}] and $\mathrm{H\alpha}$ emission line intensities in Section \hyperref[sec:result]{4}, we first compare the $\mathrm{[O\pnt{III}]\lambda5007}/\mathrm{H\alpha}$ ratios of Green Seeds and local H{\sc ii} regions in the left panel of Figure \hyperref[fig:o3ha_histo]{13} (using the line ratio of $\mathrm{[O\pnt{III}]}\lambda5007:\mathrm{[O\pnt{III}]}\lambda4959=2.97:1$ for conversion). Here, the PHANGS data are derived from the $\mathrm{H\alpha}$ emission map, while Green Seeds are extracted from [O{\sc iii}] emission map. The selection bias may introduce a higher $\mathrm{[O\pnt{III}]\lambda5007}/\mathrm{H\alpha}$ ratio in our sample. 
Thus, we also utilize the H{\sc ii} regions at $z\sim2$ from Section \hyperref[sec:dis1]{5.1}, which were first derived from $\mathrm{H\alpha}$ emission map, for additional comparison. The PHANGS-MUSE survey derived dust attenuation levels from the Balmer decrement, assuming an intrinsic Balmer ratio of $\mathrm{H\beta} = \mathrm{H\alpha}/2.86$, giving a higher $\mathrm{[O\pnt{III}]\lambda5007}/\mathrm{H\alpha}$ shown as the dashed histogram. On the other hand, considering the complexity of dust attenuation in H{\sc ii} regions at $z\sim2$ in this work, we do not correct for dust attenuation, instead comparing the observed $\mathrm{[O\pnt{III}]\lambda5007}/\mathrm{H\alpha}$ ratios for our samples in this discussion. 
We find that Green Seeds have a median ratio of $\mathrm{log([O\pnt{III}]\lambda5007}/\mathrm{H\alpha})=0.16$. Despite being selected from different emission line maps from JWST, Green Seeds and H{\sc ii} regions from the same host galaxies only show a slight discrepancy in the $\mathrm{[O\pnt{III}]\lambda5007}/\mathrm{H\alpha}$ ratio, with a larger median value by $\mathrm{\sim0.1\,dex}$ in Green Seeds. 
In contrast, Green Seeds exhibit a much higher $\mathrm{[O\pnt{III}]\lambda5007}/\mathrm{H\alpha}$ ratio than local H{\sc ii} regions by $\mathrm{\sim1.5\,dex}$ ($\sim30$ times). If we assume a fixed ratio between $\mathrm{H\alpha}$ and $\mathrm{H\beta}$ ($\mathrm{H\beta} = \mathrm{H\alpha}/2.86$), our sample is constant with the well-documented obvious enhancement in $\mathrm{[O\pnt{III}]\lambda5007}/\mathrm{H\beta}$ ratio at $z\sim2$ in previous studies \citep[e.g.,][]{Liu08,Yabe14,Steidel14,Shapley15}. 
The traditional BPT diagram \citep{Baldwin81} gives a theoretical maximum of $\mathrm{log([O\pnt{III}]\lambda5007}/\mathrm{H\beta})\simeq0.9$ for SFGs \citep{Kewley01, Steidel14, Nakajima22}, and we find that most of our Green Seeds are close to, but within, this upper limit. This ``maximum starburst” limit requires a high ionization parameter ($\mathrm{log}\,U>2$) and hard ionizing spectra ($T_\mathrm{{eff}}\sim\mathrm{50,000\,K}$) in the photoionization models from \citet{Steidel14}.
We also find that four Green Seeds exceed the ``maximum starburst” limit with their $\mathrm{[O\pnt{III}]\lambda5007}/\mathrm{H\beta}$ ratios. One possible explanation is the contamination by AGNs. Alternatively, \citet{Gutkin16} and \citet{Feltre16} used state-of-the-art stellar population and photoionization models with a wider range of parameters to model H{\sc ii} regions at higher redshift without AGNs, suggesting that $\mathrm{log([O\pnt{III}]\lambda5007}/\mathrm{H\beta})$ as high as $\sim1$ are possible in their models.

The PHANGS-MUSE H{\sc ii} region catalogs also provide $\mathrm{EW_{H\alpha}}$ for local H{\sc ii} regions. The $\mathrm{EW_{H\alpha}}$ shows the most solid correlation with stellar age and sSFR among the optical emission lines \citep{Reddy18}. We further examine the relationship between the $\mathrm{EW_{H\alpha}}$ and $\mathrm{[O\pnt{III}]\lambda5007}/\mathrm{H\alpha}$ ratio in the right panel of Figure \hyperref[fig:o3ha_histo]{13}. We find that local H{\sc ii} regions larger than $\mathrm{200\,pc}$ have $\mathrm{EW_{H\alpha}}$ comparable to Green Seeds, within $\mathrm{0.5\,dex}$, but their $\mathrm{[O\pnt{III}]\lambda5007}/\mathrm{H\alpha}$ ratios are almost $\mathrm{1.5\,dex}$ lower. This suggests that while both local H{\sc ii} regions and Green Seeds are primarily composed of young stellar population ($\sim\mathrm{10\,Myr}$) with comparable sSFR, they differ significantly in their ionization properties. The presence of top-heavy IMFs in Green Seeds may explain this phenomenon, as an increased number of high-energy photons could result in more doubly ionized oxygen ($\mathrm{O^{++}}$). For comparison, we also include the spectroscopic measurement of 43 local unresolved Green Pea galaxies from \citet{Yang17}. The distribution of $\mathrm{EW_{H\alpha}}$ and $\mathrm{[O\pnt{III}]\lambda5007}/\mathrm{H\alpha}$ for local Green Pea galaxies is nearly identical to that of our Green Seeds, as shown by the yellow star representing their median measurement in the panel. This result reveals that Green Seeds share very similar emission line properties with the local Green Pea galaxies. This finding is not surprising, as these local Green Pea galaxies have lower metallicity and higher sSFRs \citep{Amorin10}, making their properties more comparable to our higher-redshift observations than to other local observations.

The right panel of Figure \hyperref[fig:o3ha_histo]{13} also shows that the $\mathrm{EW_{H\alpha}}$ in the local H{\sc ii} regions may depend on their physical size, with larger H{\sc ii} regions ($>\mathrm{200\,pc}$) having $\mathrm{EW_{H\alpha}}$ nearly $\mathrm{1\,dex}$ higher than smaller ones. Simulations by \citet{Tamburello15} suggest that larger star-forming regions are more likely to form in galaxies with high gas fractions than in those with low gas fractions. Therefore, higher gas fractions could explain the higher sSFR and corresponding higher $\mathrm{EW_{H\alpha}}$ observed in larger H{\sc ii} regions.

Overall, the significant differences in the properties of local H{\sc ii} regions and Green Seeds, coupled with the rarity of Green Seeds in the local universe, suggest a fundamental shift in galaxy and gas properties from early epochs to the present day.

\section{The origin and fate of Green Seeds}
\label{sec:dis2}
\subsection{Formation of Green Seeds}
The commonly assumed framework, based on many observational and theoretical results, suggests that star-forming clumps have two possible formation mechanisms: (1) violent disk instability (VDI), regarded as ``in situ" origins \citep{Dekel09b,Mandelker14,Mandelker17,Dekel22}; and (2) galaxy mergers, also regarded as ``ex situ" origins \citep{DiMatteo08,Renaud15,Moreno19,Sparre22}. Considering the similarities between Green Seeds and star-forming clumps discussed in Section \hyperref[sec:compareuv]{5.2}, we propose that Green Seeds may share similar origins. However, previous studies of clumpy structures were mostly focusing on massive galaxies with stellar mass larger than $10^{10}\,M_{\odot}$ \citep[e.g.,][]{Elmegreen05, Guo12, Shibuya16}. The differing properties of their host galaxies may lead to discrepancies in the origin of Green Seeds and star-forming clumps.

In the VDI scenario, star-forming clumps are predicted to form in regions of thick, gas-rich galaxy disks, where the high surface density of gas and young stars drives the Toomre $Q$ parameter \citep{Toomre64} below unity, leading to gravitational disc instability. \citet{Guo15} and \citet{Shibuya16} found that the fraction of clumpy galaxies among SFGs is consistent with the cosmological evolution of VDI from \citet{Cacciato12}, suggesting that VDI is a major mechanism for forming star-forming clumps. Since VDI requires host galaxies to have disk-like underlying components, galaxies in HST observations from \citet{Shibuya15} exhibited a higher fraction of clumpy galaxies in systems with a low S\'{e}rsic index of $n\sim1$, supporting the VDI scenario \citep{Shibuya16}. 

\begin{figure*}[hbt!]
    \centering
    \includegraphics[width=0.47\textwidth]{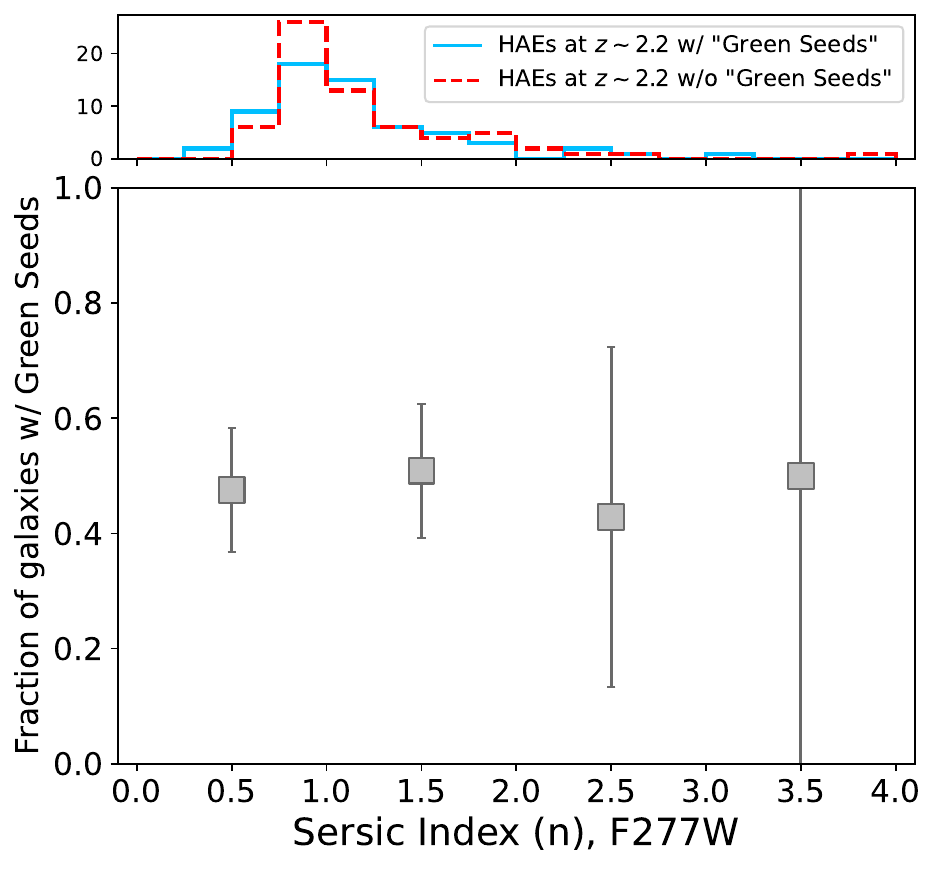}
    \includegraphics[width=0.49\textwidth]{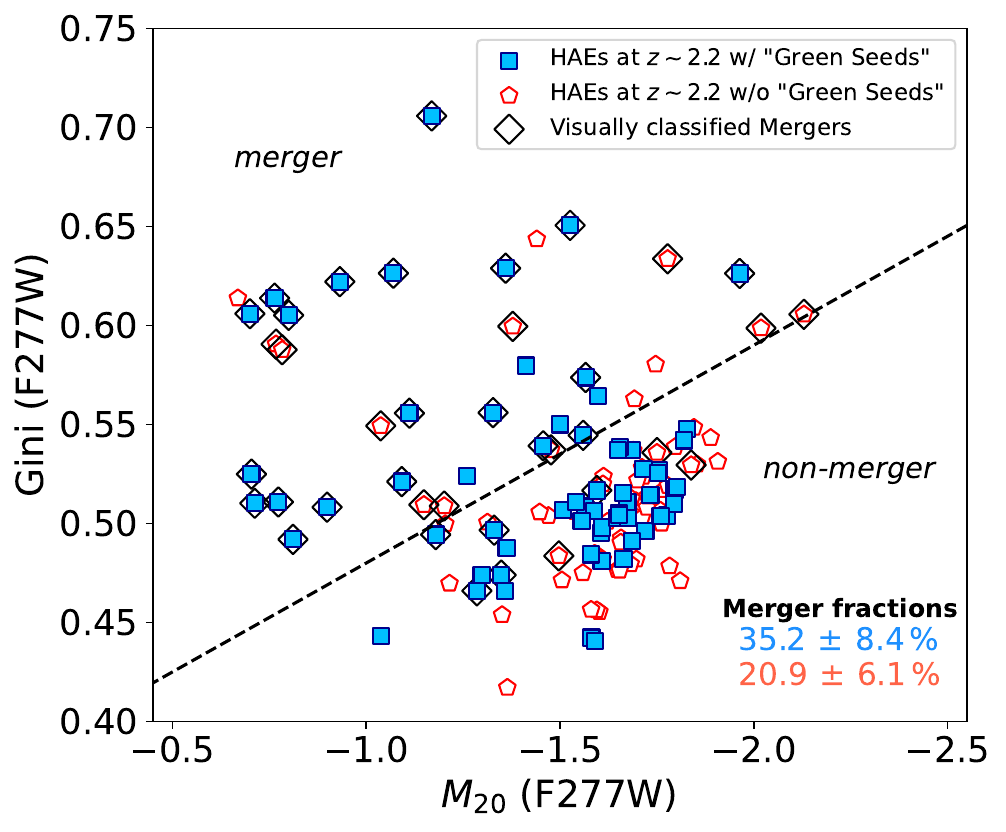}
    \label{fig:sersic}
    \caption{\textbf{Left}: Dependence of the fraction of HAEs with Green Seeds on the S\'{e}rsic index (n) measured in the F277W cutout images of all cross-matched HAEs. The error bars are given by Poisson statistics from the galaxy number counts. In the VDI scenario, a higher fraction of galaxies with Green Seeds would be expected in those with a lower S\'{e}rsic index \citep{Shibuya16}. However, we do not observe such a clear dependence in our sample, making it hard to determine whether VDI is a major mechanism for triggering Green Seeds. \textbf{Right}: Diagram of the Gini coefficient ($G$) versus the second-order moment of the brightest 20\% of the galaxy’s flux ($M_{20}$), measured in the F277W cutout images. HAEs with Green Seeds are shown as blue squares, and those without Green Seeds are represented by red open pentagons. We also conduct a visual classification of mergers directly from the cutout images, with classified mergers indicated by black squares. The dashed line represents the threshold for ongoing mergers or non-merger, as defined by \citet{Lotz04}. We observe a potentially higher fraction of ongoing mergers among HAEs with Green Seeds.}
    \vspace{0.2cm}
\end{figure*}

To investigate the structural parameters of the host galaxies in this work, we conduct the \robotoThin{GALFIT} profile fitting \citep{Peng02, Peng10} on the two-dimensional surface brightness profile from the F277W cutout images of all 135 cross-matched HAEs explained in Section \hyperref[sec:sampselect]{2.3}. 
We fit them by single S\'{e}rsic profiles centered on the galactic nuclei convolved by the F277W PSF profile from \robotoThin{WebbPSF} and optimizes the fits for $\chi^2$ minimization. The noise images are obtained from the inverse square root of the JADES weight maps \citep{Rieke23}, following \citet{Ono23}. The initial magnitude and effective radius used for the \robotoThin{GALFIT} profile fitting are taken from JADES catalog, but all parameters are allowed to vary during the profile fitting process. Based on the initial outputs, we exclude 22 unreliable or failed fits, mostly due to nearby counterparts in the 3” × 3” cutout images. 
These outliers are re-fitted with \robotoThin{GALFIT} using a double S\'{e}rsic components mode to separate the central galaxy from its counterpart. This process provides an additional good sample, but 8 HAEs are still excluded because of their complicated components (e.g., ID:9611 in Figure \hyperref[fig:rgb_main]{3}). 
Overall, the mean S\'{e}rsic index ($n$) of the HAEs with Green Seeds is slightly lower than that of the HAEs without Green Seeds by $\sim0.1$, indicating a minor difference in the Sérsic profiles of these two populations. 
In the left panel of Figure \hyperref[fig:sersic]{14}, we also present the relationship between the fraction of galaxies with Green Seeds and the S\'{e}rsic index. 
Unlike the clear correlation in \citet{Shibuya16}, where the clumpy galaxy fraction is higher in galaxies with a lower S\'{e}rsic index of $n\sim1$, our sample does not show a clear correlation between these two parameters. 
However, as displayed in the upper histogram of the same panel, a large fraction of HAEs have a S\'{e}rsic index $n\simeq1$, indicating that disk-like light profiles are generally present in the HAE sample, which could still be consistent with the VDI scenario. 
While considering our much smaller sample size compared to the $\sim190,000$ galaxies in \citet{Shibuya16}, especially the lack of galaxies with a S\'{e}rsic index $n>2$, we cannot draw a definitive conclusion about VDI being a major mechanism for triggerring the formation of Green Seeds. A larger sample is needed to further investigate this scenario.

During the \robotoThin{GALFIT} profile fitting, several cases of irregular morphology in HAEs with Green Seeds suggested the presence of galaxy mergers, as we find at least one or more bright counterparts in the 3” × 3” cutout images. We determine whether a galaxy is an ongoing galaxy mergers by Gini coefficient ($G$) and a second-order moment of a galaxy ($M_{20}$), which are one of the major non-parametric methods for quantifying galaxy morphology \citep{Lotz04}. The Gini coefficient is a statistic based on the Lorenz curve of fluxes per pixel in a galaxy and it represents the relative distribution of pixels covering the galaxy, while the other parameter $M_{20}$ is a normalized second-order moment of pixels which is measured to be the brightest 20\% flux in a galaxy. \citep{Lotz04} defined a threshold for distinguishing mergers from non-mergers, represented by the dashed line in the right panel of Figure \hyperref[fig:sersic]{14}. 
We use the \robotoThin{statmorph} Python package  \citep{RodriguezGomez19}, an affiliated package of \robotoThin{Astropy} \citep{Astropy22}, to calculate non-parametric morphological diagnostics on the cutout images of HAEs. Both HAEs with and without Green Seeds are analyzed, and the results are displayed in the same panel with respective symbols. Independently, we also carry out a visual classification of mergers directly from every cutout images, with visually classified mergers represented as black squares. We figure out that the non-parametric method and visual classification provide quite similar results, with fewer than 20\% of visually classified mergers being outliers in the $G-M_{20}$ diagram. For the following discussion, we use the merger fractions calculated from the non-parametric method.

We identify that the merger fraction of HAEs with Green Seeds is $34.8\pm8.2\,\%$, marginally higher than that of HAEs without Green Seeds of $20.9 \pm 6.1\,\%$. 
This finding raises the possibility of ``ex situ" origins for Green Seeds. In this scenario, galaxy mergers could drive external violent processes, triggering turbulent modes in the ISM that lead to rapid fragmentation of gas and the formation of clumps \citep{Bournaud11,Renaud14}. 
Observations also support a correlation between star-forming clumps and mergers; for example, \citet{Calabro19} used the same $G-M_{20}$ diagrams from HST images and found a factor of three higher clumpiness in mergers compared to the rest of the population from nearly 200 massive galaxies. Mergers and galaxy–galaxy interactions could also drive extreme emission lines, which are directly correlated with the physical background of Green Seeds. In an unresolved view, \citet{Gupta23} found out that EELGs are mostly surrounded by massive companion galaxies and are more likely to have recently experienced strong interactions. The merger-induced bursty star formation histories produce extreme emission lines on a significantly short timescale. Notably, due to the limited sample size, the $1\sigma$ lower limit of the merger fraction of HAEs with Green Seeds is still lower than the $1\sigma$ upper limit of the merger fraction of HAEs without Green Seeds, indicating that we cannot make a definitive conclusion about a higher merger fraction in HAEs with Green Seeds. Furthermore, our HAEs exhibit a wide diversity on their morphology, from the smooth disk-like structure to the irregular and merger-like structure. Since these merger systems constitute less than half of the full HAE sample, it is difficult to conclude that galaxy mergers are the primary mechanism for triggering the formation of Green Seeds. 

Recently, \citet{Dekel22} simulated a wide range of clumps, from in situ clumps to ex situ clumps, and identified distinguishing features that ex situ clumps tend to populate the outer disc, while in situ clumps are expected at all radii. These simulated results generally correspond to Green Seeds through visual inspection. Although we cannot quantitatively determine which mechanism is more important for forming Green Seeds, we suggest that both mechanisms are active in our sample.

Additionally, C24 noted that a significant fraction of low-mass HAEs lie above the SFMS in the COSMOS and UDS fields, while this sample is lacking in the GOOD-South field, as shown in the middle panel of Figure \hyperref[fig:samp]{2}. This enhanced SFR in host galaxies may be related to environment factors, such as the presence of galaxy clusters \citep{Spitler12,Shimakawa18a, Momose21}. It remains unclear whether Green Seeds are more frequently found in clustered environments, where galaxy mergers are more common in higher galaxy overdensity \citep[e.g.,][]{Lotz13,Liu23,Shibuya24}. Therefore, a larger sample of HAEs with deep JWST images in the COSMOS and UDS fields is needed to address this question.

\subsection{Fate of Green Seeds}
While the origins of Green Seeds may be linked to VDI and/or galaxy mergers, their ultimate fate remains a topic of debate despite various theoretical studies and numerical simulations have been done. Early simulations by \citet{Elmegreen08} showed that clumpy structures interact strongly with disk stars, gradually lose their angular momentum, migrate toward the gravitational centers of their host galaxies, and eventually coalesce into a young bulge within $\mathrm{500\,Myr}$. 
In Section \hyperref[sec:result]{4}, we have compared various properties of Green Seeds and Red Seeds. Previous studies based on HST images mainly focused on rest-frame UV-bright structures, lacking observations of these redder systems. Red Seeds are located closer to the galaxy center (central pixel) than Green Seeds (a median distance of $\mathrm{0.85\pm0.36\,kpc}$ vs. $\mathrm{1.44\pm0.79\,kpc}$). Moreover, we compare the stellar age of Green Seeds and Red Seeds from SED fitting estimation, and reveal a clear discrepancy: the median stellar age of Red Seeds ($\mathrm{251\pm35\,Myr}$) is much older than that of Green Seeds ($\mathrm{74\pm50\,Myr}$). These findings probably suggest that these ``red” and massive clumps are likely migrating toward the gravitational center, supporting the so-called ``migration" scenario. 

Along with this migration scenario, \citet{Shapiro10} proposed that these massive gas clumps may be the formation site of globular clusters \citep[GCs;][]{Brodie06}. As the clumps migrate towards the centre, a
small fraction of the mass is being stripped off during this process and remaining in the galaxy disc to form the metal-rich GCs. These metal-rich GCs are preferentially close to the galaxy center, typically located in the galactic thick disc.

Notably, the above simulation results supporting the migration scenario are based on giant clumps with stellar mass of $10^{8-9}\,M_{\odot}$, which applies to most Red Seeds but only a few Green Seeds. In Figure \hyperref[fig:mass_histo]{9}, we show that Green Seeds have stellar masses ranging from $10^{6.5}\,M_{\odot}$ to $10^{8.5}\,M_{\odot}$. As a result, Green Seeds with stellar mass lower than a few $10^7\,M_{\odot}$ may not be suitable for the migration scenario. Also, because the accretion of gas from the surrounding disk and gaseous outflows from stellar feedback are happening simultaneously \citep{Bournaud14}, it is difficult for low-mass Green Seeds to rapidly accumulate their stellar masses. As suggested by \citet{Dekel22}, less massive Green Seeds are typically short-lived, in most cases $<\mathrm{200\,Myr}$. In this simulation, the cold gas reservoirs are expelled through stellar feedback, supernova feedback, and tidal torques due to their shallower potential wells. The short-lived nature means that low-mass Green Seeds may be disrupted during or before migration toward galactic centers, eventually becoming part of the disk of their host galaxies.

From the continuum map and emission map in Figure \hyperref[fig:emimap]{8} and Appendix \hyperref[sec:appendixsample]{B}, we also observe several faint, low-mass Green Seeds located away from galaxy disks, which do not seem to fit the two scenarios mentioned above, likely only applicable within galaxy disks. We consider them to be an outlier populations that may follow a different evolution path. One possible assumption is that these isolated Green Seeds may be the progenitors of metal-poor GCs observed in the local universe. The metal-poor GCs are thought to be formed from young massive clusters in the halos of their host galaxies \citep{Forbes08, Kruijssen14, Forbes18}. Cosmological zoom-in simulations by \citet{Mandelker18} presented a model in which cold filamentary accretion forms GCs. It is shown that cold streams can fragment into dense clusters not associated with the disk structure, whereas star formation occurs. The largest cluster in the simulation has a radius of $\sim\mathrm{1\,kpc}$ and stellar mass of $4\times10^{6}\,M_{\odot}$, corresponding to the very low-mass Green Seeds in Figure \hyperref[fig:mass_histo]{9}. The typical stellar masses of local GCs are in the range $\sim10^{4}\,M_{\odot} - 10^{6}\,M_{\odot}$ \citep{Brodie06}, while these surviving GCs are likely have undergone mass loss since their formation, requiring them to have been roughly 20 times more massive than their present-day masses in the assumed models \citep{Kruijssen14}. Therefore, the stellar masses of low-mass Green Seeds, which are $\sim10^{7}\,M_{\odot}$, could be reasonable if we consider them to be proto-GCs that have experienced extensive mass loss to evolve into the local metal-poor GCs. 

Another possible scenario is that these isolated low-mass Green Seeds could become typical satellite galaxies, and then evolve into ultracompact dwarf galaxies \citep[UCDs;][]{Drinkwater00,Phillipps01} in the nearby Universe. The stellar masses of UCDs are ranging within $M_*=10^{6}-10^{8}\,M_{\odot}$, quite close to low-mass Green Seeds. While UCDs typically hold sizes less than $\mathrm{100\,pc}$, recent detections of extended stellar envelopes around a number of luminous UCDs suggest that UCDs are the remnants of nucleated dwarf galaxies that have survived tidal stripping \citep{Liu15,Liu20,Wang23}. The stellar envelopes and expected tidal radius for these objects are extended to several $\mathrm{100\,pc}$, which is close to typical sizes of isolated low-mass Green Seeds.

Some of the isolated Green Seeds away from the host galaxies might be foreground or background contaminants at different redshifts. However, from the images, more than half of them seem to be connected to their host galaxies with possible traces of tidal structures. Therefore, we think that foreground or background contamination does not significantly impact our results.

Overall, we list several possible scenarios for the evolution path of Green Seeds mainly based on their average stellar mass and size. These results are driven from the multi-band photometric data from JWST/NIRCam. However, there is a lack of critical kinematic information on Green Seeds to further support these scenarios. To address these limitations and make further progress, we anticipate that the JWST/NIRSpec IFU spectroscopy will play a crucial role.

\section{Summary}
\label{sec:con}
In this work, we investigate the highly resolved and deep NIRCam imaging from the JWST Advanced Deep Extragalactic Survey \citep[JADES][]{Rieke23} on a cross-matched sample of 135 $\mathrm{H\alpha}$ emitters (HAEs) at $z\sim2.2$, originally from the ZFOURGE-CDFS field \citep{Straatman16}. These HAEs cover a wide range of stellar masses, from $10^8\,M_{\odot}$ to $10^{10}\,M_{\odot}$. From the convolved RGB (F115W+F150W+F277W) images, we identify a large number of kiloparsec-scale resolved [O{\sc iii}]+$\mathrm{H\beta}$ emission line regions, exhibiting prominent green colors (flux excesses in F150W) due to their high equivalent widths (EWs). We apply an EW-limited algorithm to extract these regions and designate them as Green Seeds. We further derive the physical properties of Green Seeds, including stellar mass and star formation rate (SFR), based on the spatially resolved photometry from F090W to F277W. To explore the origin and fate of Green Seeds, we compare them with other resolved structures at similar redshift and from the local universe, incorporating simulation results. Our main findings and conclusions are as follows:

\begin{enumerate}
    \item Utilizing the resolved stellar continuum map and emission line map in the F150W filter, we successfully identify 187 [O{\sc iii}]+$\mathrm{H\beta}$ emission line regions with $\mathrm{EW_{[O\pnt{III}]+H\beta}}>200\mathrm{\AA}$. Based on a series of SED model grids, we develop a color selection to separate them into two populations: 128 Green Seeds from 68 HAEs, characterized by their distinctly green colors, and 59 Red Seeds. The substantial number of Green Seeds enable us to statistically study their physical properties.
    \item We estimate the stellar mass ($M_*$) of each Green Seed, using both color-mass-to-light ratio and SED fitting methods, which provide closely aligned estimations. The stellar masses of Green Seeds are mostly ranging from $10^{6.5}\,M_{\odot}$ to $10^{8.5}\,M_{\odot}$, with a median stellar mass of $10^{7.4}\,M_{\odot}$. In contrast, Red Seeds are more massive than Green Seeds by $\sim\mathrm{1\,dex}$.
    \item Most Green Seeds also exhibit significant flux excesses in the F200W filter, driven by $\mathrm{H\alpha}$ emission lines. We derive the SFRs of Green Seeds from their intrinsic $\mathrm{H\alpha}$ luminosities. The $M_*$-SFR relation of Green Seeds has a higher normalization than that of host galaxies by $\sim0.6\,\mathrm{dex}$, indicating $\sim4\times$ more rapid stellar mass assembly in Green Seeds.
    \item Green Seeds have a median equivalent width of $\mathrm{EW_{[O\pnt{III}]},med} = 431\mathrm{\AA}$,  due to the intense ionizing radiation fields in Green Seeds. We identify 17 Green Seeds with extremely large rest-frame $\mathrm{EW_{[O\pnt{III}]}}>1000\mathrm{\AA}$, potentially indicating resolved LyC leakage regions in galaxies at $z\sim2$.
    \item We compare Green Seeds with UV star-forming clumps (from F090W) and H{\sc ii} regions (from F200W) in the same host galaxies. As all these spatially-resolved structures are associated with star-forming activities, we observe the co-location among them in most cases. While Green Seeds are sometimes ``off-peak" relative to UV star-forming clumps, which could be explained by differences in the timescales of star formation or ionized gas outflows driven by star formation feedback.
    \item A comparison between Green Seeds and large H{\sc ii} regions in the local universe from the PHANGS-MUSE survey \citep{Groves23} reveals a much higher $\mathrm{[O\pnt{III}]\lambda5007}/\mathrm{H\alpha}$ ratios by $\sim\mathrm{1.5\,dex}$ in Green Seeds, but comparable $\mathrm{EW_{H\alpha}}$ between the two populations. This suggests that Green Seeds may have lower metallicity and top-heavy IMFs, indicating the discrepancy in the gas and galaxy properties between the local universe and Cosmic Noon.
    \item Our analysis on the morphology of host galaxies supports two different formation mechanisms proposed by theoretical works, that Green Seeds may form through violent gravitational instability (in situ) and/or galaxy mergers (ex situ). Still, we cannot definitively determine which scenario plays the dominant role in the formation of Green Seeds.
    \item Comparing with Green Seeds and Red Seeds on galaxy disks, we observe radial and stellar age variations between them. This may suggest the migration scenario, in which Green Seeds gradually migrate toward the centers of their host galaxies, evolve into Red Seeds, and eventually coalesce to build the central bulge of their host galaxies. While those low-mass ($<10^8\,M_{\odot}$) Green Seeds are more likely to dissipate within the galactic disks. Additionally, we connect Green Seeds to the formation of local galactic substructures. These isolated Green Seeds could be considered progenitors of GCs or UCDs in the local universe. Our observational results of Green Seeds align well with theoretical models on stellar mass and size.
\end{enumerate}

Ongoing observations with the JWST are allowing us to explore the early universe in unprecedented detail. The NIRCam images are transforming our approach to study galaxies, enhancing our understanding of their internal structures and mass assembly processes, especially in the critical low-mass regime and at rest-frame optical or even longer wavelengths. The systematic study of emission line regions conducted in this work has the potential to significantly deepen our understanding of galaxy structure and evolution at Cosmic Noon. To further investigate and refine our knowledge of these Green Seeds, follow-up high-resolution spectroscopic observations are essential to uncover the detailed kinematics and ISM properties within them.

\section*{Acknowledgement}
The authors thank referee for the insightful suggestions and comments that helped improve this article. We are grateful for enlightening conversations with Rebecca L. Davies, Kimihiko Nakajima, Daichi Kashino, Akio Inoue and Chengze Liu. This work is mainly based on observations made with the NASA/ESA/CSA James Webb Space Telescope.
The JWST JADES data presented in this paper are obtained from the Mikulski Archive for Space Telescopes (MAST) at the Space Telescope Science Institute \citep{JADES_data,JEMS_data}. We thank the PHANGS-MUSE collaboration for providing their data releases, and ``Green Pea" catalog from Huan Yang, which include the measurements of emission line fluxes in the local universe. Data analysis was in part carried out on the Multiwavelength Data Analysis System operated by the Astronomy Data Center (ADC), National Astronomical Observatory of Japan. This work was supported by the SPRING GX Program from the University of Tokyo, JSPS Core-to-Core Program (grant number: JPJSCCA20210003), and MEXT/JSPS KAKENHI grant Nos. 20H00171 and 22K21349.

\section*{\textbf{Appendix A\\The cutout images of remaining cross-matched HAEs}}
\label{sec:noo3apx}
\begin{figure*}[hbt!]
    \centering
    \includegraphics[width=1\textwidth]{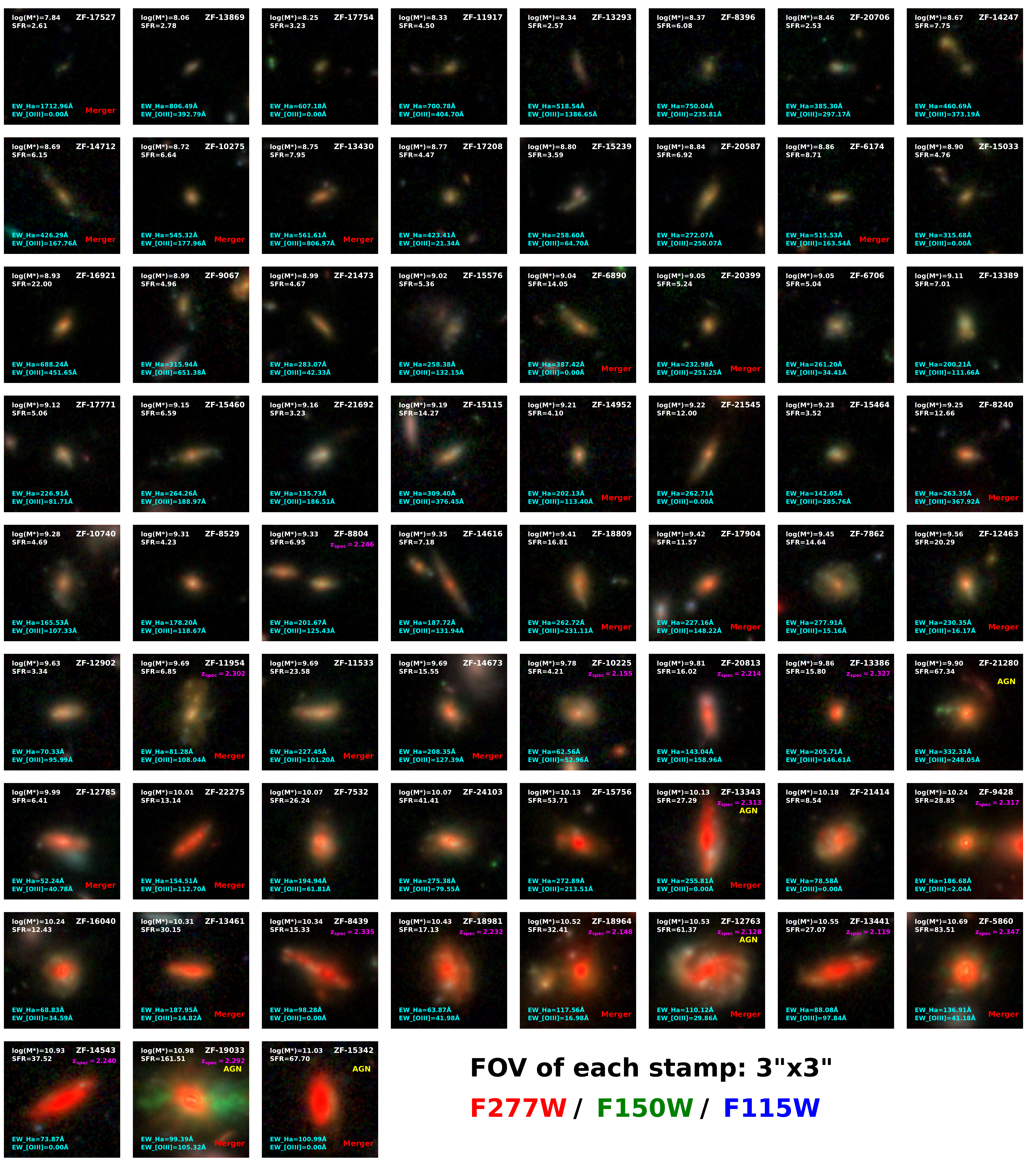}
    \label{fig:rgb_other}
    \vspace{-0.2cm}
    \caption{JWST/NIRCam cutout images of the other 67 HAEs with size of 3” × 3”, which are not exhibited in Figure \hyperref[fig:rgb_main]{3}. Outlines as in Figure \hyperref[fig:rgb_main]{3}. We do not extract any Green Seeds from these HAEs. This sub-sample of HAEs have a higher average stellar mass than that contains Green Seeds. The AGN identified in \citet{Cowley16} are labeled at the bottom-right corner. The HAEs in Figure \hyperref[fig:rgb_main]{3} contain no AGN. Notably, an AGN (ZF-19033) may exhibit a pair of strong warm ionized outflow traced by the [O{\sc iii}] emissions, and another AGN (ZF-21280) also exhibit a possible ionized outflow. These two AGNs are beyond this work, thus we do not further discuss their galaxy properties.}
\end{figure*}

In this work, a total of 135 HAEs is cross-matched as shown in Figure \hyperref[fig:samp]{2}. In Section \hyperref[sec:method]{3}, we also illustrate how we extract Green Seeds from the combination of resolved stellar continuum map and emission line map, and we present the cutout color images of a series of HAEs built combining the F115W, F150W, and F277W filters which contains Green Seeds in Figure \hyperref[fig:rgb_main]{3}. Still, from the whole sample of host galaxies, we have a number of 67 HAEs that do not contains Green Seeds and we present the RGB images of these rest HAEs in Figure \hyperref[fig:rgb_other]{15}. As is shown here, these sample do not reveal a prominent resolved compact green structures.

Considering the integrated galaxy properties of HAEs in C24, we found about two thirds of the HAEs exhibit strong [O{\sc iii}] emission lines, which is slightly higher than the fraction of the HAEs which contains Green Seeds. One reason could be the errors on the photometric redshifts of the HAEs from the ZFOURGE survey \citep{Straatman16}. Although ZFOURGE is able to constrain such errors at $\sim3\%$ at $z>2$, some [O{\sc iii}] emission lines may still drop out from the F150W filter of JWST as outliers.

We also mention one special object: ID:19033 in Figure \hyperref[fig:rgb_other]{15}, which actually contains strong [O{\sc iii}] emission lines. This object is also classified as an AGN in \citet{Cowley16}. Interestingly, the [O{\sc iii}] emission on both sides exhibits a symmetric filament structure, indicating a warm ionized outflow originating from the galaxy centre and extending over $\mathrm{10\,kpc}$. This structure is similar to the large ionized Hanny’s Voorwerp features in nearby AGN, which have strong associations with mergers and radio jets/outflows. Furthermore, the JWST imaging uncovers an ongoing merger system in the galaxy’s nuclear region, revealing two closely separated point sources by an distance of 4 pixels, equating to a physical distance of $\sim\mathrm{1.5\,kpc}$. These two new findings outstand the uniqueness of this AGN, especially at high redshift. However, a detailed analysis on this object is beyond the scope of this paper.

\section*{\textbf{Appendix B\\The Continuum and Emission-line Map for the remaining HAEs with Green Seeds}}
\label{sec:appendixsample}
\begin{figure*}[p]
    \centering
    \includegraphics[width=0.32\textwidth]{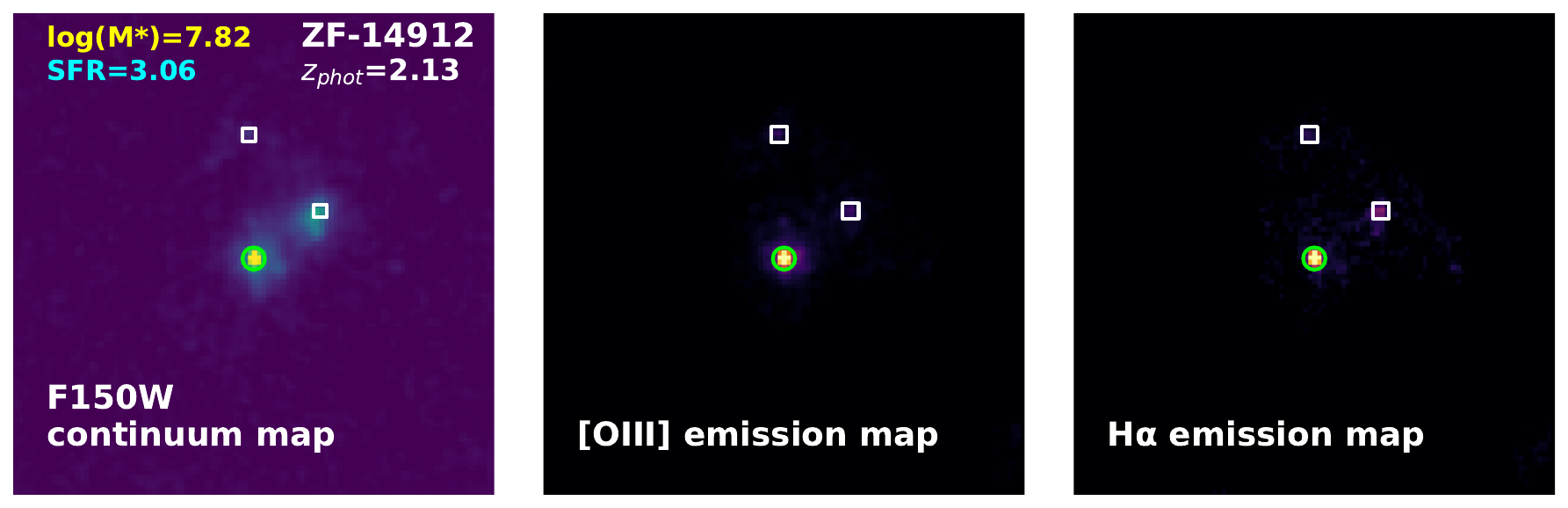}
    \hspace{0.1cm}
    \includegraphics[width=0.32\textwidth]{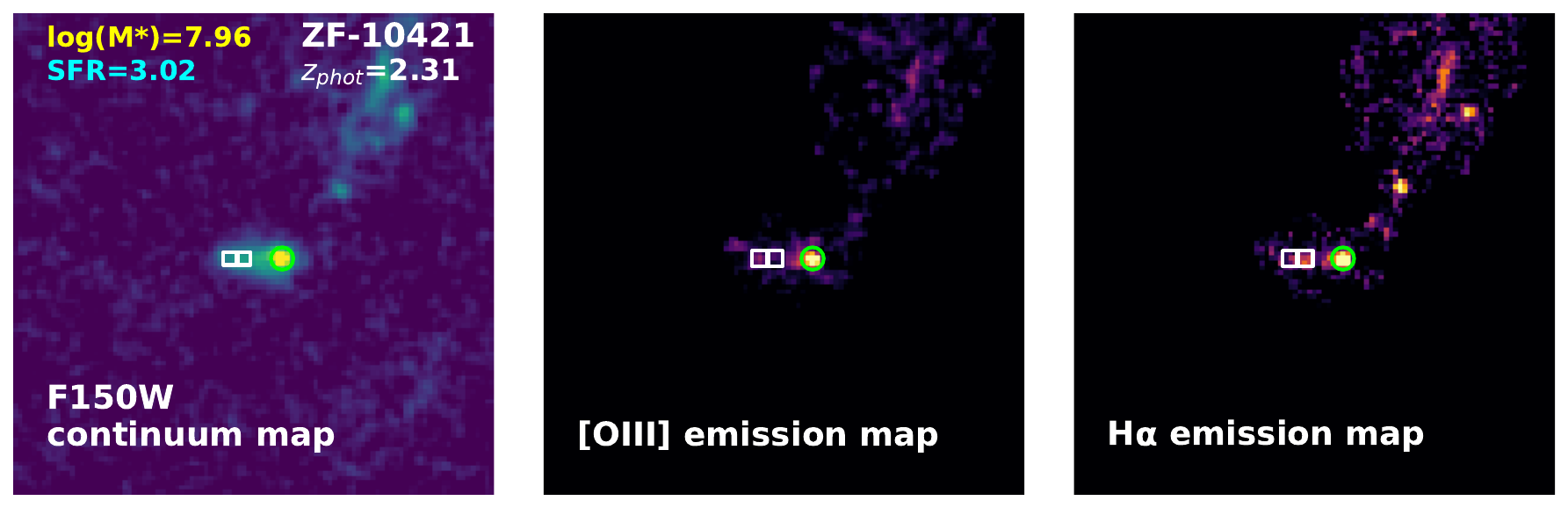}
    \hspace{0.1cm}
    \includegraphics[width=0.32\textwidth]{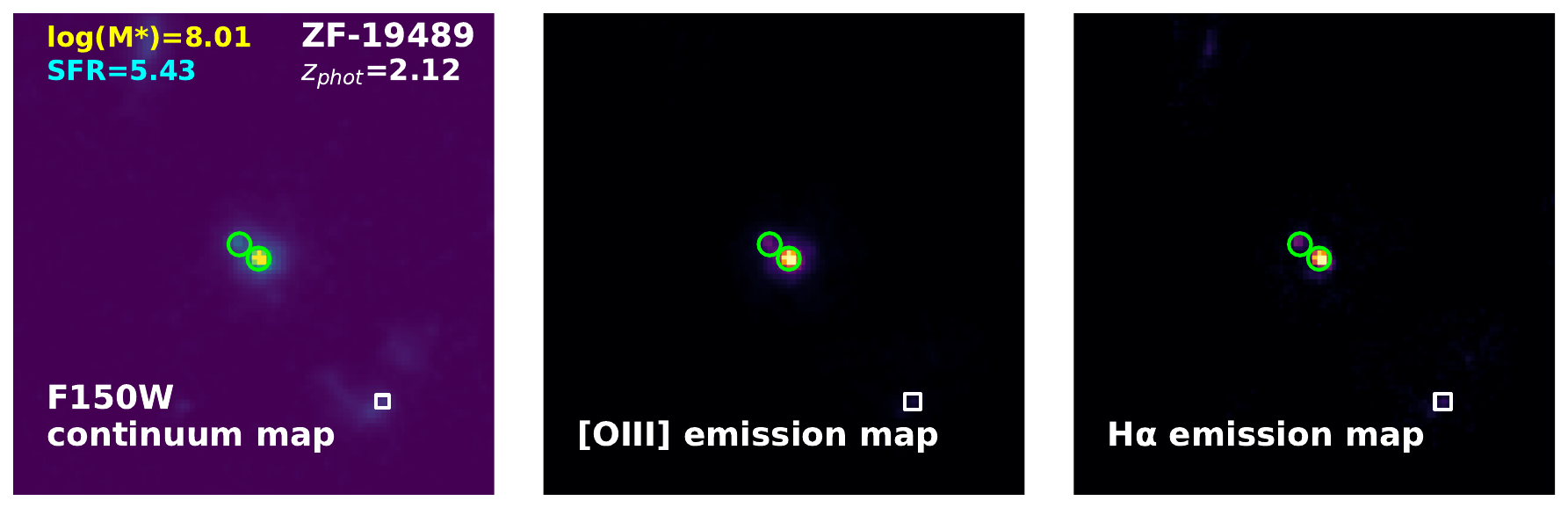}
    \hspace{0.1cm}
    \includegraphics[width=0.32\textwidth]{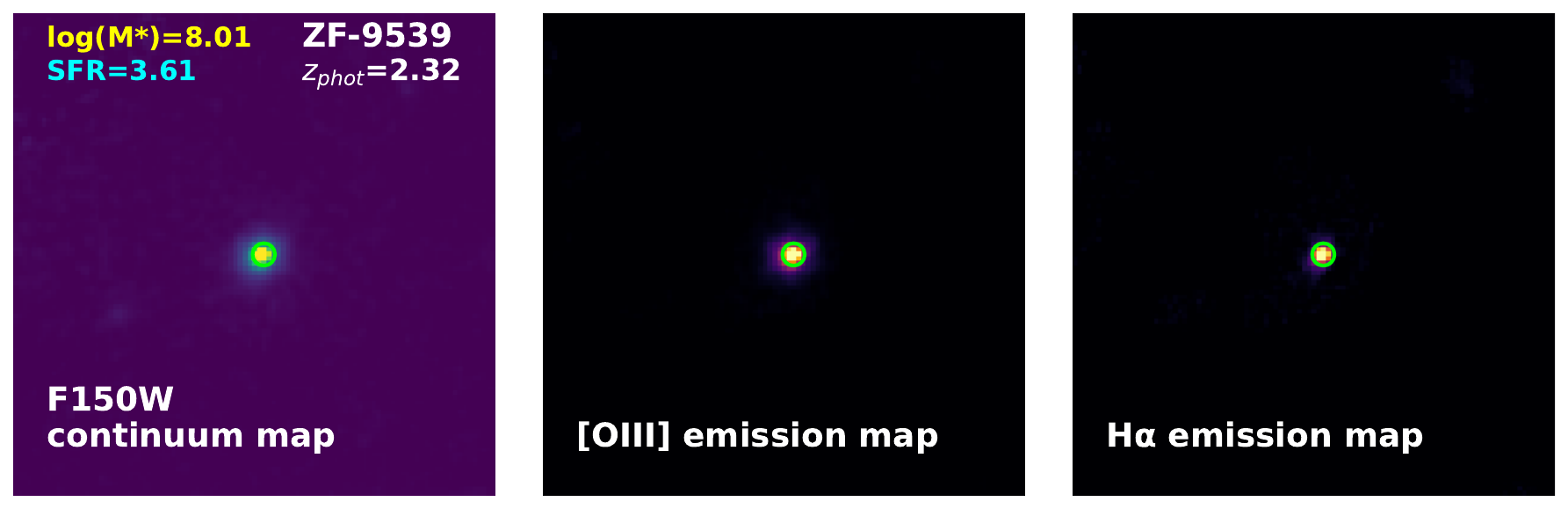}
    \hspace{0.1cm}
    \includegraphics[width=0.32\textwidth]{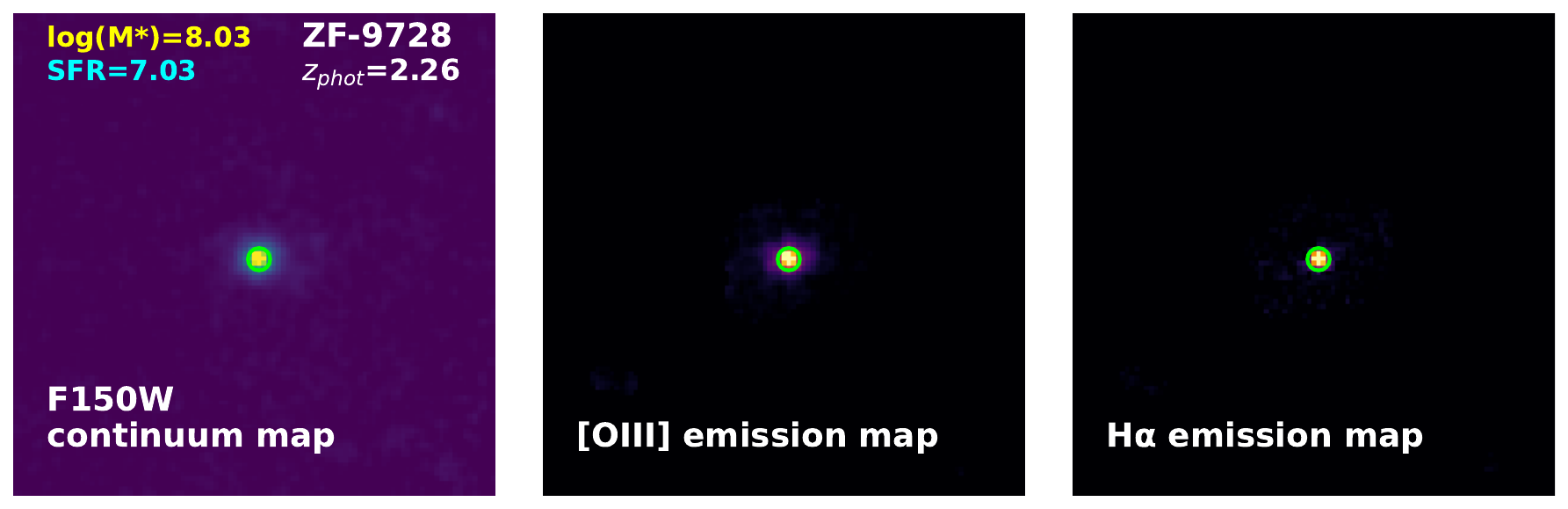}
    \hspace{0.1cm}
    \includegraphics[width=0.32\textwidth]{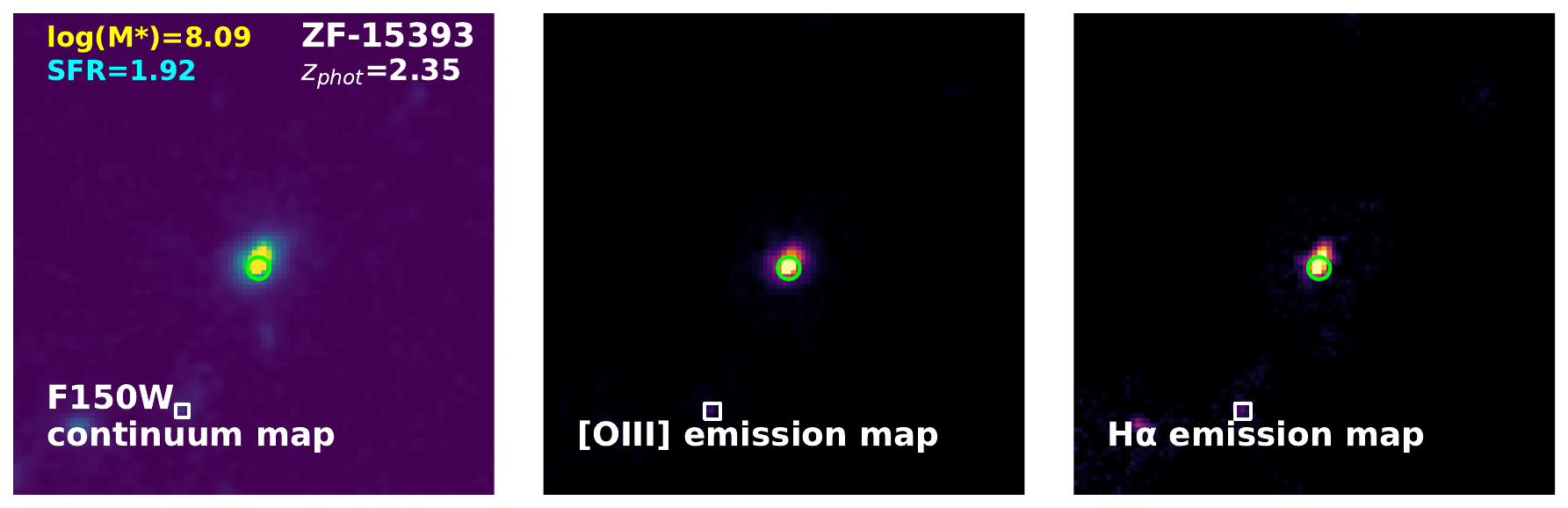}
    \hspace{0.1cm}
    \includegraphics[width=0.32\textwidth]{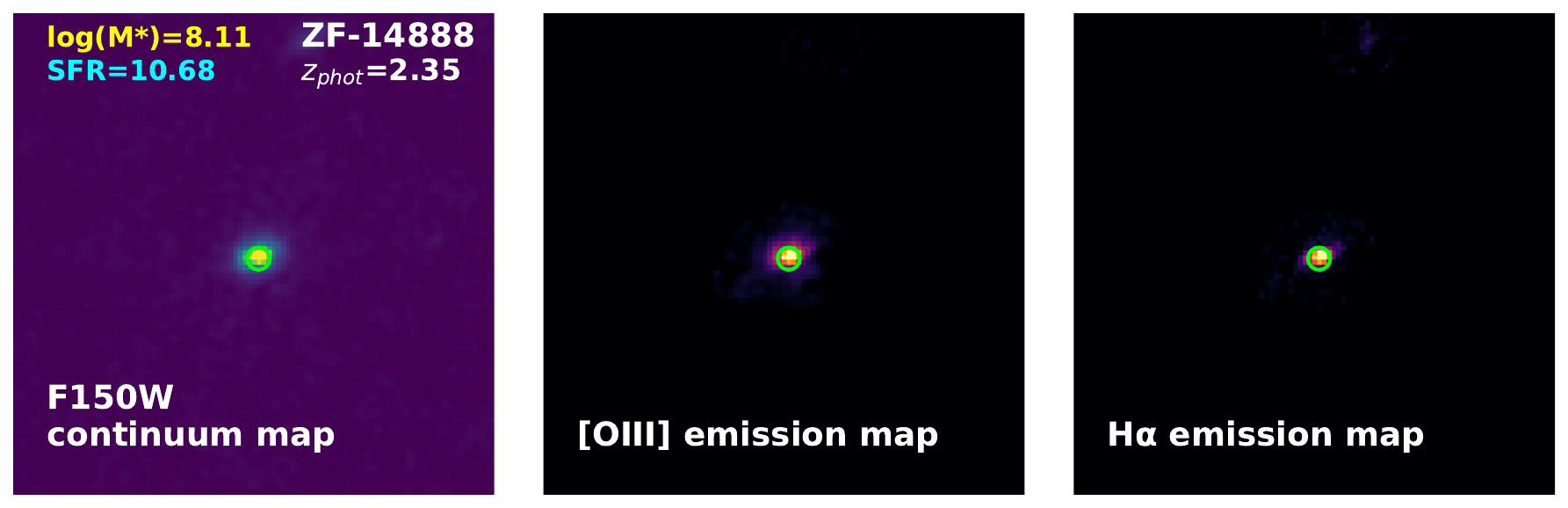}
    \hspace{0.1cm}
    \includegraphics[width=0.32\textwidth]{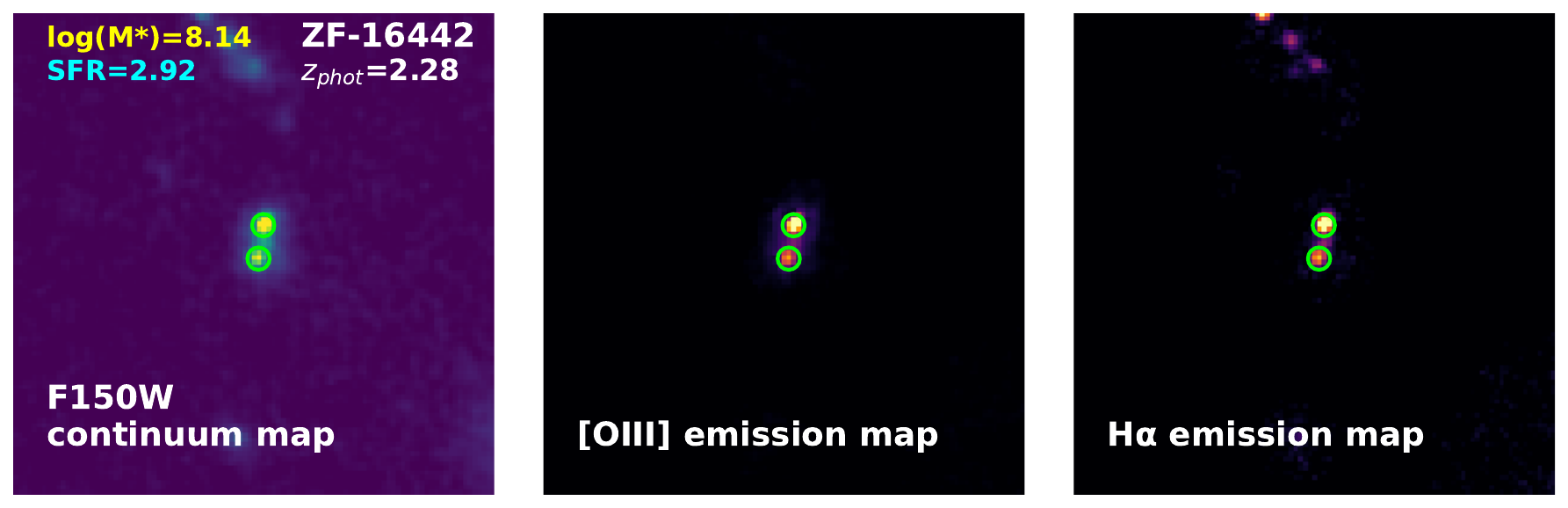}
    \hspace{0.1cm}
    \includegraphics[width=0.32\textwidth]{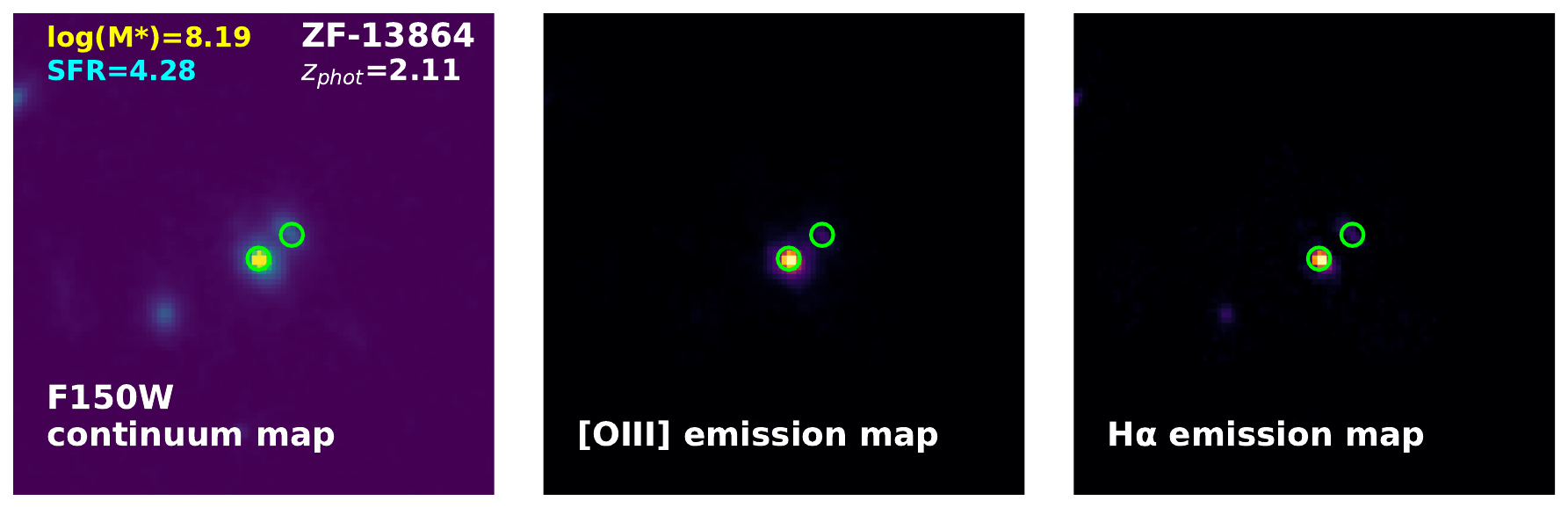}
    \hspace{0.1cm}
    \includegraphics[width=0.32\textwidth]{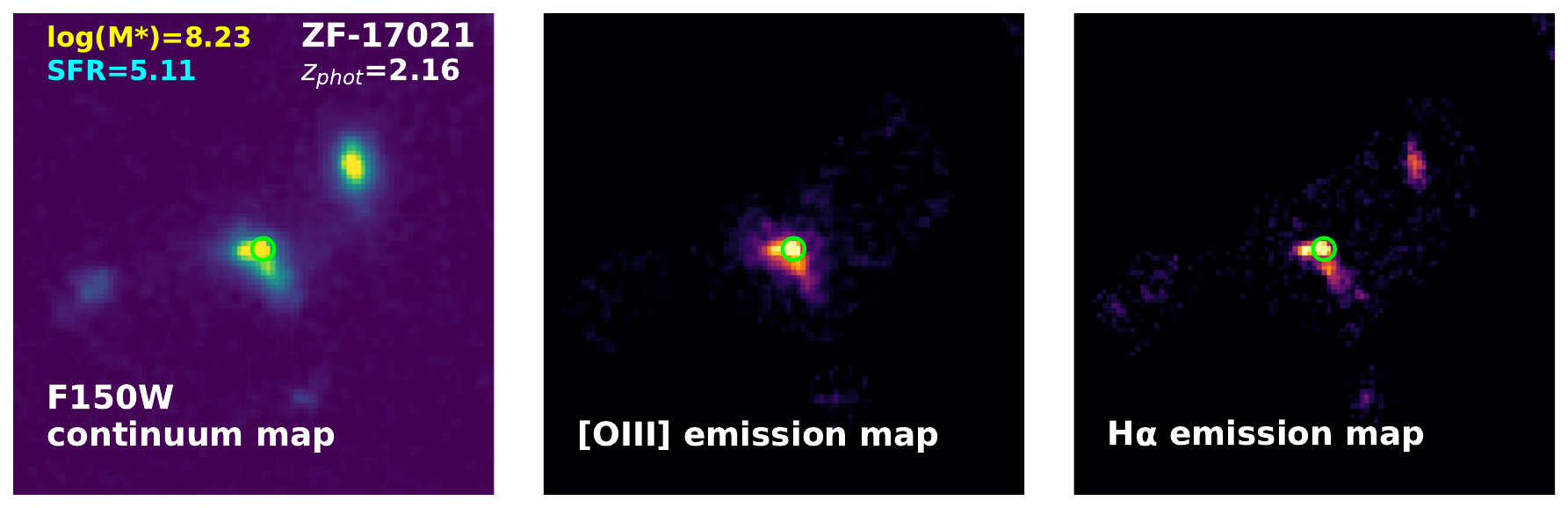}
    \hspace{0.1cm}
    \includegraphics[width=0.32\textwidth]{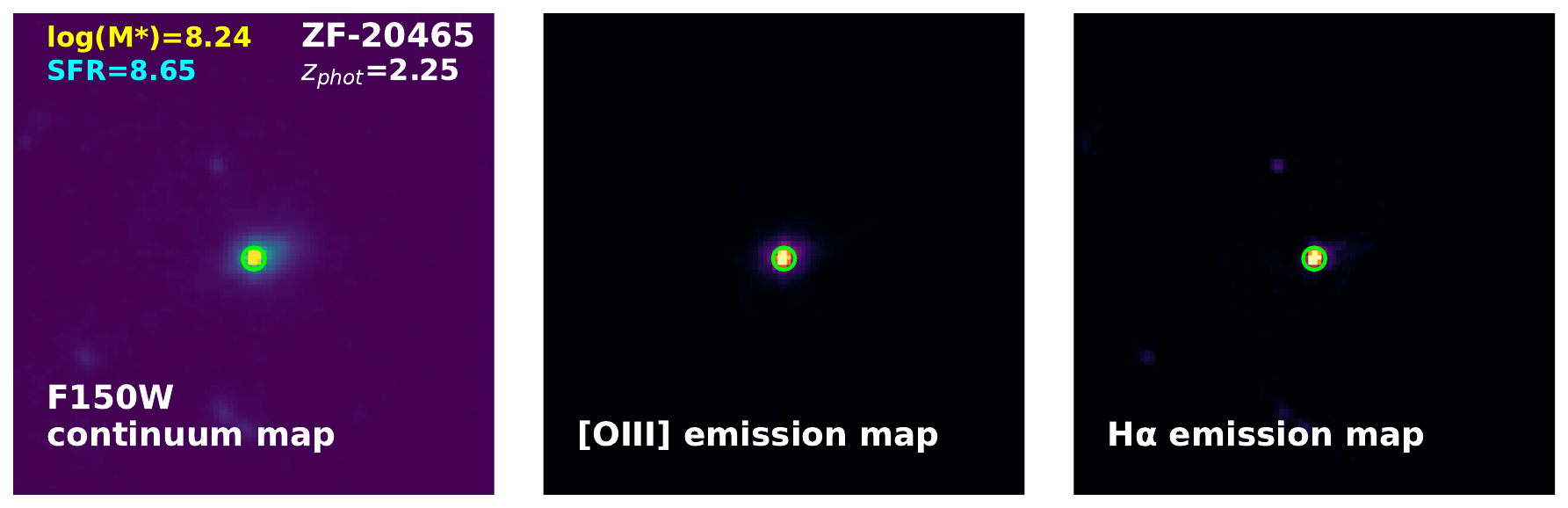}
    \hspace{0.1cm}
    \includegraphics[width=0.32\textwidth]{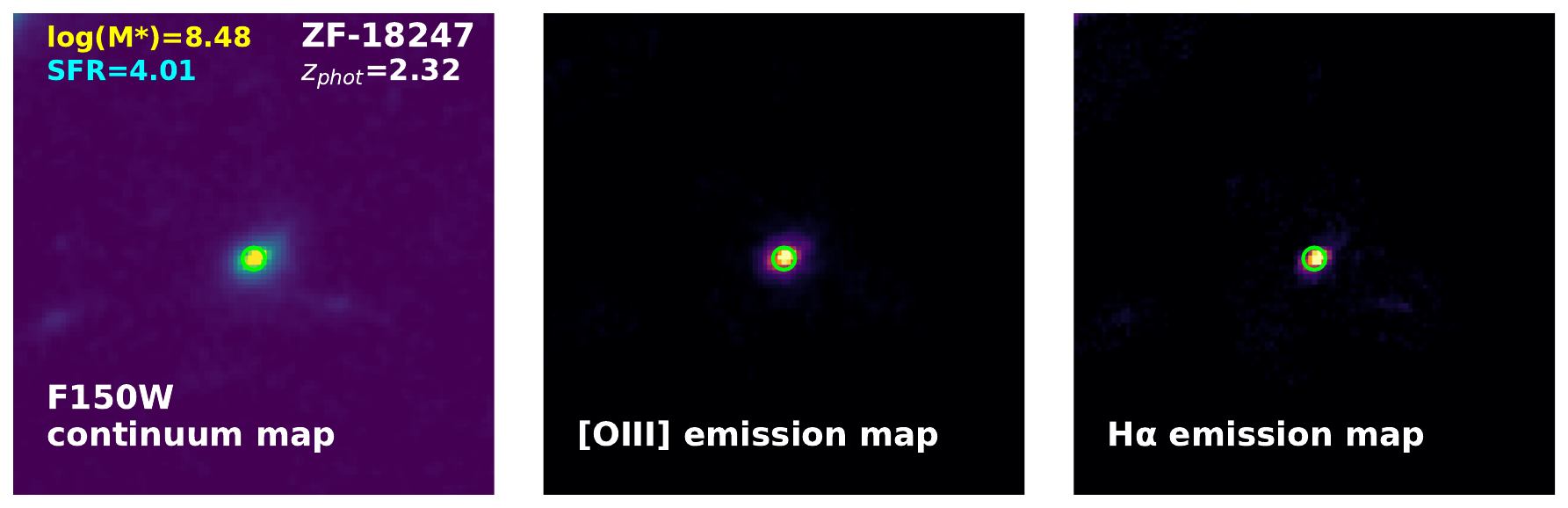}
    \hspace{0.1cm}
    \includegraphics[width=0.32\textwidth]{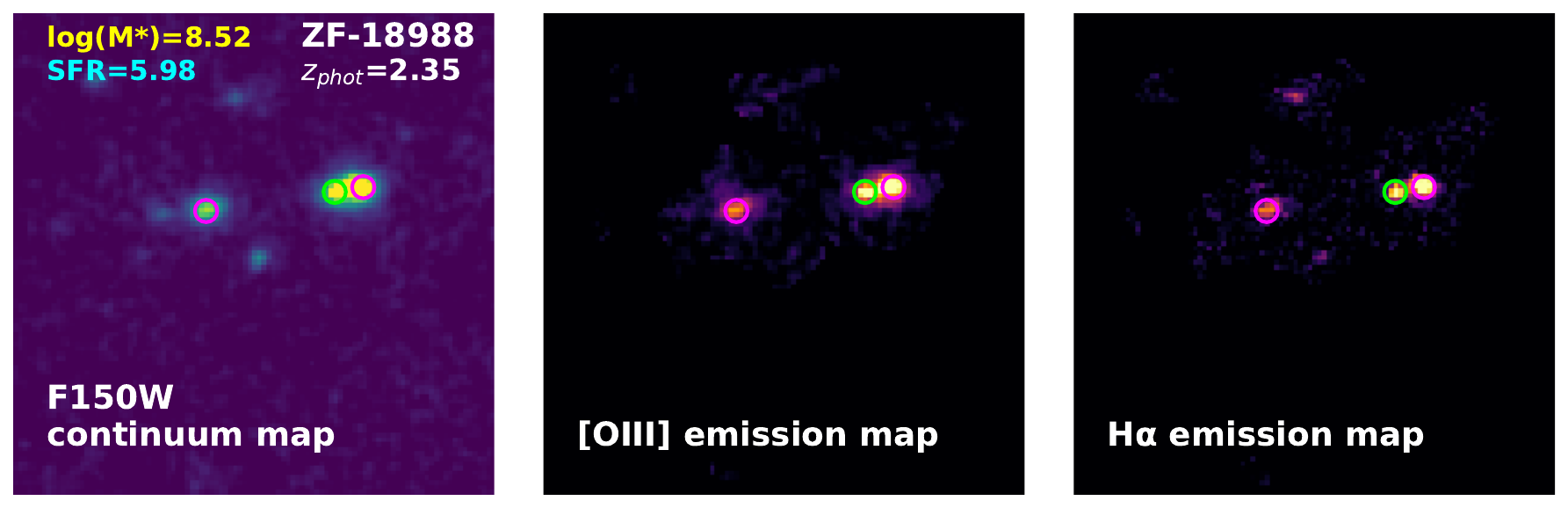}
    \hspace{0.1cm}
    \includegraphics[width=0.32\textwidth]{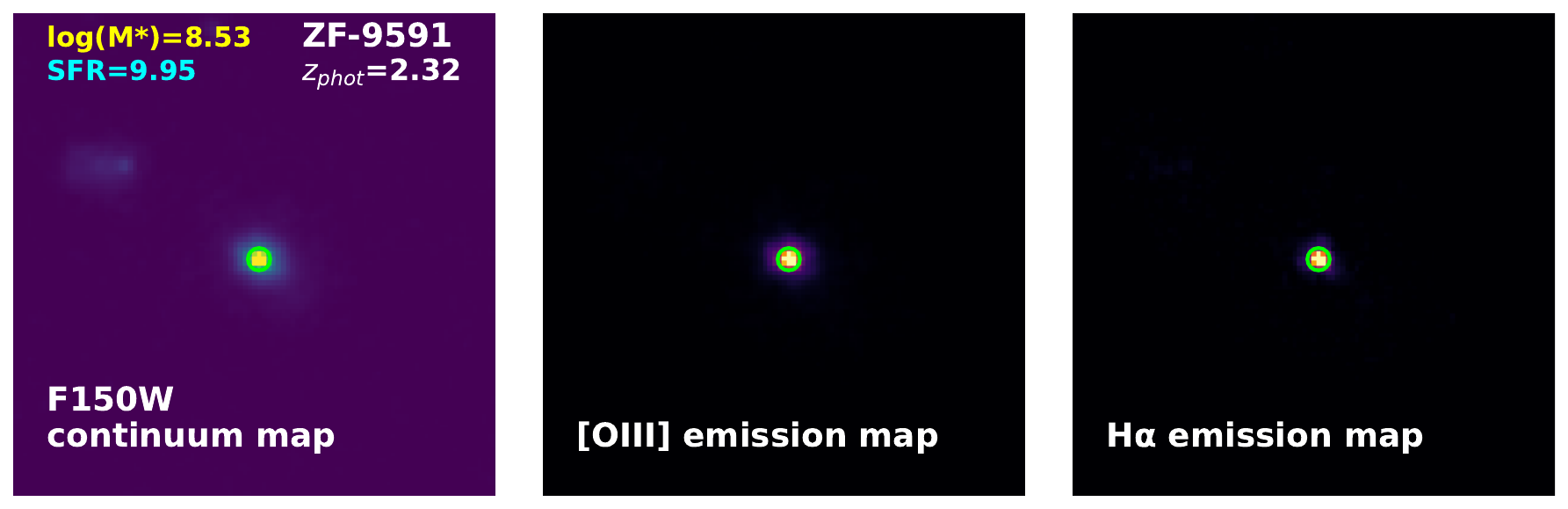}
    \hspace{0.1cm}
    \includegraphics[width=0.32\textwidth]{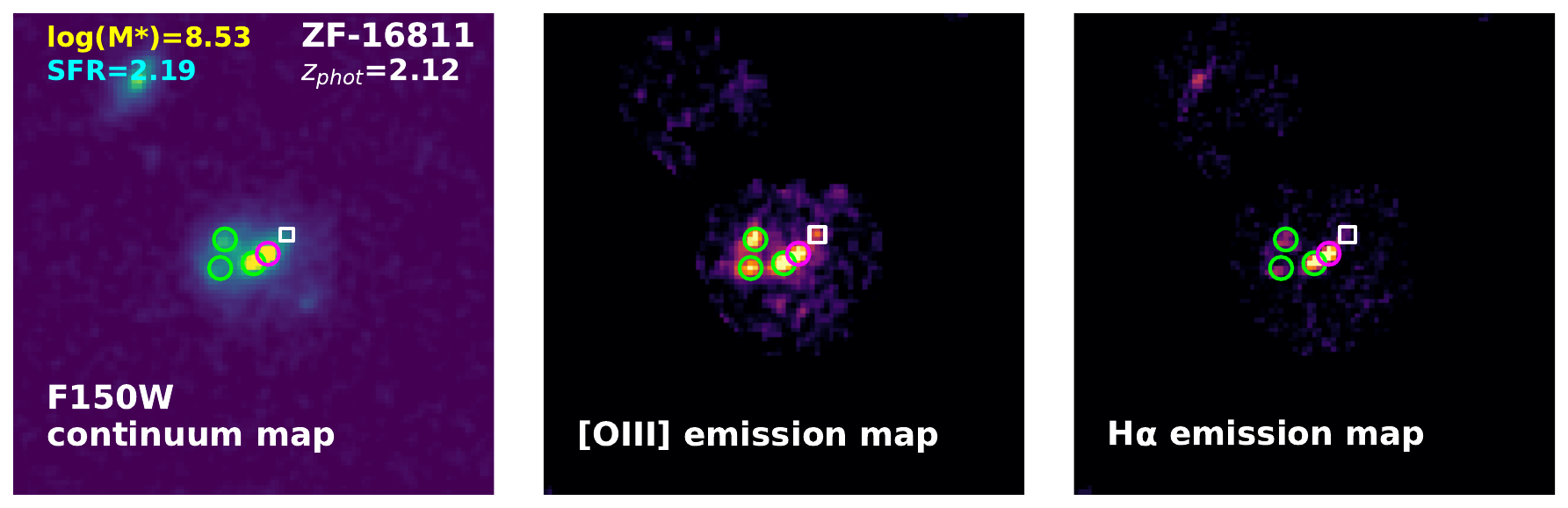}
    \hspace{0.1cm}
    \includegraphics[width=0.32\textwidth]{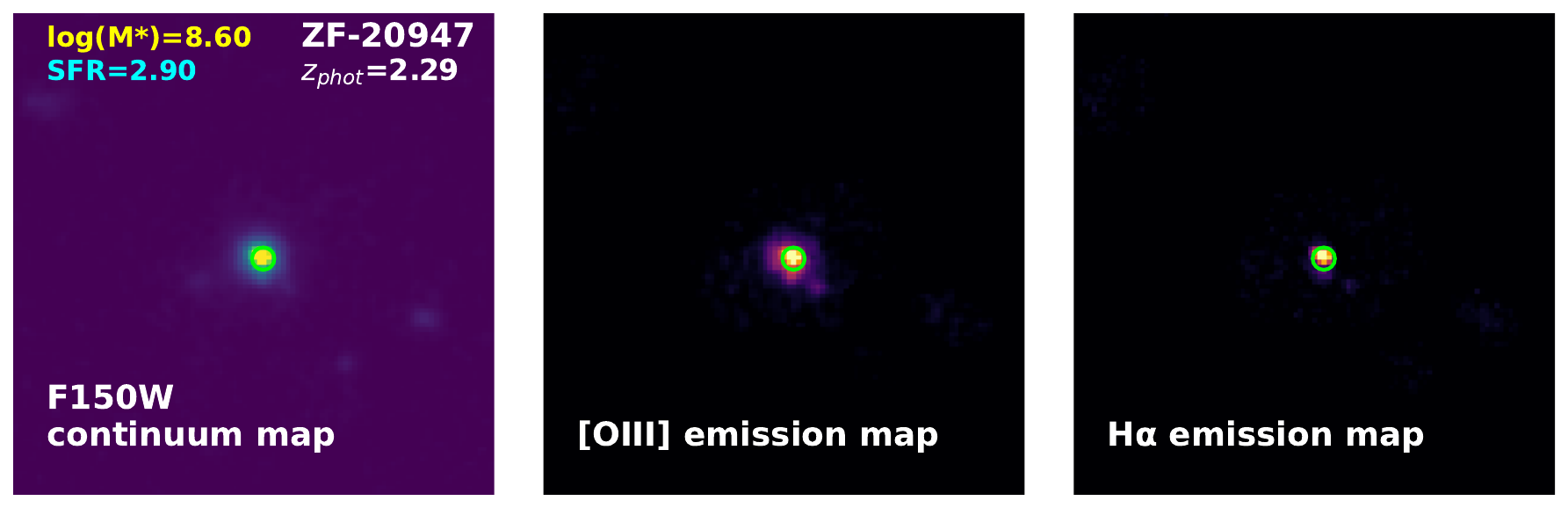}
    \hspace{0.1cm}
    \includegraphics[width=0.32\textwidth]{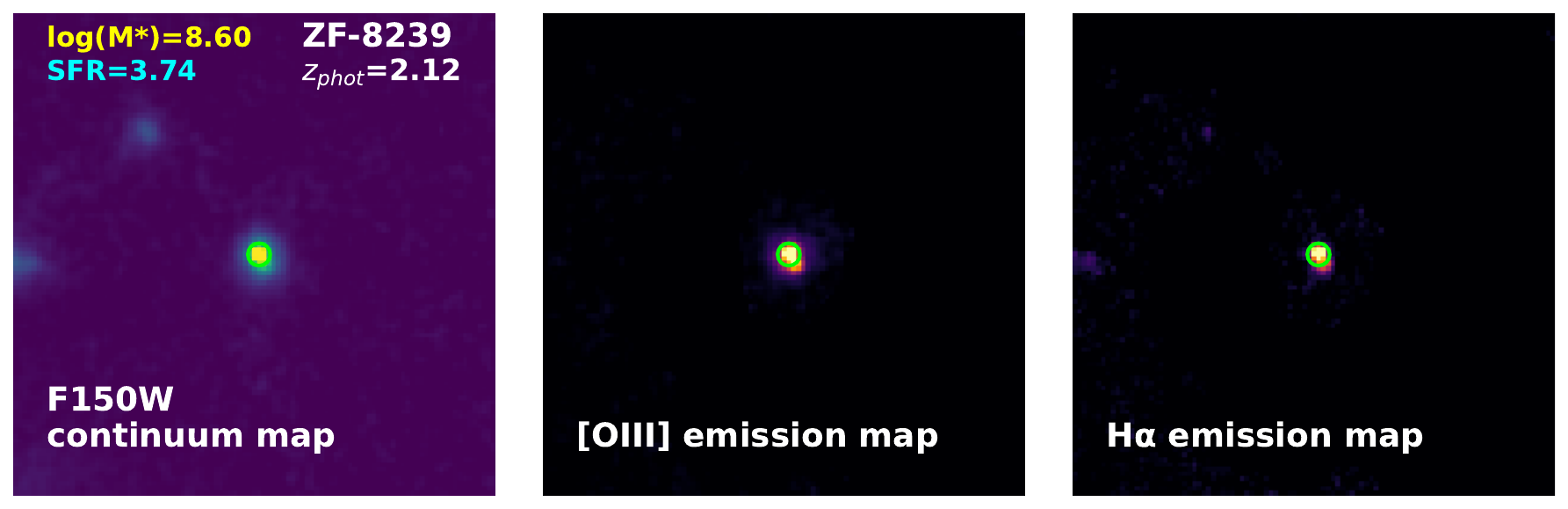}
    \hspace{0.1cm}
    \includegraphics[width=0.32\textwidth]{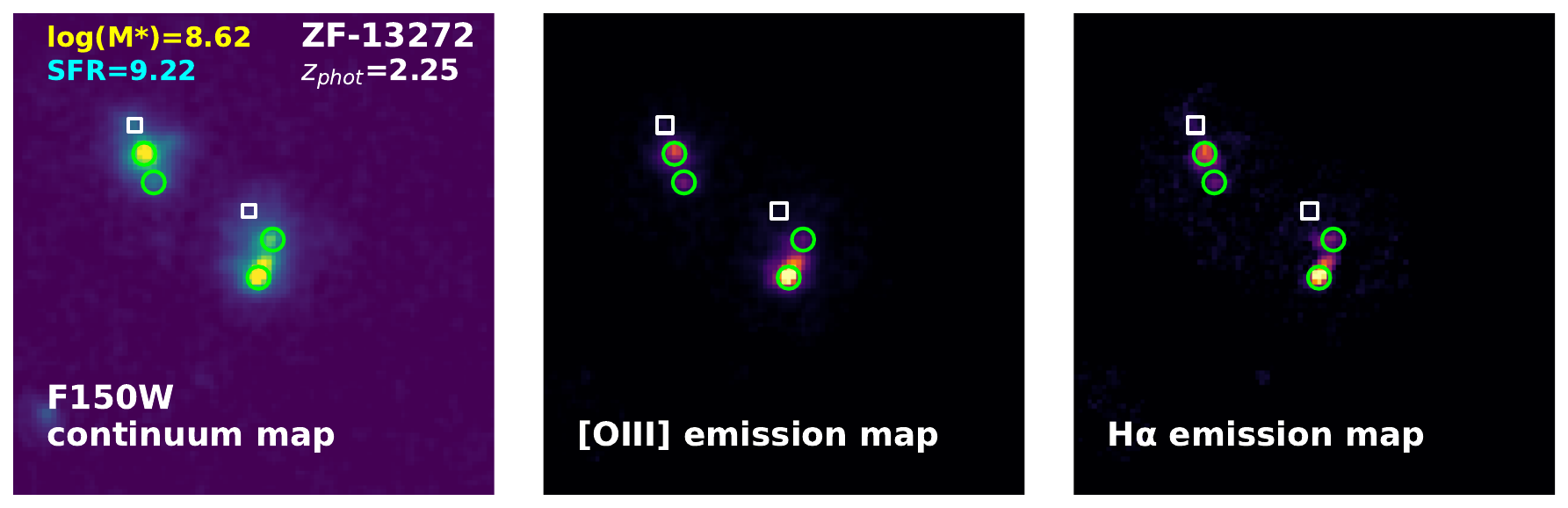}
    \hspace{0.1cm}
    \includegraphics[width=0.32\textwidth]{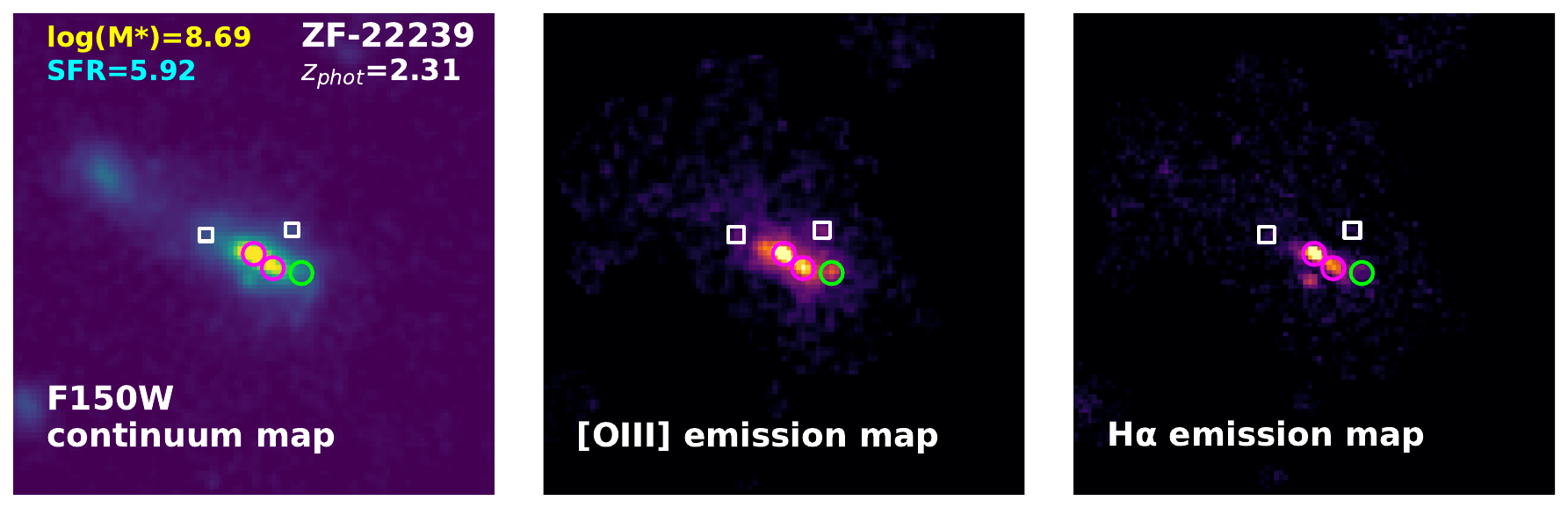}
    \hspace{0.1cm}
    \includegraphics[width=0.32\textwidth]{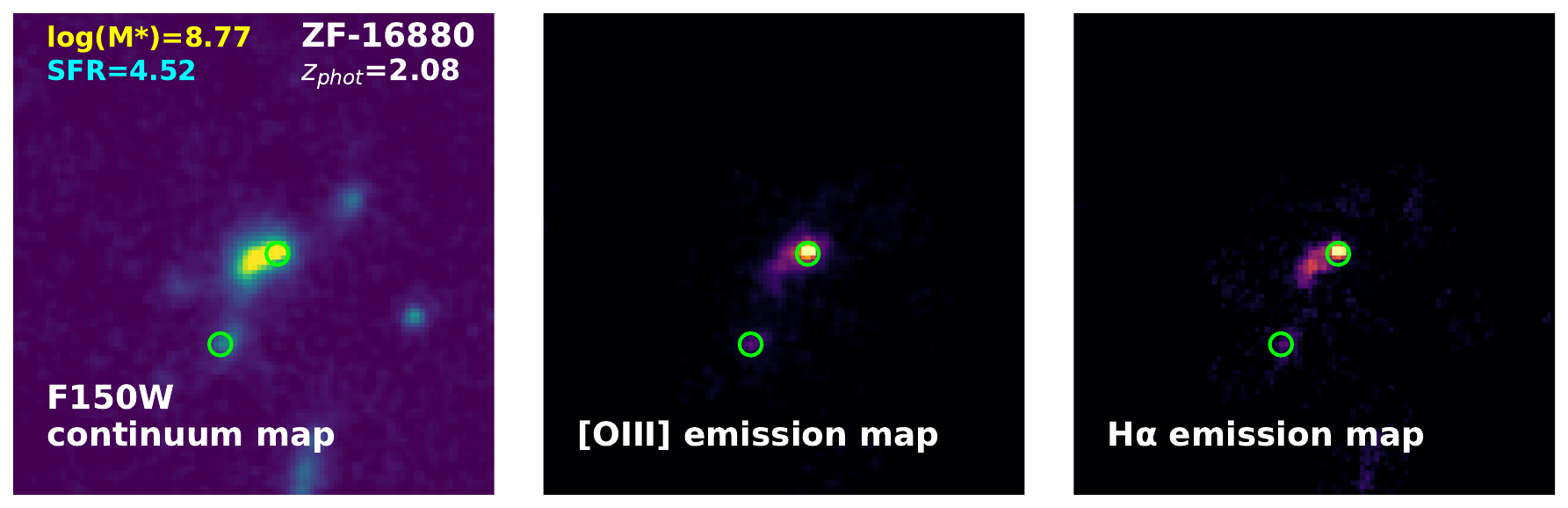}
    \hspace{0.1cm}
    \includegraphics[width=0.32\textwidth]{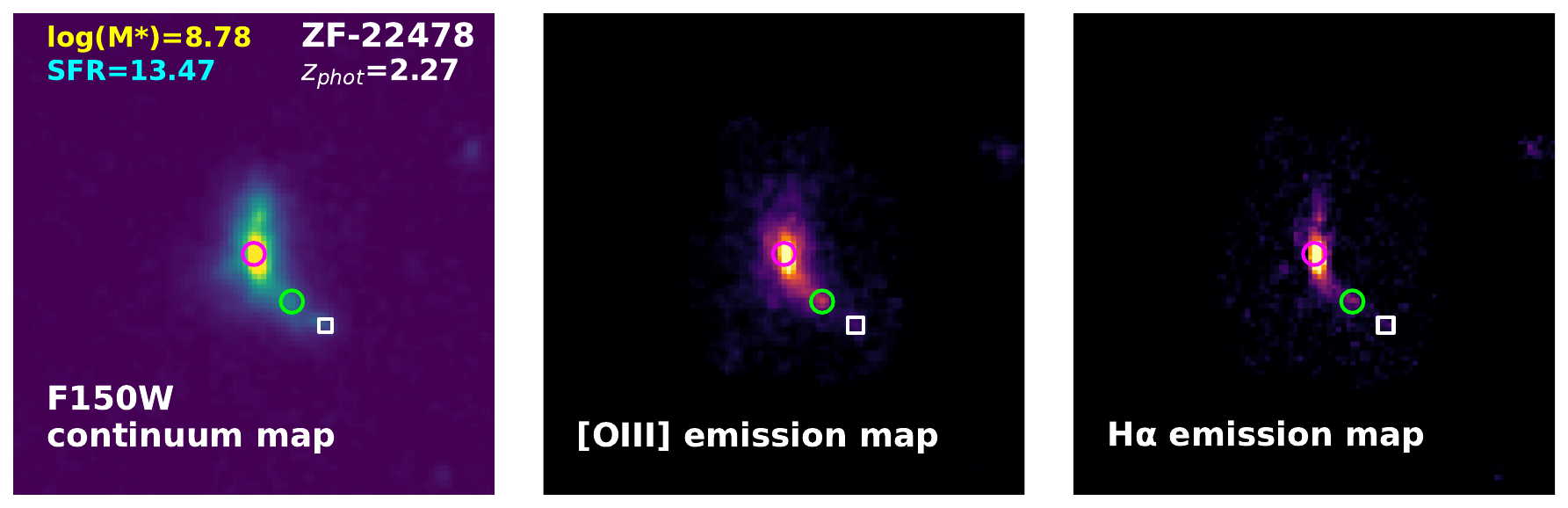}
    \hspace{0.1cm}
    \includegraphics[width=0.32\textwidth]{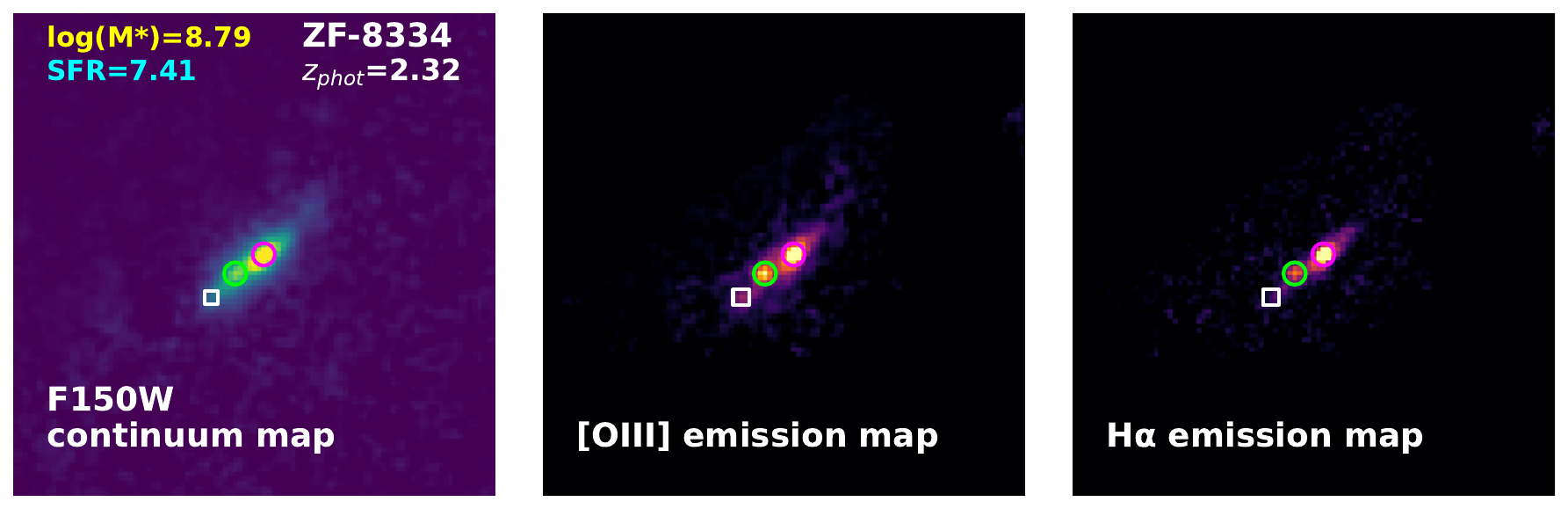}
    \hspace{0.1cm}
    \includegraphics[width=0.32\textwidth]{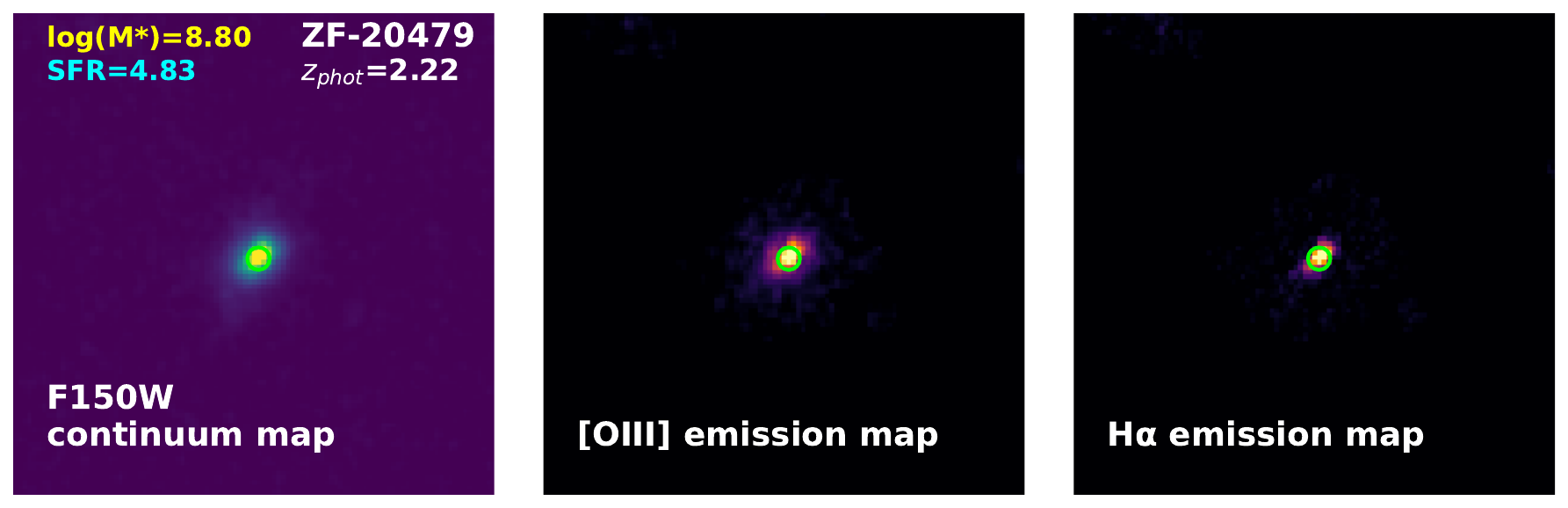}
    \hspace{0.1cm}
    \includegraphics[width=0.32\textwidth]{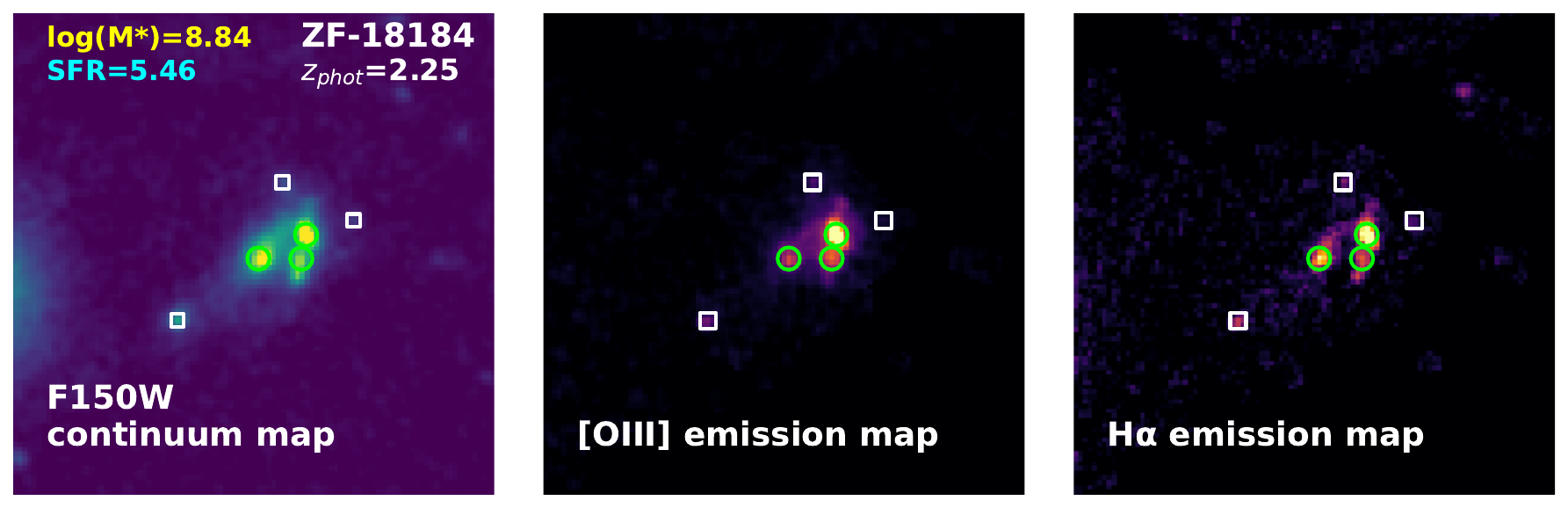}
    \hspace{0.1cm}
    \includegraphics[width=0.32\textwidth]{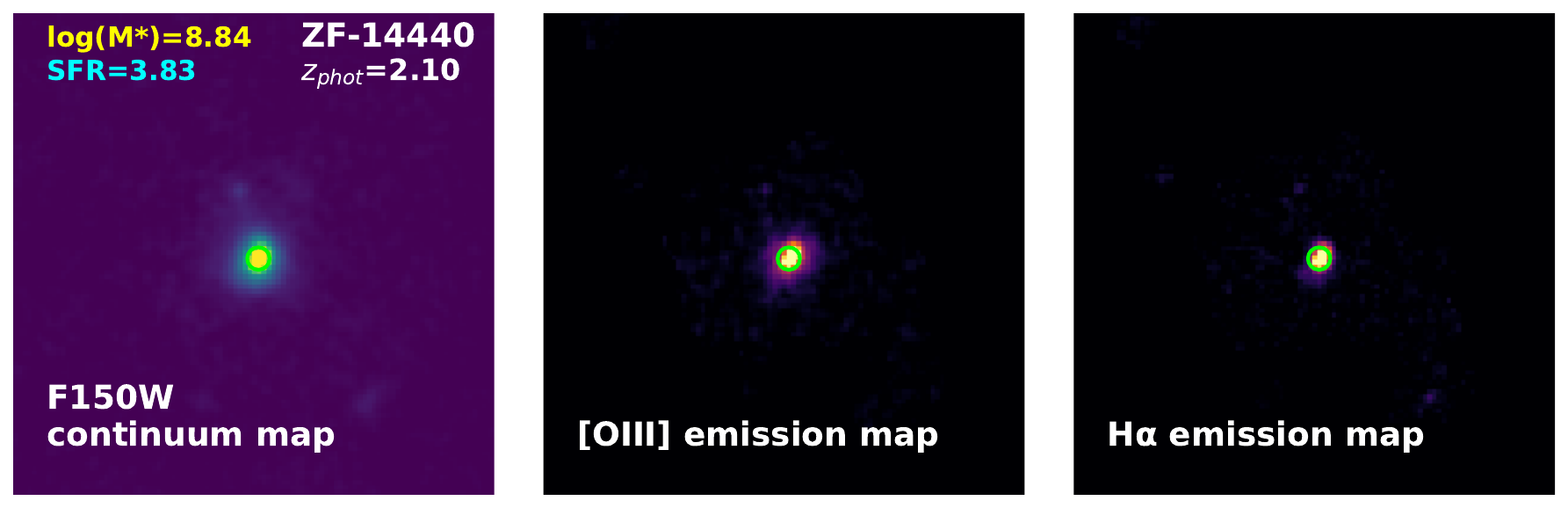}
    \hspace{0.1cm}
    \includegraphics[width=0.32\textwidth]{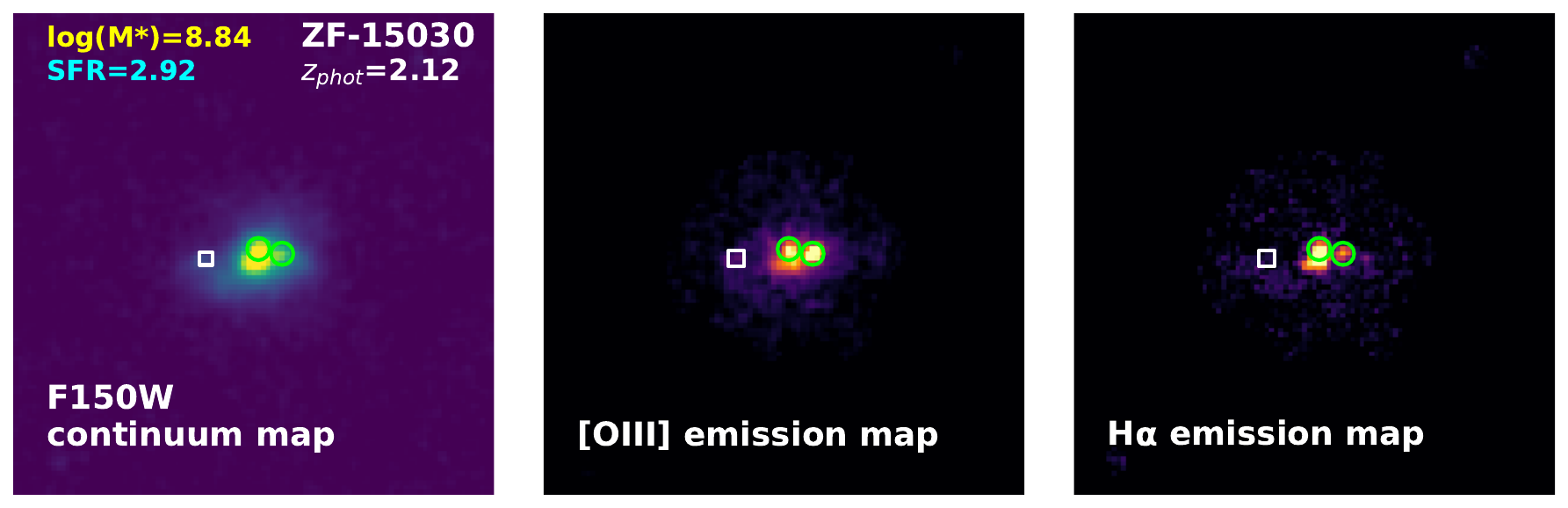}
    \hspace{0.1cm}
    \includegraphics[width=0.32\textwidth]{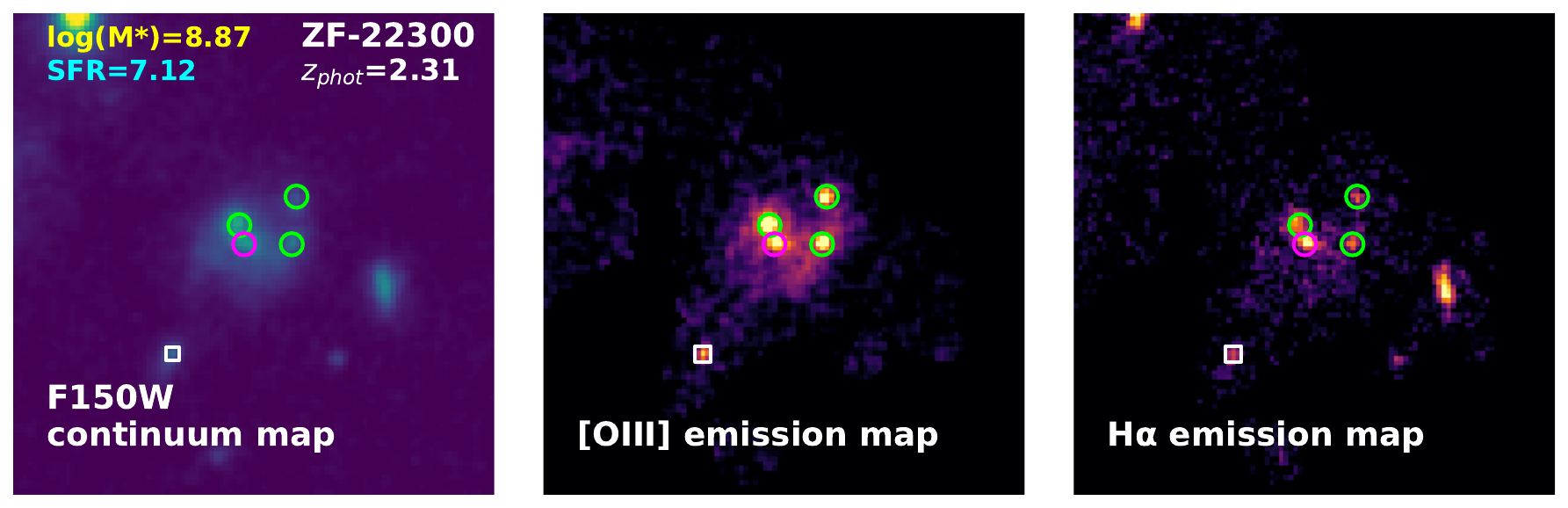}
    \hspace{0.1cm}
    \includegraphics[width=0.32\textwidth]{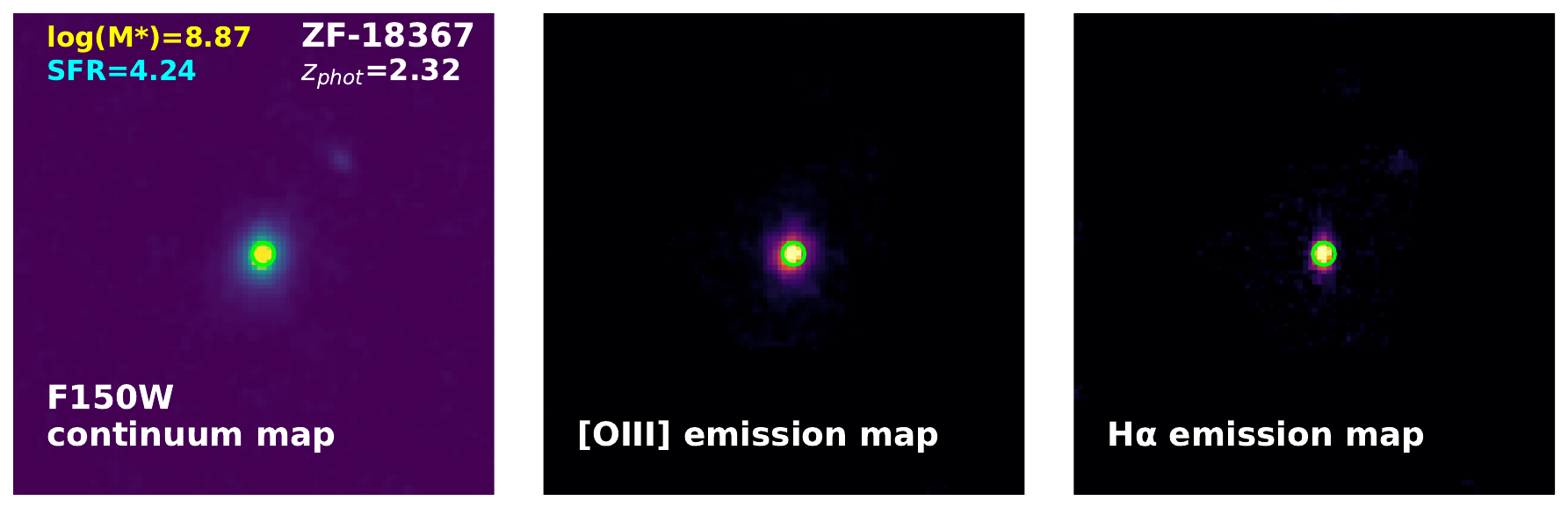}
    \hspace{0.1cm}
    \includegraphics[width=0.32\textwidth]{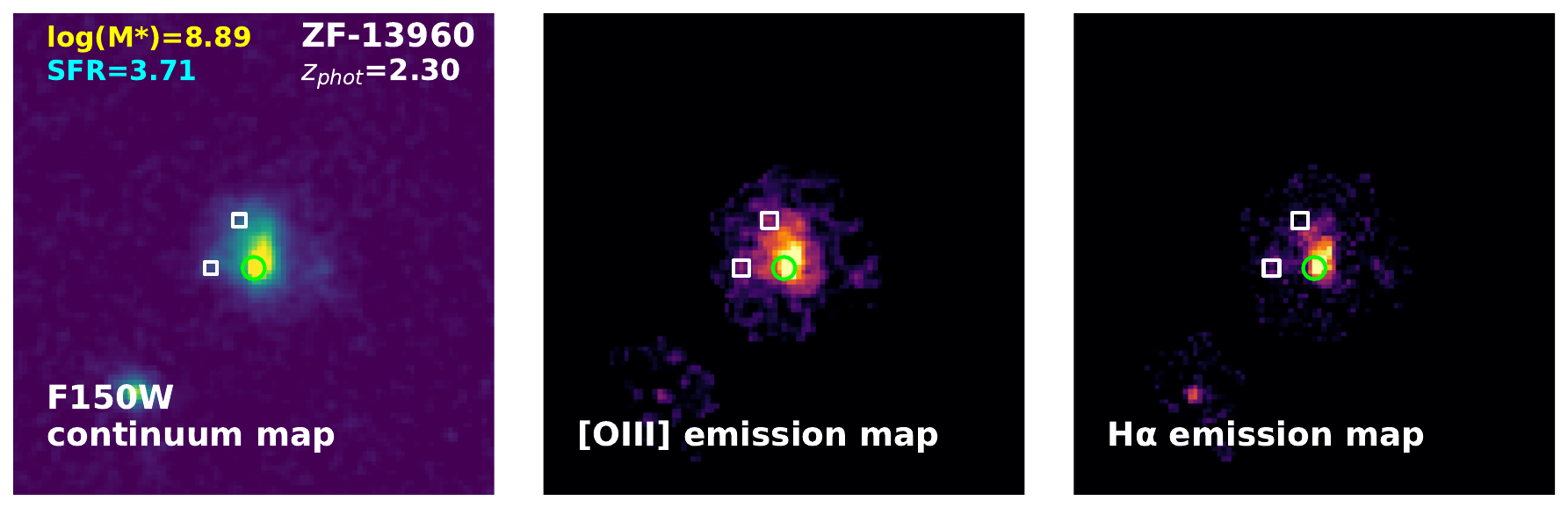}
    \hspace{0.1cm}
    \includegraphics[width=0.32\textwidth]{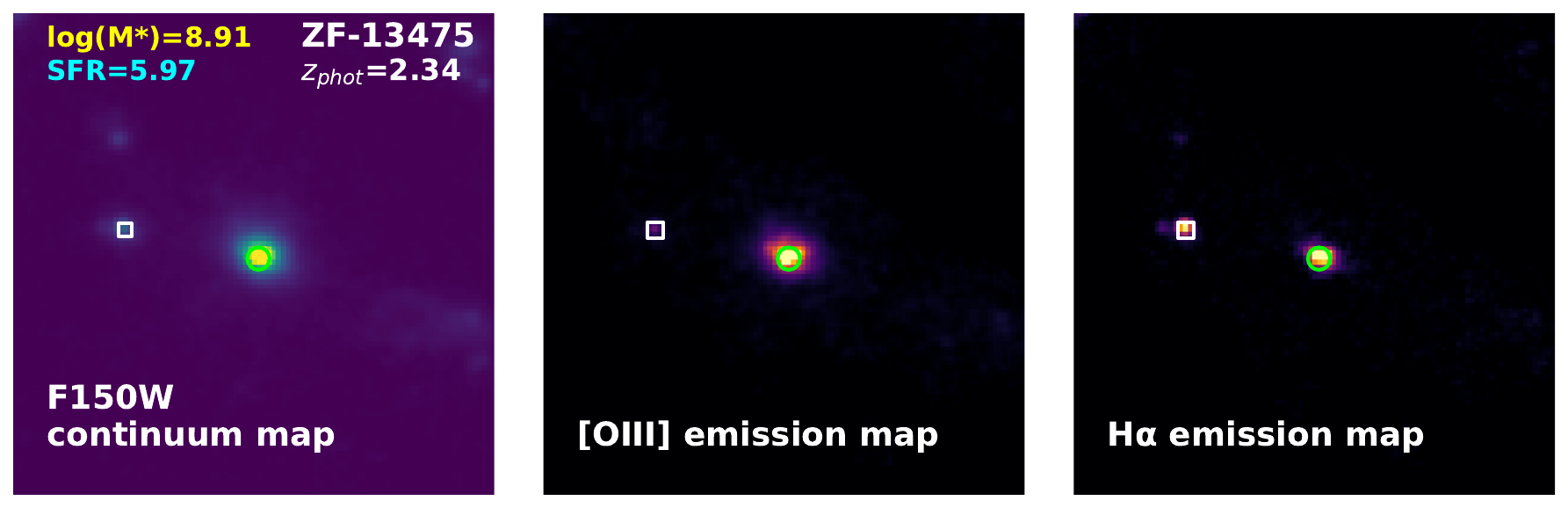}
    \hspace{0.1cm}
    \includegraphics[width=0.32\textwidth]{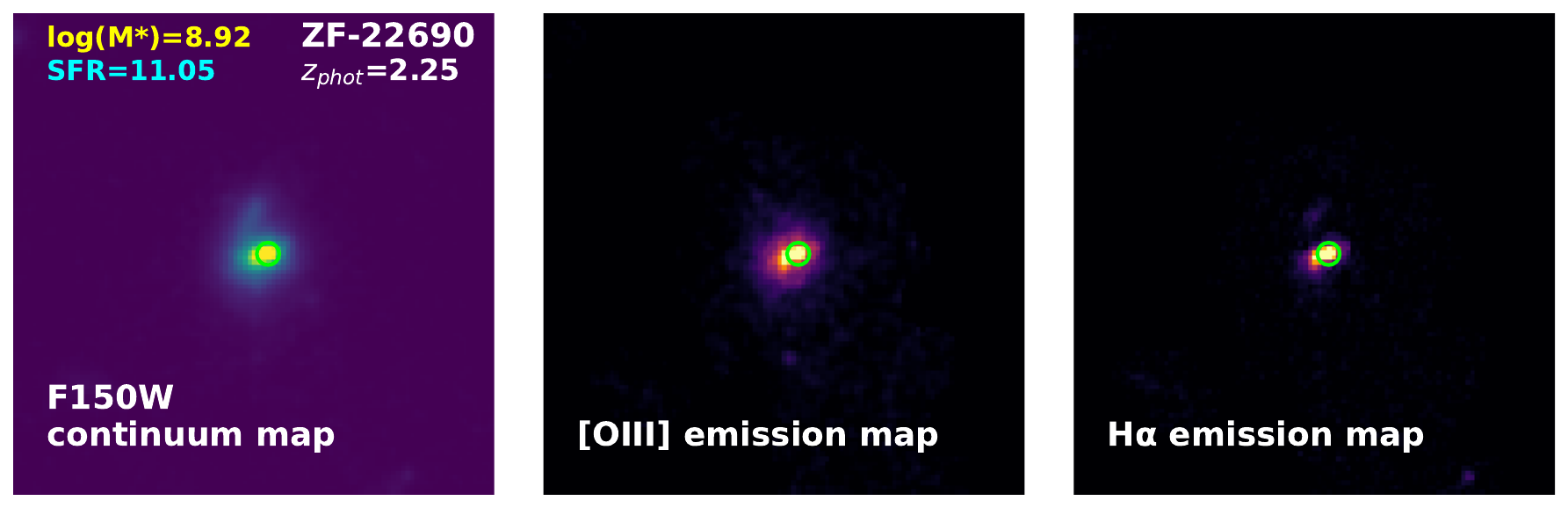}
    \hspace{0.1cm}
    \includegraphics[width=0.32\textwidth]{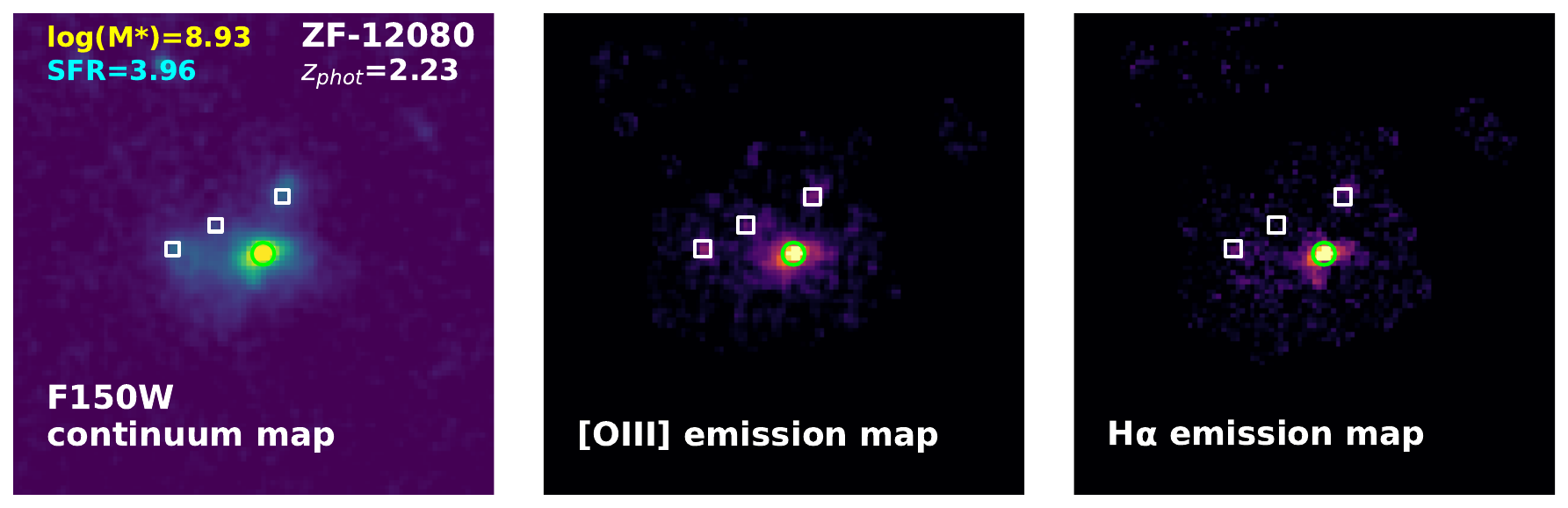}
    \hspace{0.1cm}
    \includegraphics[width=0.32\textwidth]{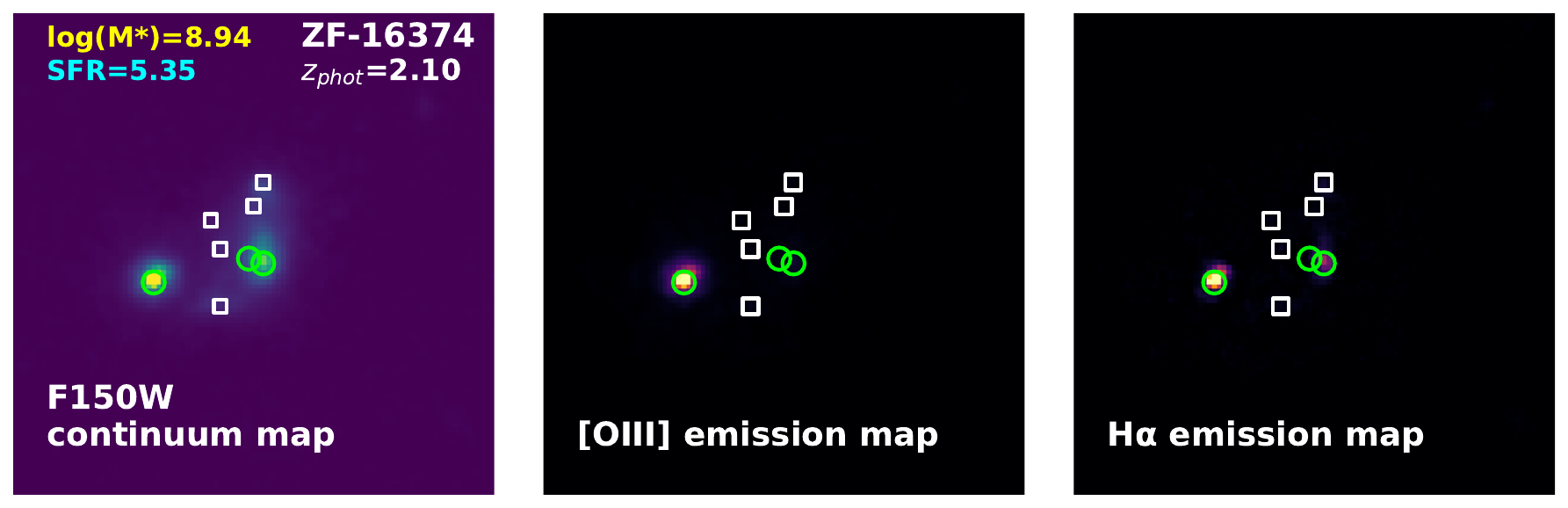}
    \hspace{0.1cm}
    \label{fig:emimap1}
    \caption{The stellar continuum map at $1.5\,\mu m$, the [O{\sc iii}]+$\mathrm{H\beta}$ emission line map, and the $\mathrm{H\alpha}$ emission line map of the rest HAEs which contain Green Seeds. Outlines as in Figure \hyperref[fig:emimap]{8}. Those sample detected with only one Green Seeds may also be considered as ``Green Pea" galaxies at $z\sim2$.}
    \vspace{0.3cm}
\end{figure*}

\begin{figure*}[p]
    \centering
    \includegraphics[width=0.32\textwidth]{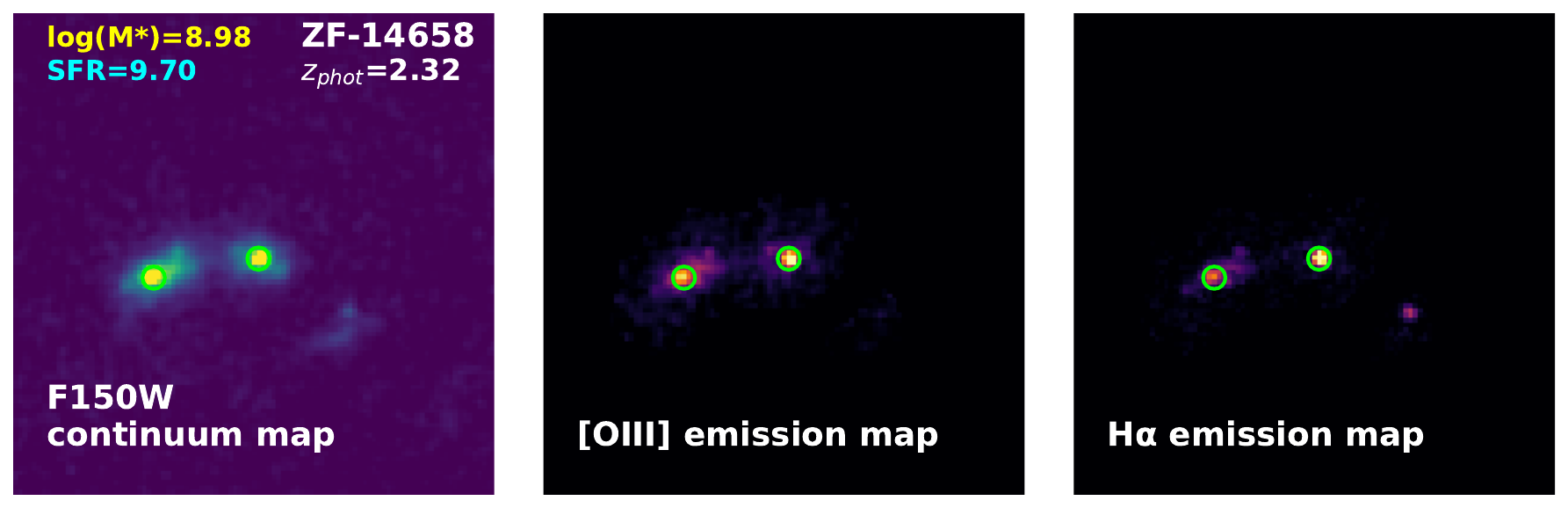}
    \hspace{0.1cm}
    \includegraphics[width=0.32\textwidth]{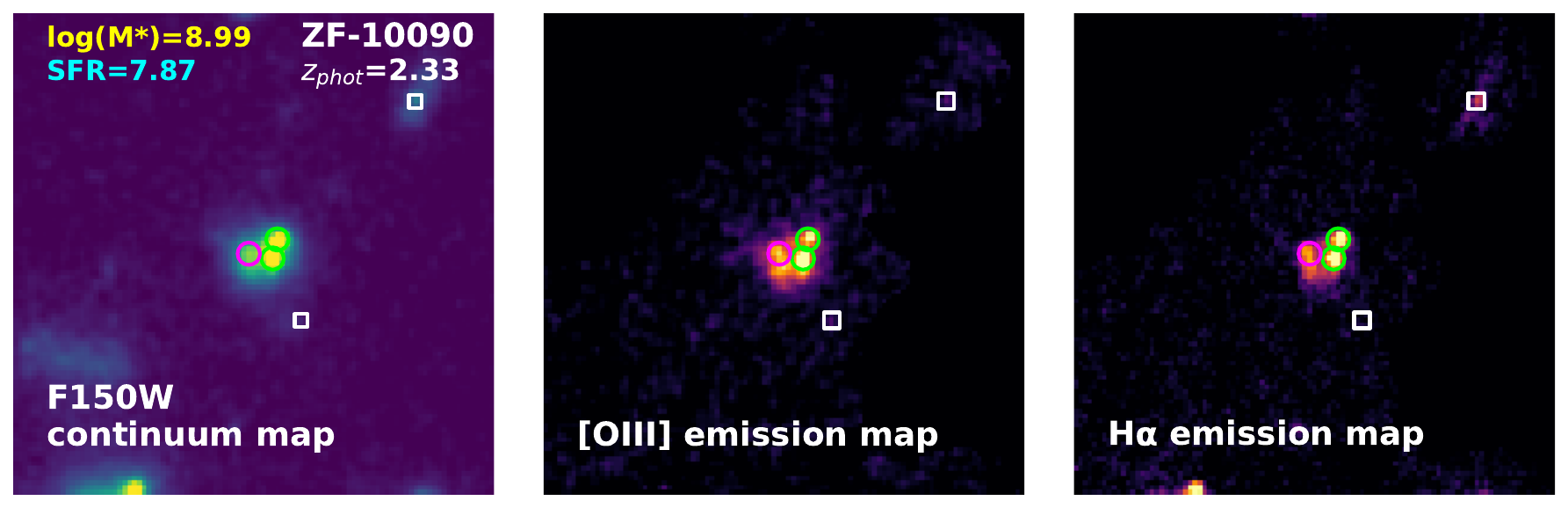}
    \hspace{0.1cm}
    \includegraphics[width=0.32\textwidth]{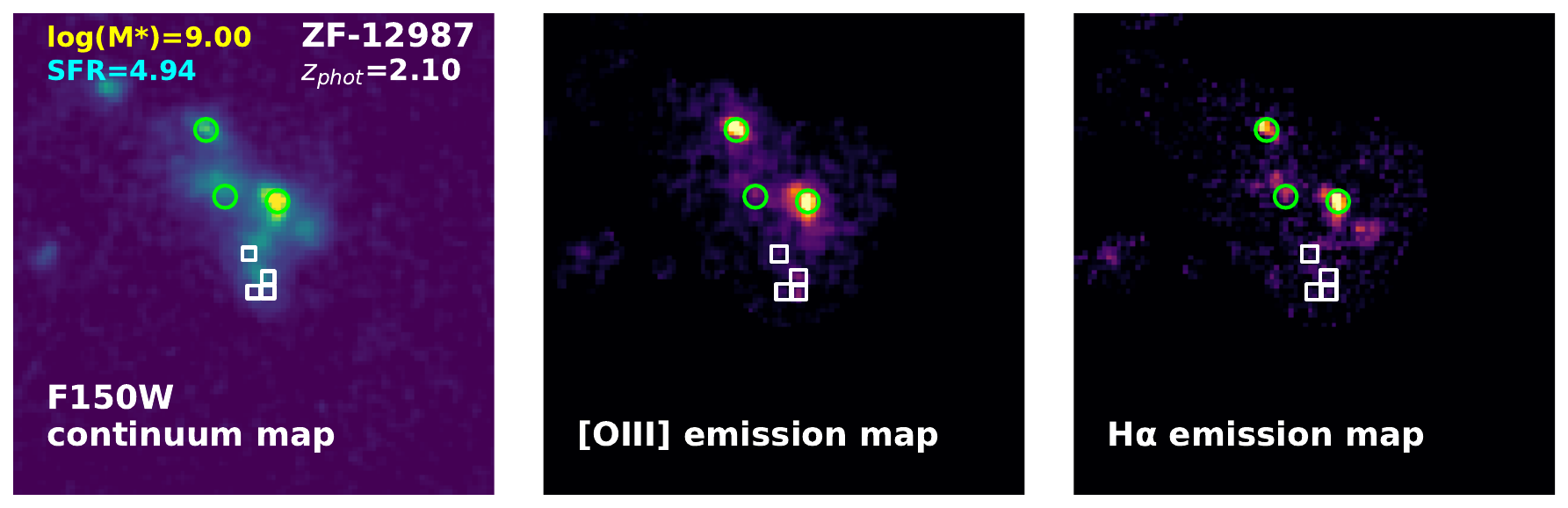}
    \hspace{0.1cm}
    \includegraphics[width=0.32\textwidth]{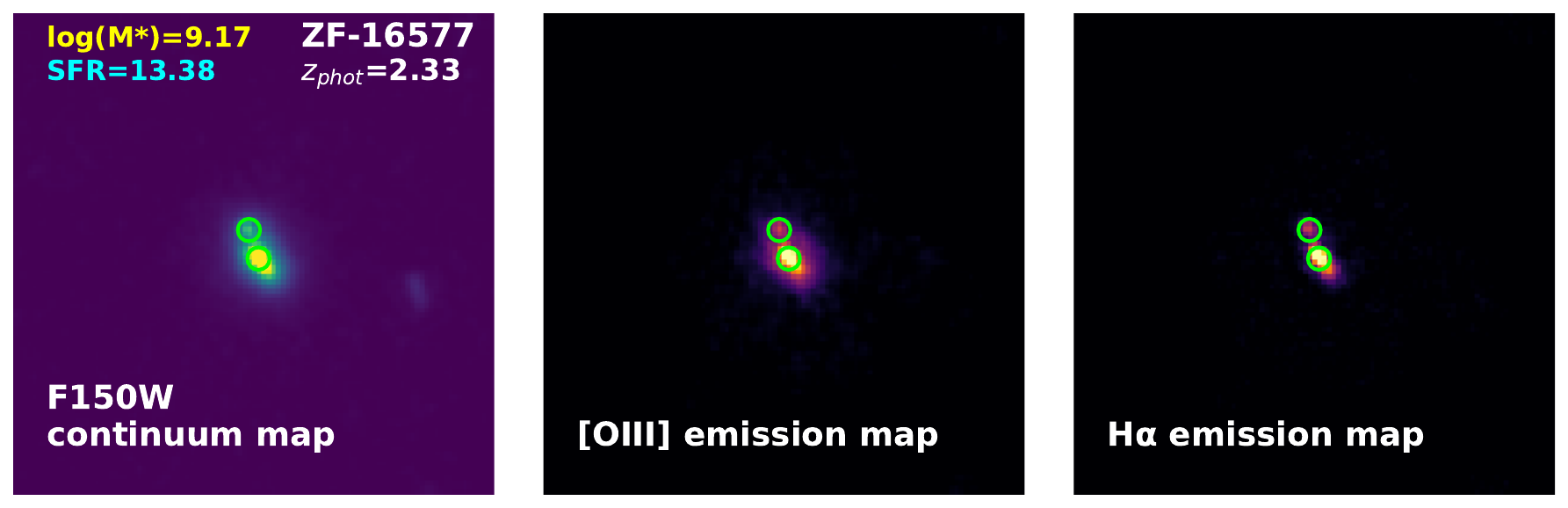}
    \hspace{0.1cm}
    \includegraphics[width=0.32\textwidth]{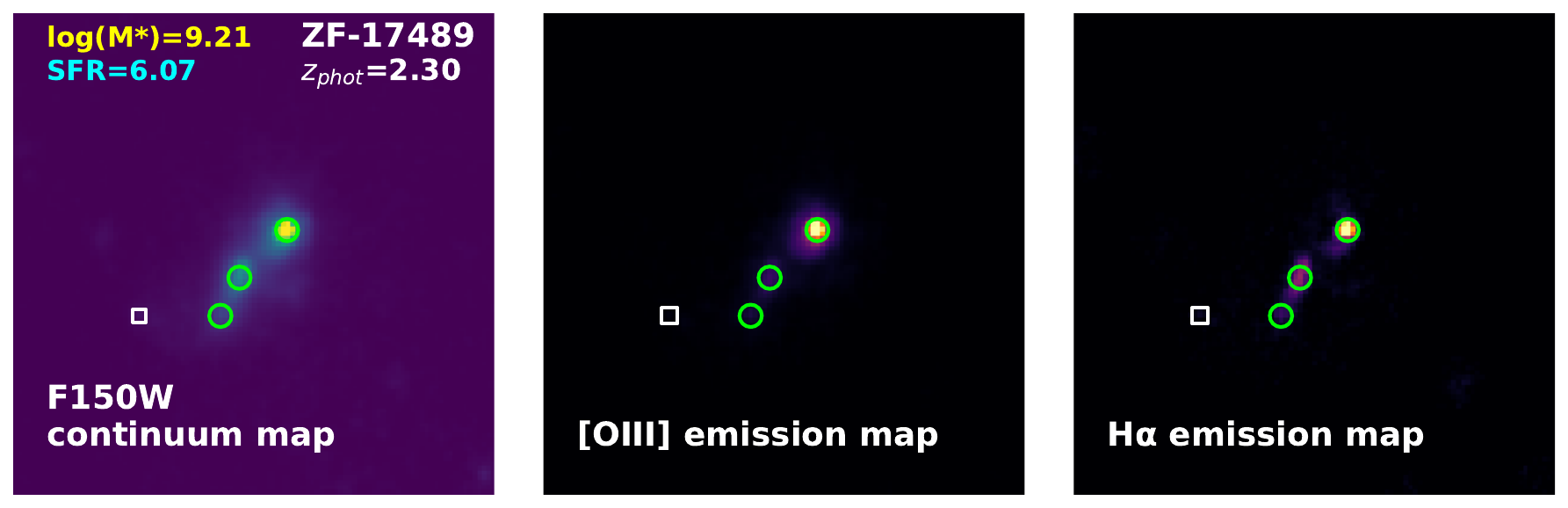}
    \hspace{0.1cm}
    \includegraphics[width=0.32\textwidth]{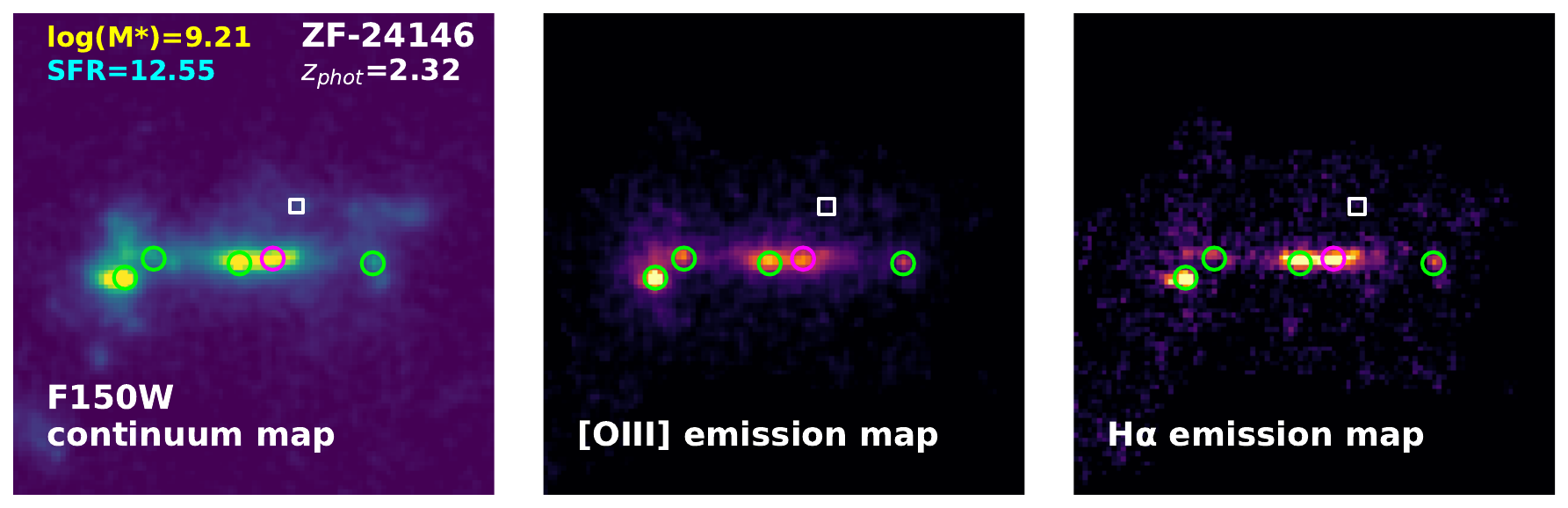}
    \hspace{0.1cm}
    \includegraphics[width=0.32\textwidth]{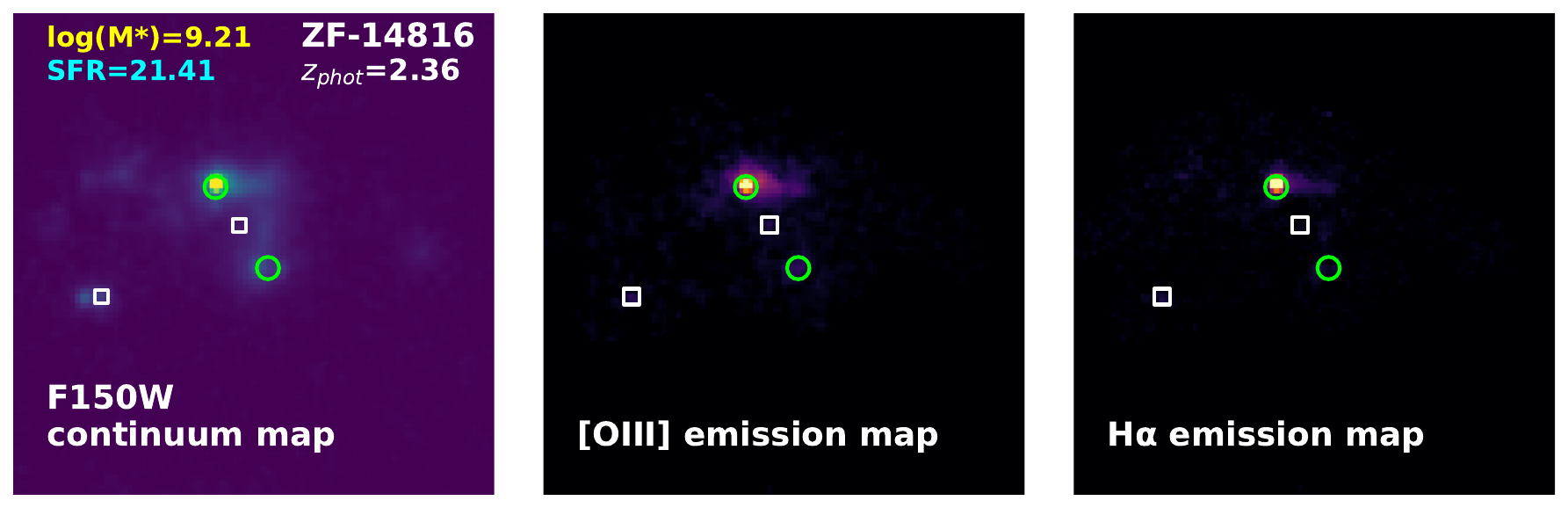}
    \hspace{0.1cm}
    \includegraphics[width=0.32\textwidth]{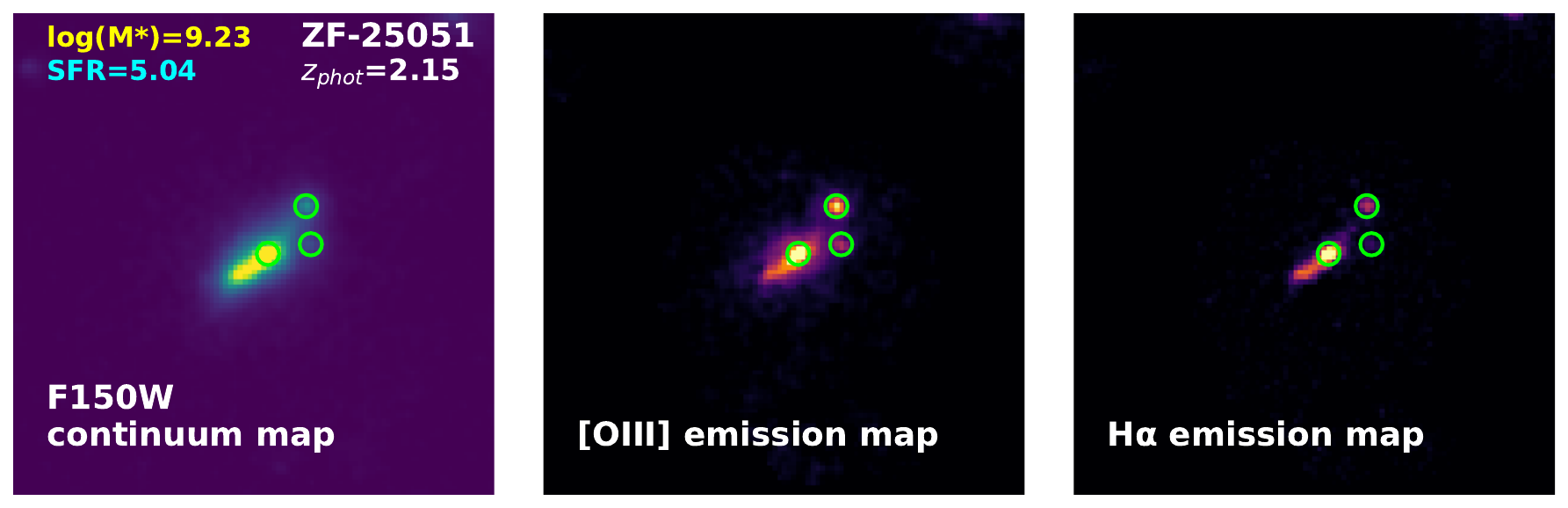}
    \hspace{0.1cm}
    \includegraphics[width=0.32\textwidth]{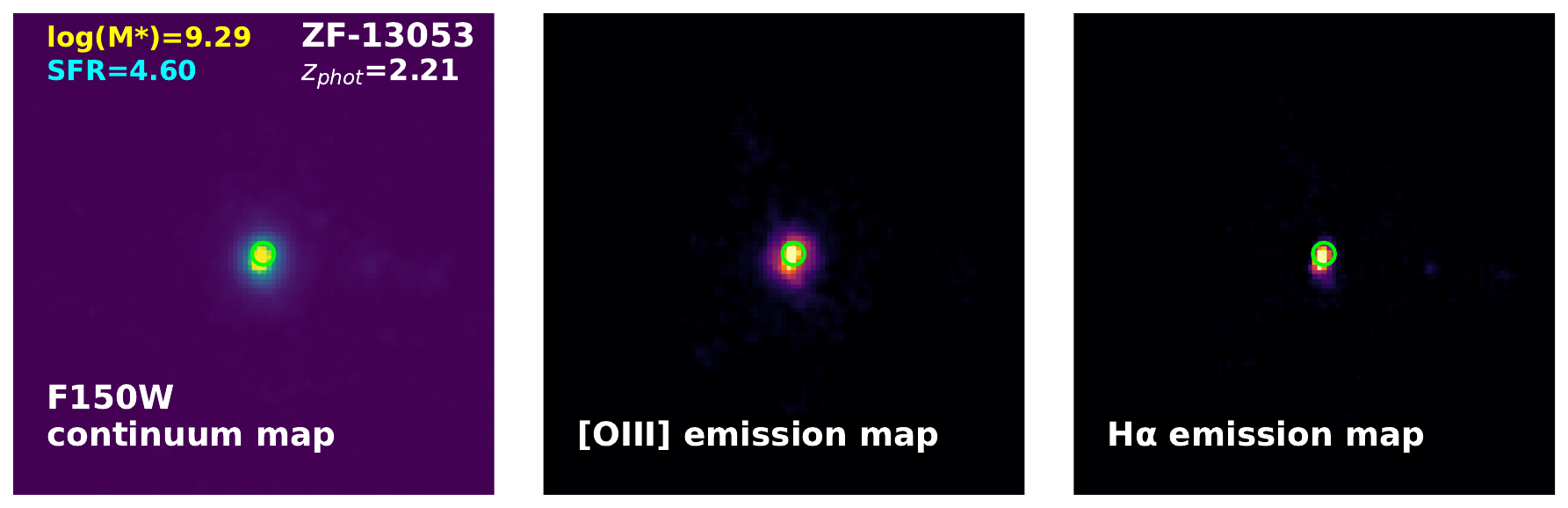}
    \hspace{0.1cm}
    \includegraphics[width=0.32\textwidth]{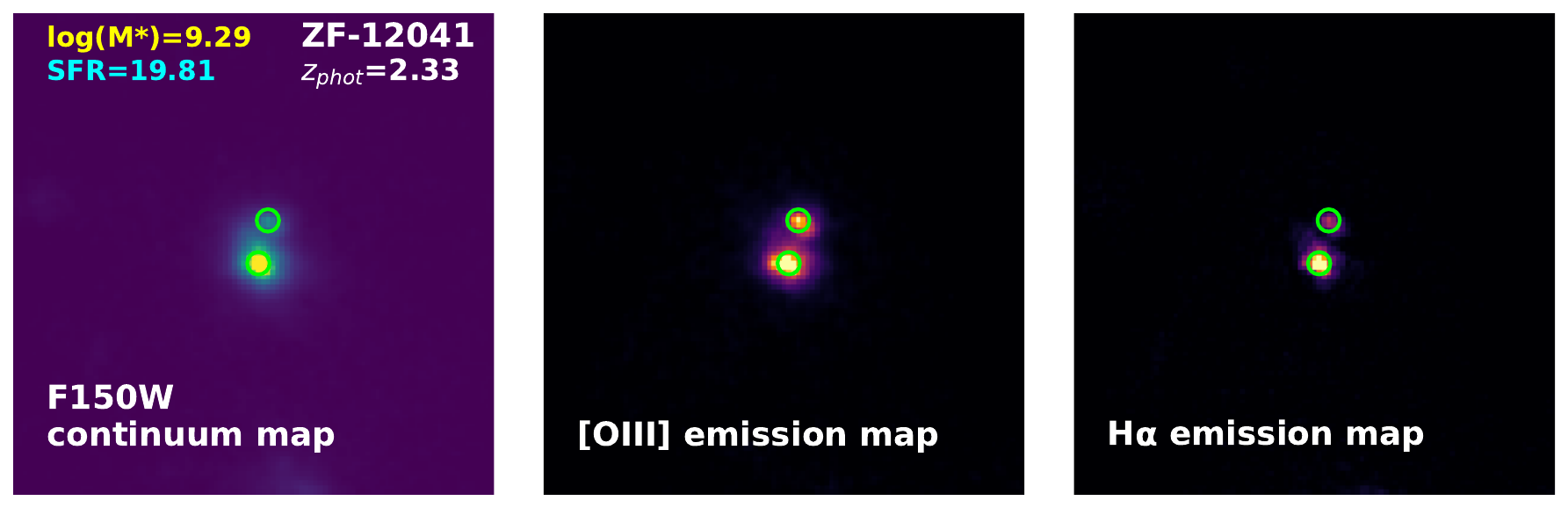}
    \hspace{0.1cm}
    \includegraphics[width=0.32\textwidth]{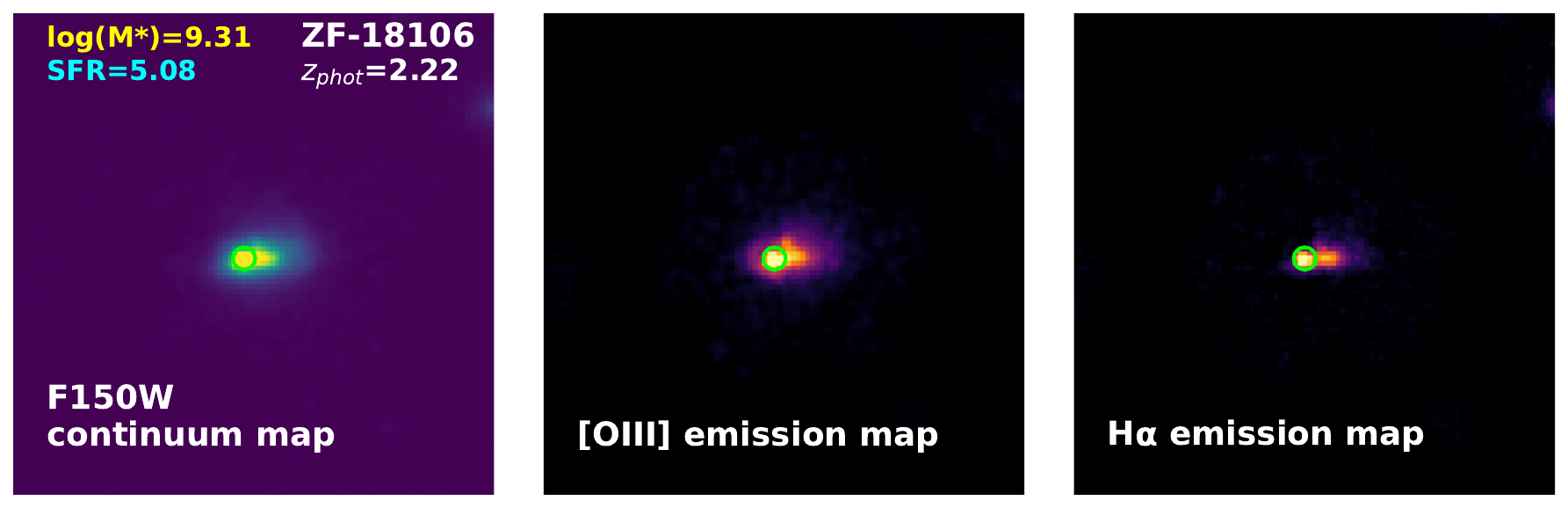}
    \hspace{0.1cm}
    \includegraphics[width=0.32\textwidth]{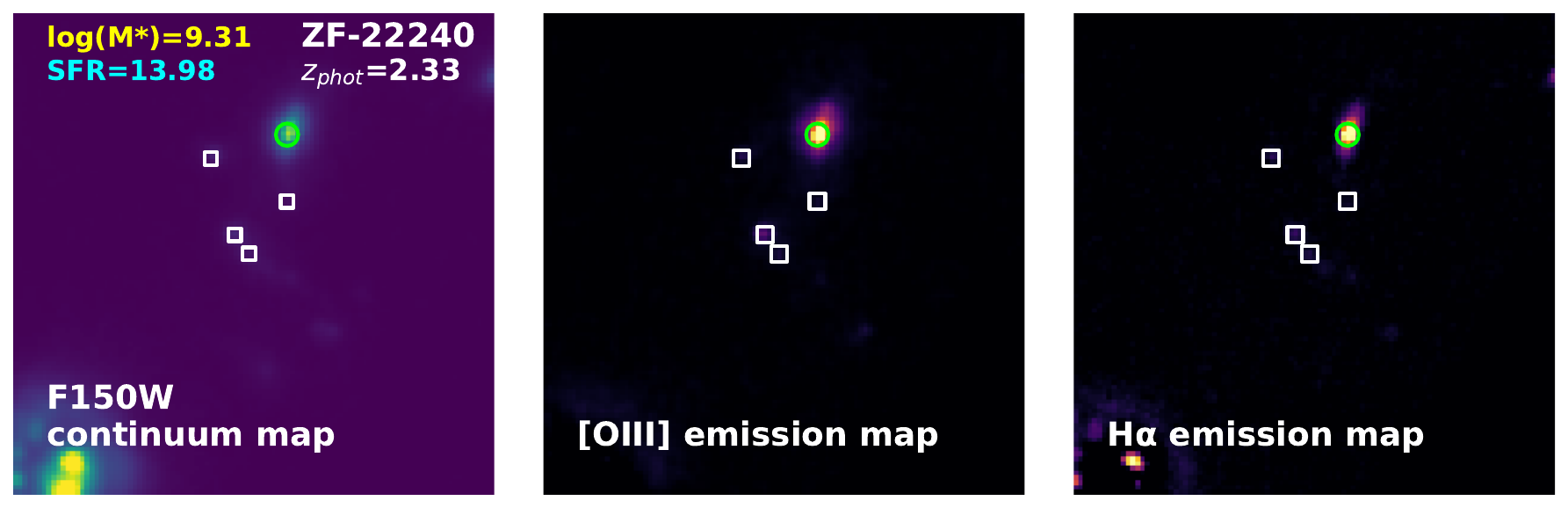}
    \hspace{0.1cm}
    \includegraphics[width=0.32\textwidth]{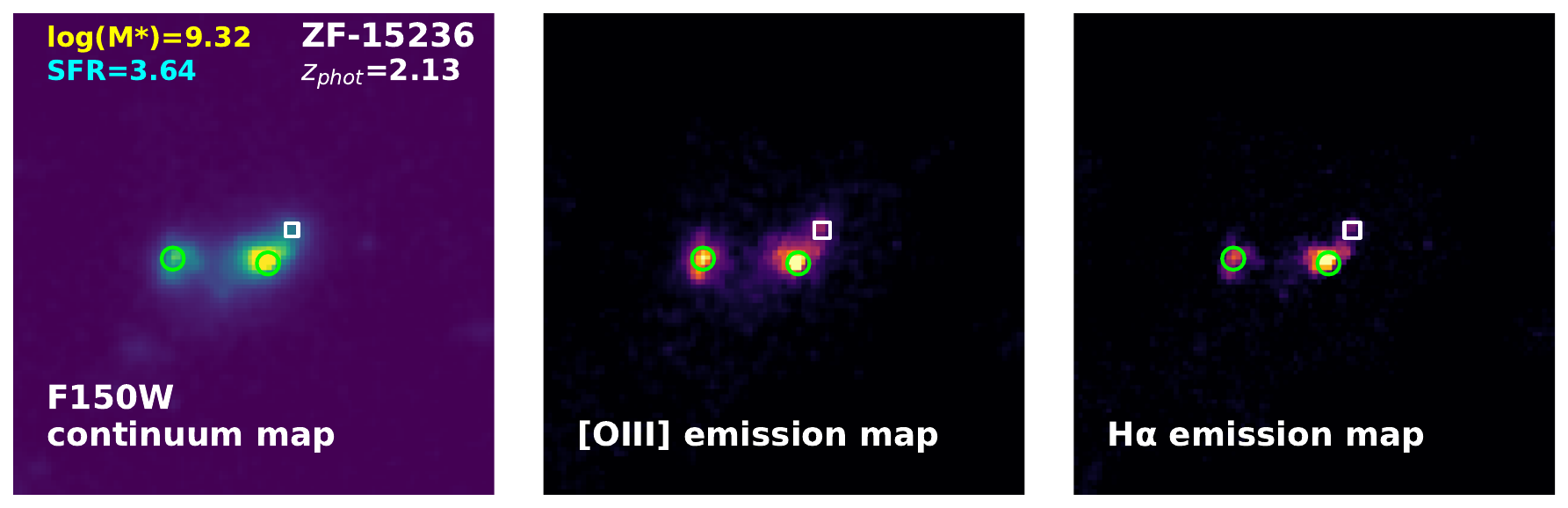}
    \hspace{0.1cm}
    \includegraphics[width=0.32\textwidth]{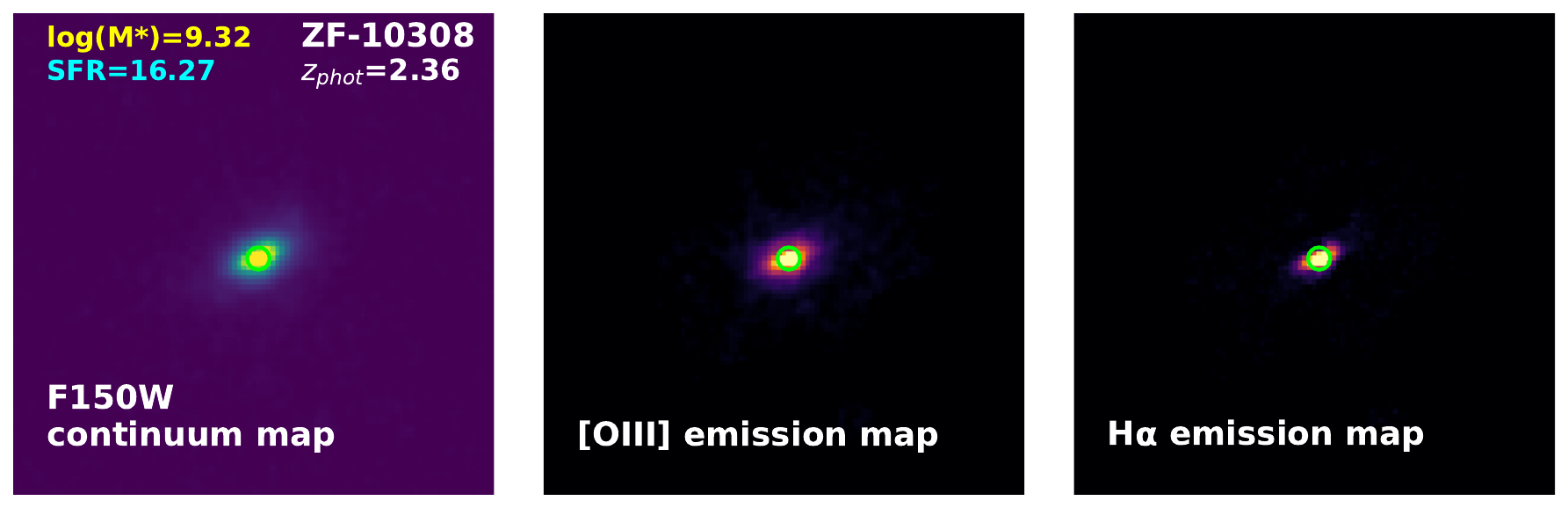}
    \hspace{0.1cm}
    \includegraphics[width=0.32\textwidth]{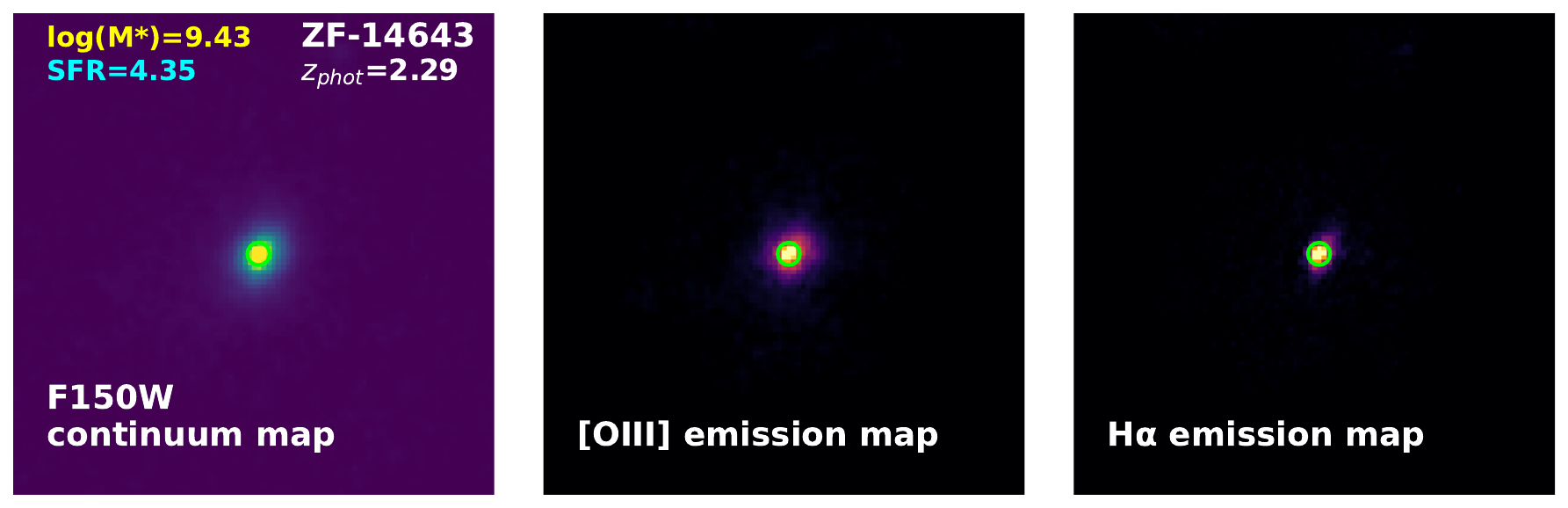}
    \hspace{0.1cm}
    \includegraphics[width=0.32\textwidth]{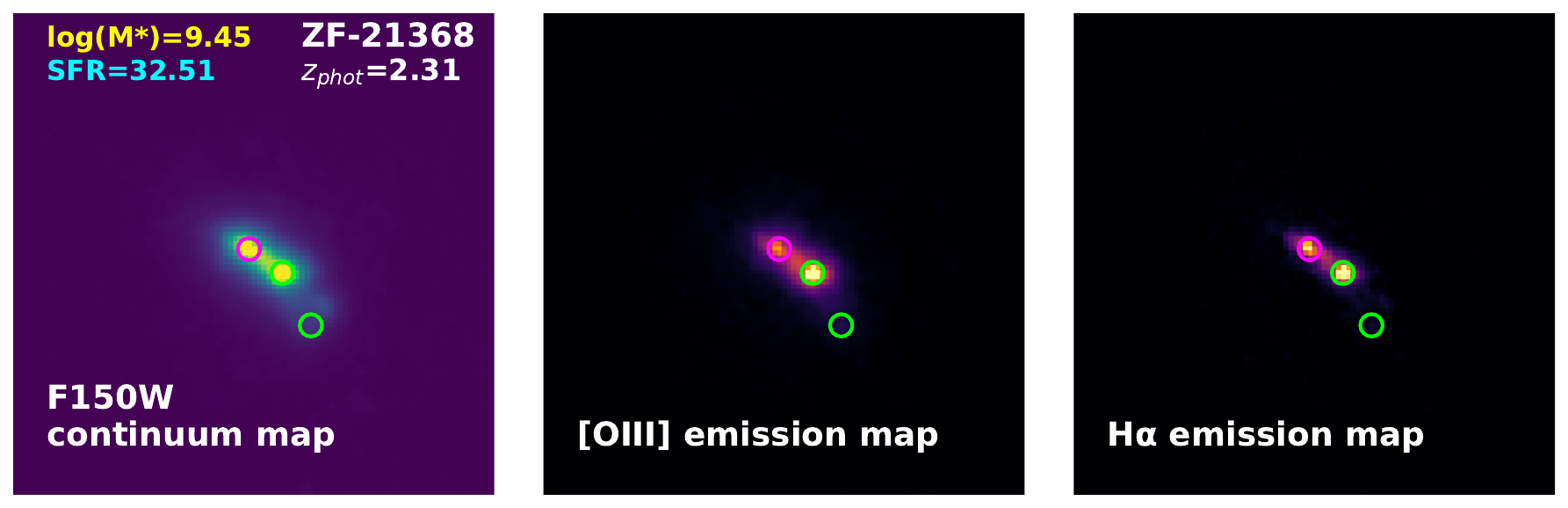}
    \hspace{0.1cm}
    \includegraphics[width=0.32\textwidth]{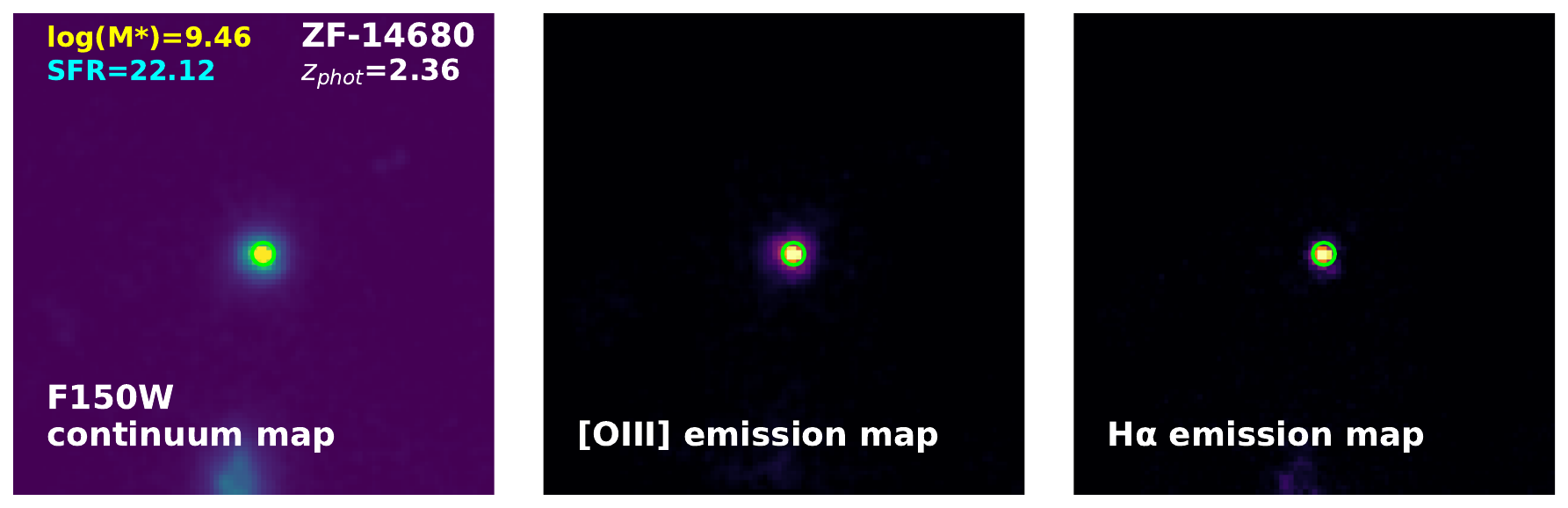}
    \hspace{0.1cm}
    \includegraphics[width=0.32\textwidth]{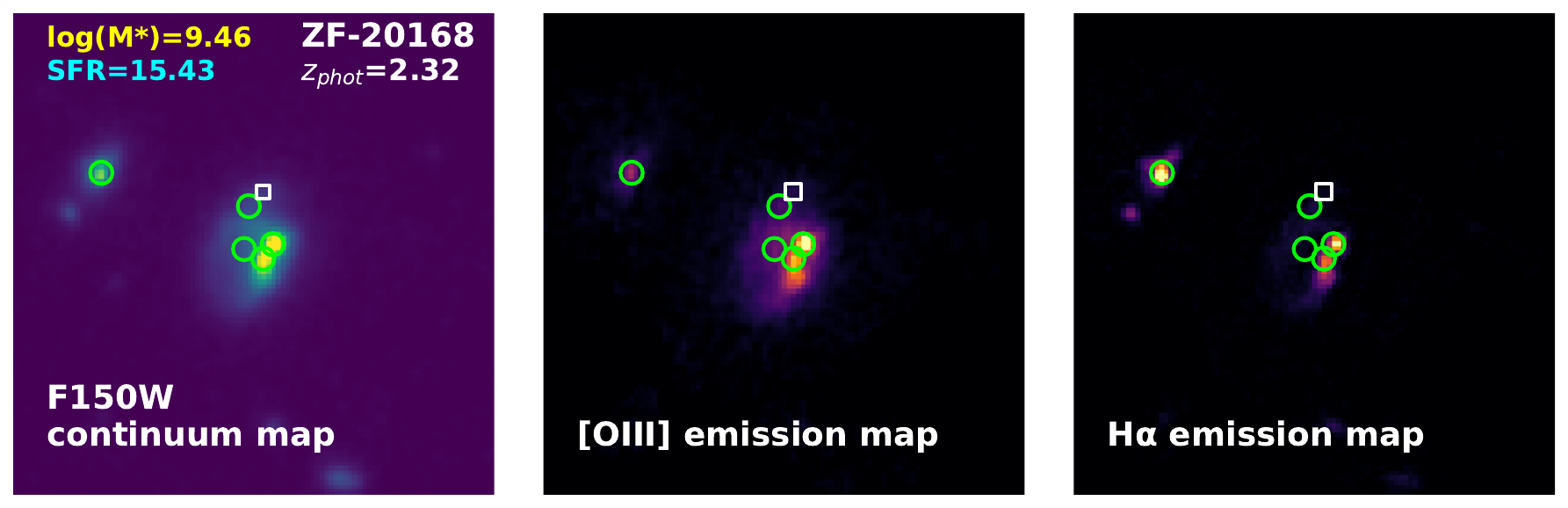}
    \hspace{0.1cm}
    \includegraphics[width=0.32\textwidth]{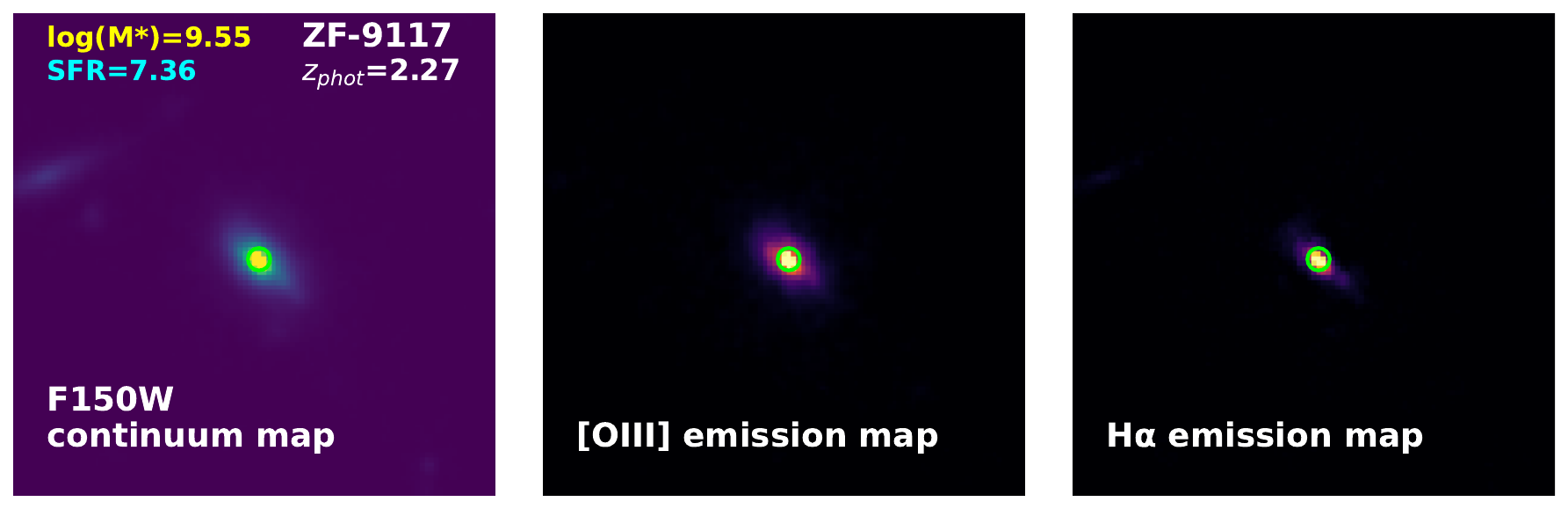}
    \hspace{0.1cm}
    \includegraphics[width=0.32\textwidth]{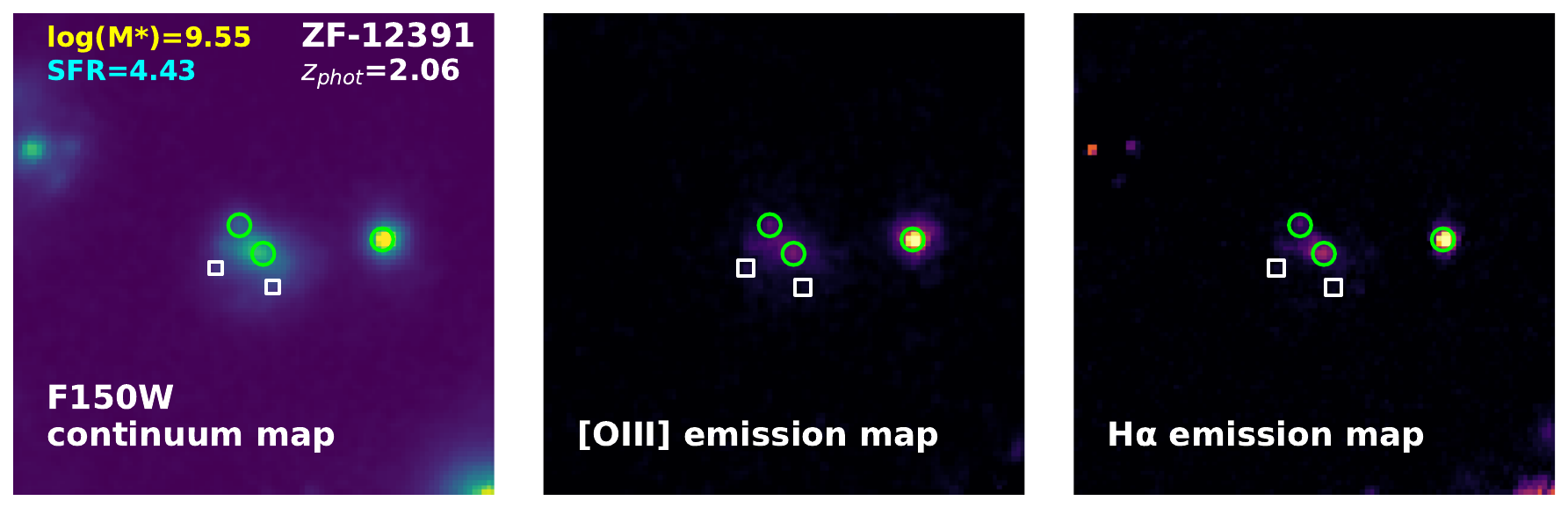}
    \hspace{0.1cm}
    \includegraphics[width=0.32\textwidth]{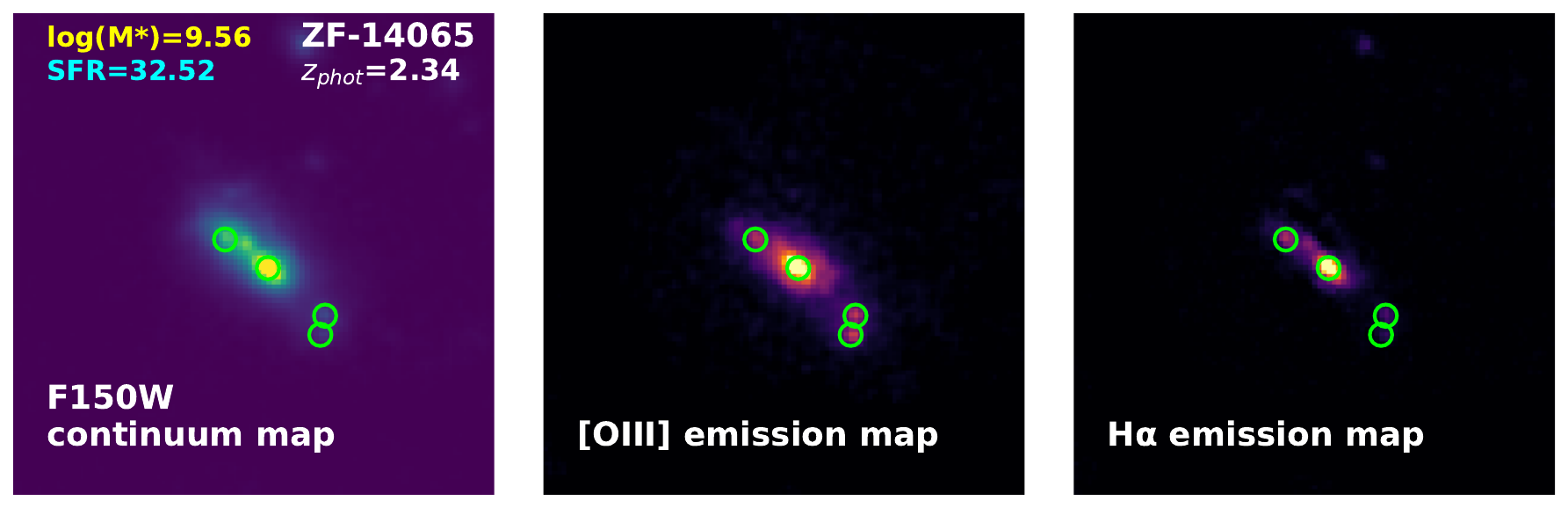}
    \hspace{0.1cm}
    \includegraphics[width=0.32\textwidth]{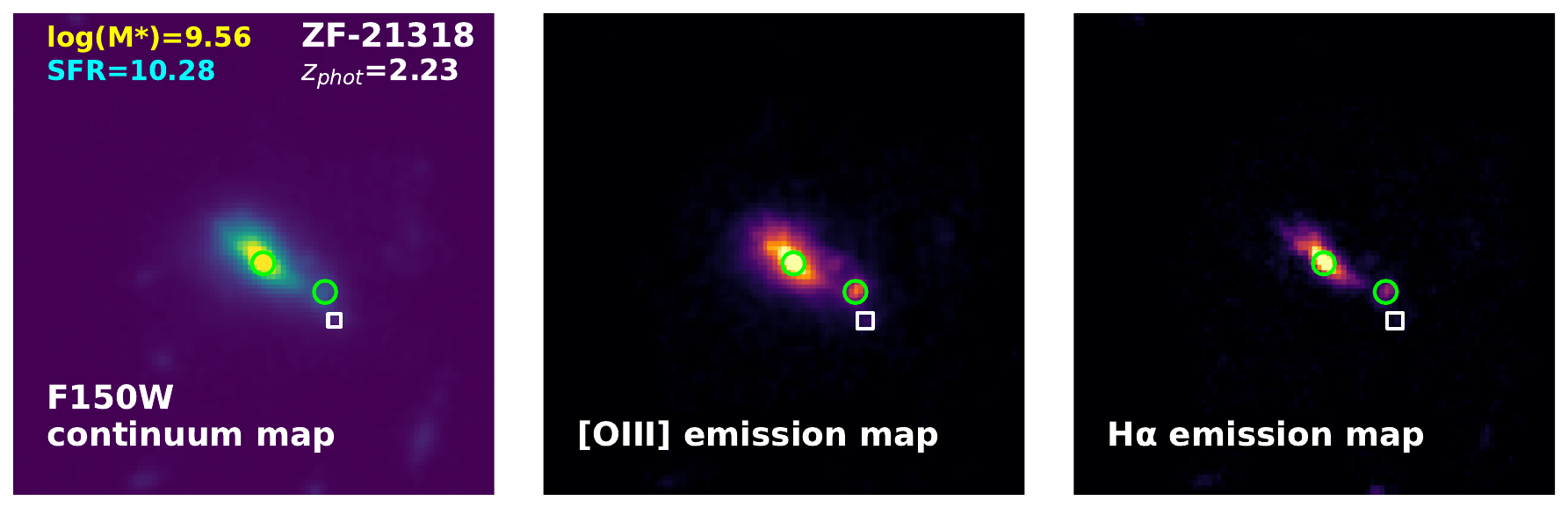}
    \hspace{0.1cm}
    \includegraphics[width=0.32\textwidth]{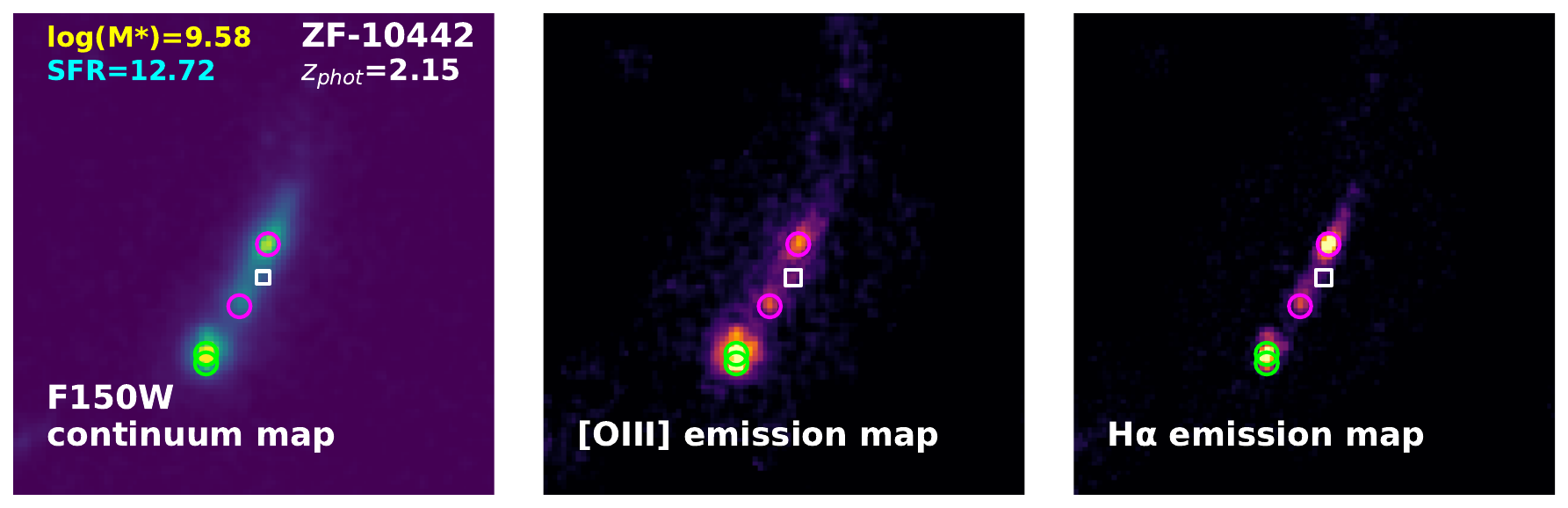}
    \hspace{0.1cm}
    \includegraphics[width=0.32\textwidth]{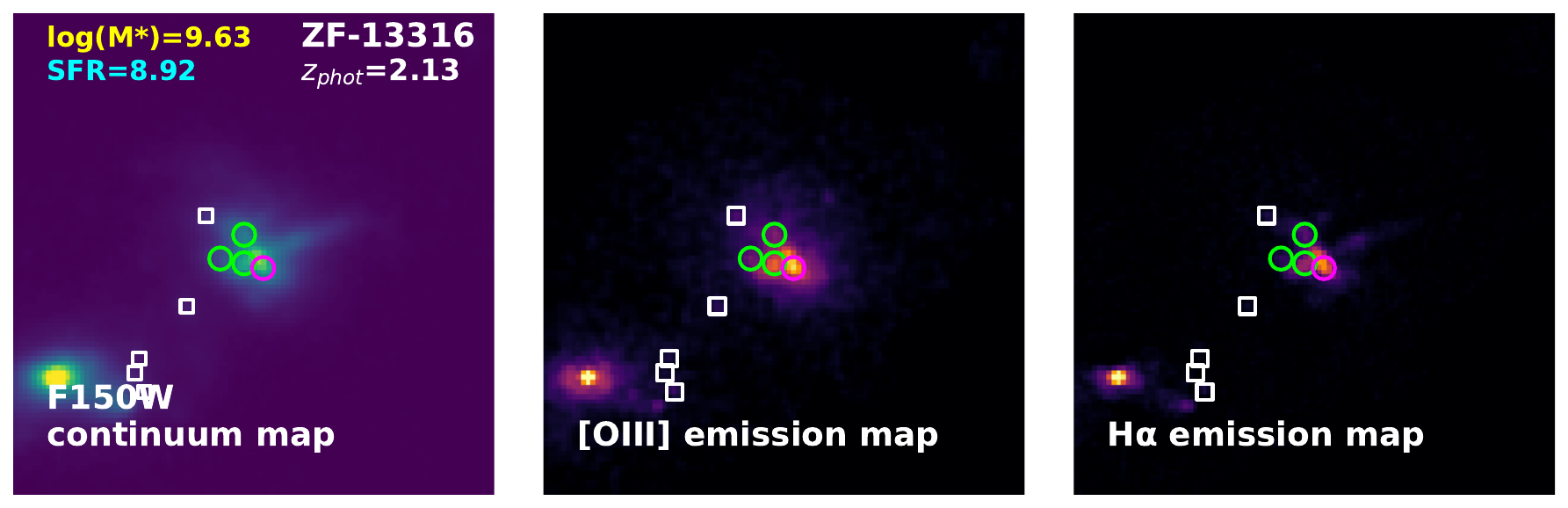}
    \hspace{0.1cm}
    \includegraphics[width=0.32\textwidth]{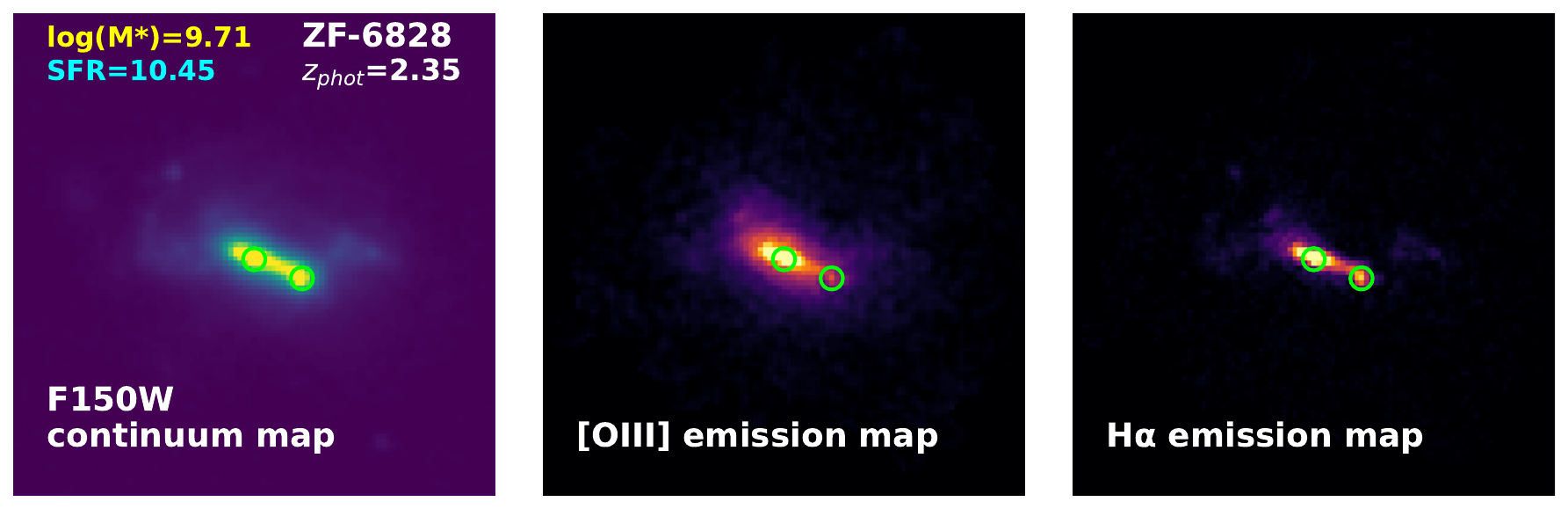}
    \hspace{0.1cm}
    \includegraphics[width=0.32\textwidth]{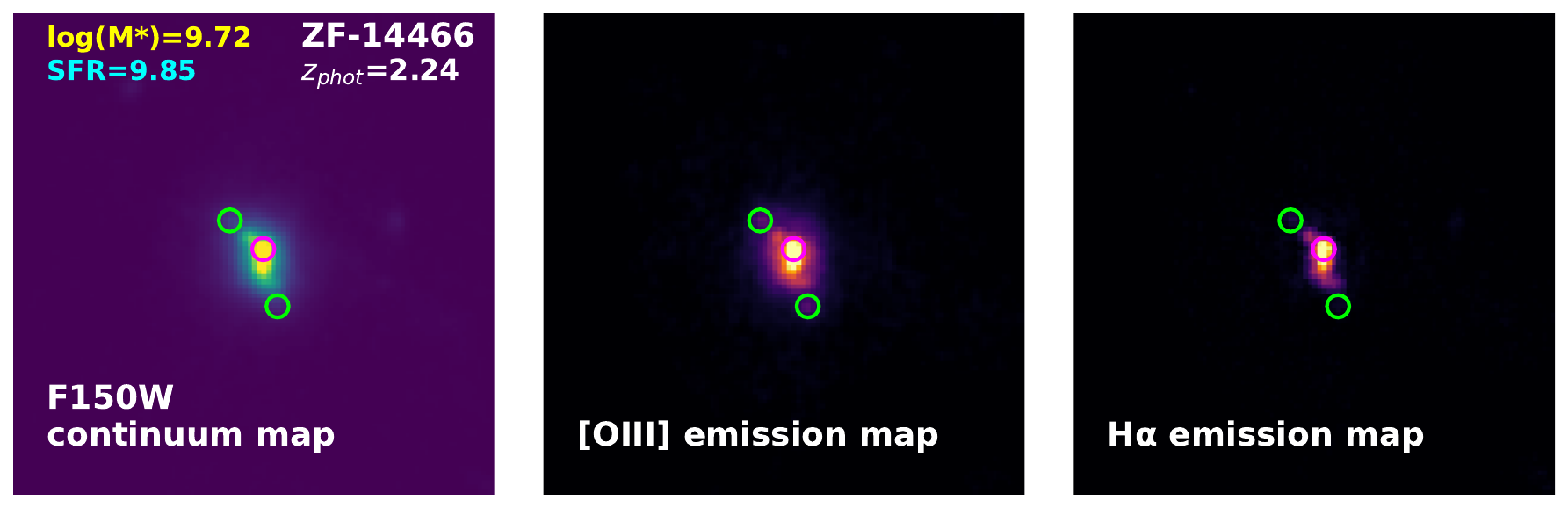}
    \hspace{0.1cm}
    \includegraphics[width=0.32\textwidth]{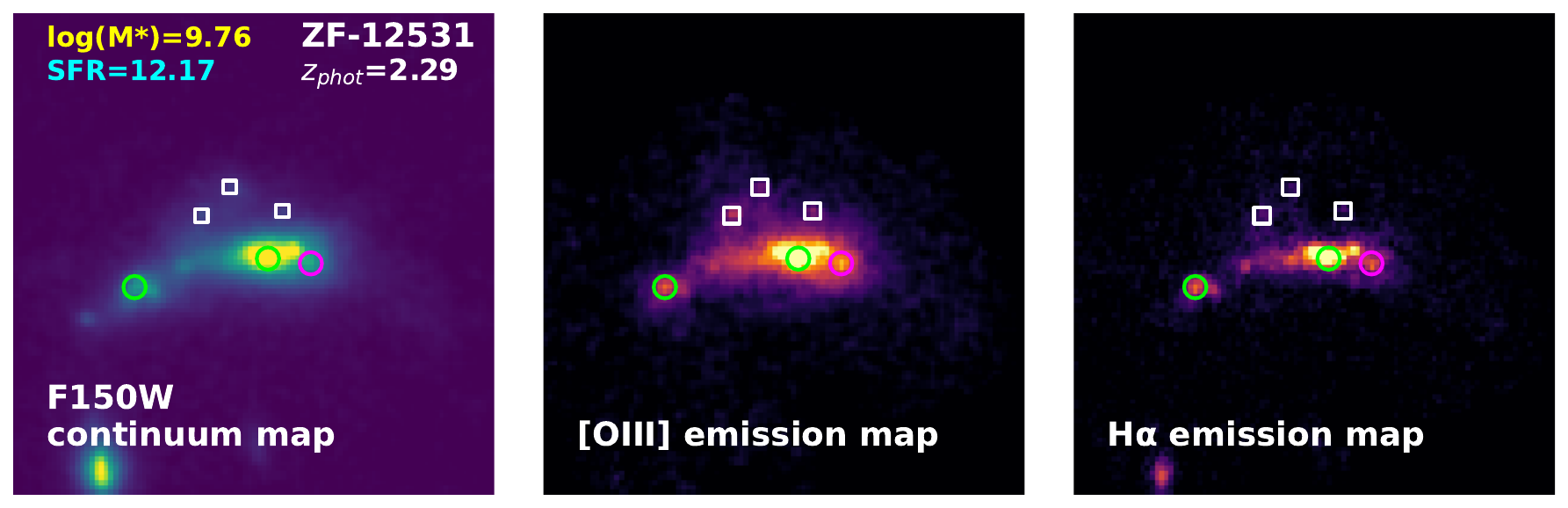}
    \hspace{0.1cm}
    \includegraphics[width=0.32\textwidth]{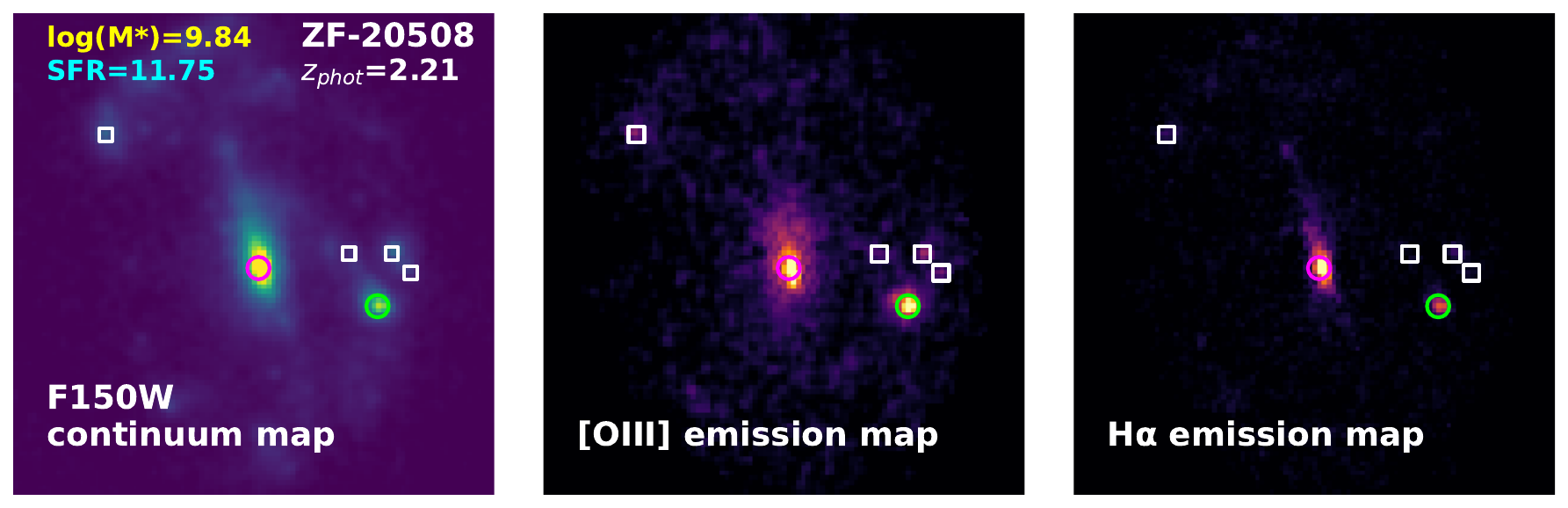}
    \hspace{0.1cm}
    \includegraphics[width=0.32\textwidth]{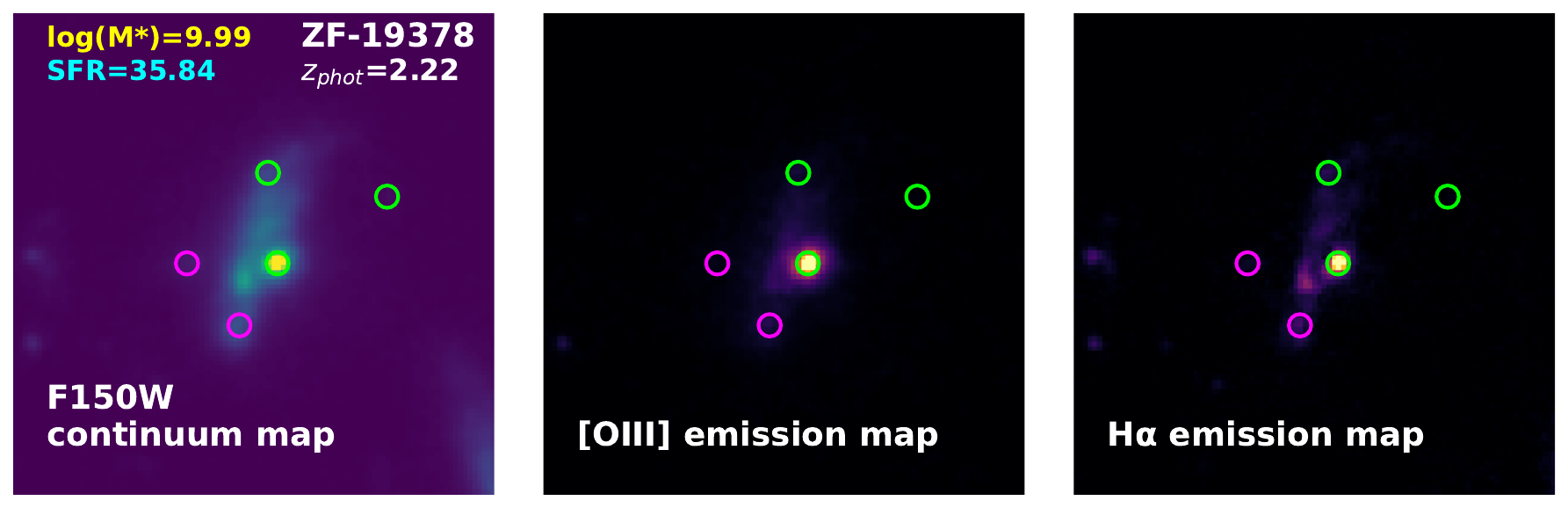}
    \hspace{5.9cm}
    \label{fig:emimap2}
        \caption{Continue of Figure \hyperref[fig:emimap1]{16}. Outlines as in Figure \hyperref[fig:emimap]{8}.}
\end{figure*}

In Figure \hyperref[fig:emimap]{8}, we present the stellar continuum map at $1.5\,\mu m$, the [O{\sc iii}]+$\mathrm{H\beta}$ emission line map, and the $\mathrm{H\alpha}$ emission line map of only six samples. Here we provide the extended versions in Figure \hyperref[fig:emimap1]{16} and Figure \hyperref[fig:emimap2]{17}, which include the results of all 68 HAEs that contain Green Seeds in this work.

\section*{\textbf{Appendix C\\The remaining UV clumps in combination with Green Seeds}}
\label{sec:appendixsample2}
\begin{figure*}[p]
    \centering
    \includegraphics[width=1\textwidth]{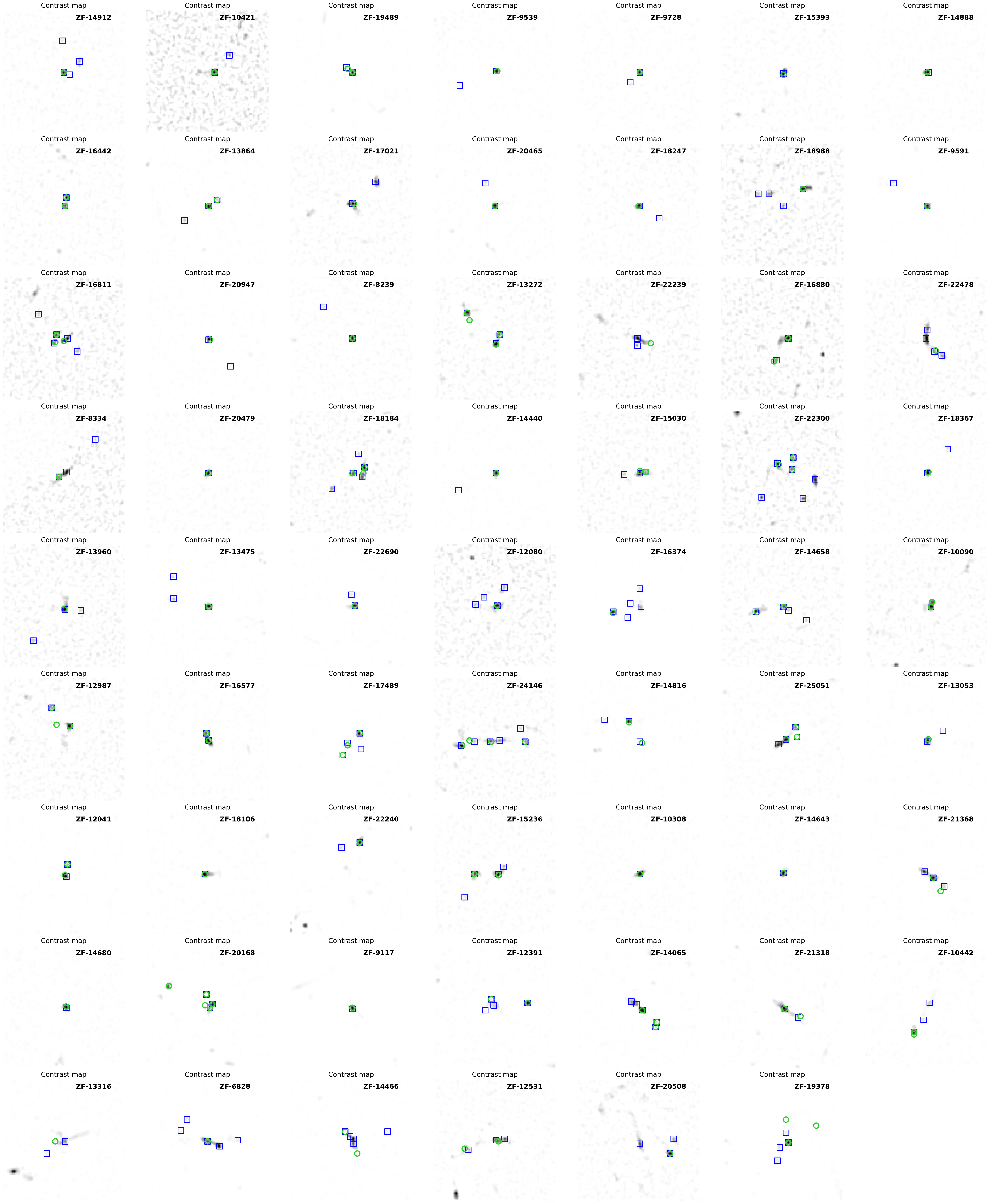}
    \label{fig:clump_other}
    \vspace{-0.2cm}
    \caption{The rest contrast maps and the UV star-forming clumps of HAEs which contain Green Seeds. Outlines as in Figure \hyperref[fig:clumpmap]{12}.}
    \vspace{0.4cm}
\end{figure*}

Similarly, in Figure \hyperref[fig:clumpmap]{12}, we illustrate the UV star-forming clumps of only six HAEs identical to the sample in Figure \hyperref[fig:emimap]{8}. Here we provide the extended versions in Figure \hyperref[fig:clump_other]{18}, which include the clump-detection of all 68 HAEs that contain Green Seeds in this work.

\bibliography{main}{}
\bibliographystyle{aasjournal}



\end{document}